\documentclass{aa}

\usepackage{natbib}
\bibpunct{(}{)}{;}{a}{}{,} 

\usepackage{placeins}
\usepackage{soul} 
\usepackage{float}
\usepackage{subfloat}
\usepackage{graphicx}
\usepackage{caption}
\DeclareCaptionFormat{cont}{#1 (cont.)#2#3\par} 
\usepackage{txfonts}
\usepackage{siunitx}  
\usepackage{color}
\usepackage{amsmath}
\usepackage[babel,german=quotes]{csquotes}
\usepackage{dblfloatfix}    
\usepackage{etoolbox}

\newcommand{\tablenote}[1]{\parbox{18.3cm}{\indent \footnotesize{#1}}}
\newcommand{\tablenotea}[1]{\parbox{  8.9cm}{\indent \footnotesize{#1}}}

\begin{document} 

\title{The abundance of S- and Si-bearing molecules in O-rich circumstellar envelopes of AGB stars\thanks{Based on observations carried out with the IRAM 30m Telescope. The Institut de Radioastronomie Millim\'etrique (IRAM) is supported by INSU/CNRS (France), MPG (Germany), and IGN (Spain).}}

\titlerunning{SiO, CS, SiS, SO, and SO$_2$ abundances in oxygen star envelopes}
\authorrunning{Massalkhi et al.}

\author{S.~Massalkhi\inst{1}, M.~Ag\'undez\inst{1}, J.~Cernicharo\inst{1} and L. Velilla-Prieto\inst{2}}

\institute{Instituto de F\'isica Fundamental, CSIC, C/ Serrano 123, E-28006, Madrid, Spain \and Dpt. of Space, Earth and Environment, Chalmers University of Technology, Onsala Space Observatory, 439 92 Onsala, Sweden}

\date{Received; accepted}


\abstract
{}
{We aim to determine the abundances of SiO, CS, SiS, SO, and SO$_2$ in a large sample of oxygen-rich asymptotic giant branch (AGB) envelopes covering a wide range of mass loss rates to investigate the potential role that these molecules could play in the formation of dust in these environments.}
{We surveyed a sample of 30 oxygen-rich AGB stars in the $\lambda$ 2 mm band using the IRAM 30m telescope. We performed excitation and radiative transfer calculations based on the large velocity gradient (LVG) method to model the observed lines of the molecules and to derive their fractional abundances in the observed envelopes.}
{We detected SiO in all 30 targeted envelopes, as well as CS, SiS, SO, and SO$_2$ in 18, 13, 26, and 19 sources, respectively. Remarkably, SiS is not detected in any envelope with a mass loss rate below $10^{-6}$ M$_{\odot}$ yr$^{-1}$, whereas it is detected in all envelopes with mass loss rates above that threshold. From a comparison with a previous, similar study on C-rich sources, it becomes evident that the fractional abundances of CS and SiS show a marked differentiation between C-rich and O-rich sources, being two orders of magnitude and one order of magnitude more abundant in C-rich sources, respectively, while the fractional abundance of SiO turns out to be insensitive to the C/O ratio. The abundance of SiO in O-rich envelopes behaves similarly to C-rich sources, that is, the denser the envelope the lower its abundance. A similar trend, albeit less clear than for SiO, is observed for SO in O-rich sources.}
{The marked dependence of CS and SiS abundances on the C/O ratio indicates that these two molecules form more efficiently in C- than O-rich envelopes. The decline in the abundance of SiO with increasing envelope density and the tentative one for SO indicate that SiO and possibly SO act as gas-phase precursors of dust in circumstellar envelopes around O-rich AGB stars.}


\maketitle

\section{Introduction}

When an evolved star with a mass lower than $\sim$ 8 M$_{\odot}$ is on the asymptotic giant branch (AGB), it experiences extensive mass loss up to rates of $\sim$ 10$^{-4}$ M$_{\odot}$ yr$^{-1}$ that dominates the evolution of that stage. AGB stars are considered the main providers of dust and enriched material to the interstellar medium \citep{geh1989}. The copious amount of material released gives rise to an expanding circumstellar envelope (CSE), which provides favorable thermodynamic conditions for the formation of simple molecules and dust grains. At the start of the AGB phase, the element mixture at the stellar photosphere has a carbon-to-oxygen ratio C/O $<1$, making the stars oxygen-rich (O-rich). In the CSE of these stars, O-bearing molecules, such as H$_2$O and SiO \citep{eng1979}, and S-bearing species, such as SO, SO$_2$, and H$_2$S \citep{omo1993}, are observed to be abundant. Dredge-up events experienced by the AGB star mix carbon from the interior helium-burning shell to the surface such that the C/O ratio becomes $>1$ and carbon-bearing molecules become abundant in the CSE. The synthesis of dust in AGB CSEs is evidenced by the identification of typical dust features in the spectral energy distribution (SED) of AGB stars, such as the \SI{9.7}{\micro\metre} and \SI{18}{\micro\metre} emission features of silicate dust in O-rich AGB stars and the \SI{11.3}{\micro\metre} feature of SiC dust in carbon stars. 

Despite the important role dust plays in many astrophysical phenomena, the mechanisms responsible for its formation and production are still poorly understood. In general, the main scenario for dust formation around an AGB star involves a two-step process. First, some gas-phase precursors condense to produce seed nuclei with sizes on the order of nanometers, which then grow by processes of coagulation and accretion to form a macroscopic dust particle. Yet this picture remains poorly constrained. In particular, how the transition between gas-phase molecules and solid phases occurs and which molecules act as precursors of seed nuclei are questions that have yet to be answered. 

Various observational studies have provided hints as to which molecules could act as precursors of dust in the circumstellar envelopes of evolved stars. Silicon monoxide, SiO, is known to be a candidate or precursor of dust. \cite{gon2003}, \cite{sch2006}, \cite{ram2009}, and \cite{mas2019} observed and modeled the SiO emission in the three chemical types of AGB stars: M-, S-, and C-type. Those studies found a trend of decreasing SiO abundance with increasing wind density, most notably for the O-rich and C-rich AGB stars, which is thought to be due to an increased depletion of SiO onto dust grains. Similar studies on SiS were less conclusive as to the role of this molecule as a gas-phase precursor of dust (\citealt{sch2007}, \citealt{dan2018}, \citealt{mas2019}). The molecules SiC$_2$ and CS were also found to show a similar behavior in C-rich CSEs as what was found for SiO, that is to say, an abundance decline with increasing envelope density, which suggests that these molecules are playing an important role in the formation of silicon carbide (SiC) and magnesium sulfide (MgS) dust in the envelopes of C-rich AGB stars, respectively \citep{mas2018,mas2019}. 

In this paper, we focus on potential gas-phase precursors of dust in O-rich AGB stars. Some metal oxides recently detected have been suggested to act as precursors of seed nuclei, for example, TiO and TiO$_2$ \citep{gai1998} and AlO \citep{gob2016}. However, observational constraints are still not conclusive \citep{ban2012,kam2013,kam2016,kam2017,dec2017,deb2017}. The formation of the first condensation nuclei must necessarily occur from gas-phase species present in the precondensation region, and the bulk of dust must be formed at the expense of gaseous species during the phase of grain growth. Since the gas around AGB stars is largely molecular, molecules are good candidates to serve as precursors of dust. Previous observational studies done on large samples of O-rich AGB stars to investigate potential precursors of dust are meagre. To investigate which gas-phase molecules could play a role in the formation of dust around O-rich AGB stars, in this paper we carry out a study of the abundance of five molecules, SiO, CS, SiS, SO, and SO$_2$, in 30 oxygen-rich AGB stars. In Sec.~\ref{sec:sample}, we outline the sample. In Sec.~\ref{sec:observations} we describe the observations carried out and in Sec.~\ref{sec:observation_results} we present the main results from the observations. In Sec.~\ref{sec:model}, we describe the radiative transfer model, the molecular data, and the procedure adopted for the derivation of the molecular abundances. In Sec.~\ref{sec:results_model} we describe the results from the radiative transfer model and comment on a few peculiar cases that stood out during the modeling. Finally, in Sec.~\ref{sec:discussion} we discuss the main results of our study and present our conclusions in Sec.~\ref{sec:conclusion}. 

\section{The sample}\label{sec:sample}

\begin{table*}
\caption{Sample of oxygen stars}\label{table:parameters}
\centering
\resizebox{\linewidth}{!}{
\begin{tabular}{lrrcrcrccccc}
\hline \hline
\multicolumn{1}{l}{Name} & \multicolumn{1}{c}{R.A.} & \multicolumn{1}{c}{Decl.} & \multicolumn{1}{c}{$V_{\rm LSR}$} & \multicolumn{1}{c}{D} & \multicolumn{1}{c}{$T_{\star}$} & \multicolumn{1}{c}{$L_{\star}$} & \multicolumn{1}{c}{$\dot{M}$} & \multicolumn{1}{c}{$V_{\rm exp}$} &   \multicolumn{1}{c}{$T_{\rm d}(r_c)$} & \multicolumn{1}{c}{$r_c$} & $\Psi$ \\
\multicolumn{1}{c}{}          & \multicolumn{1}{c}{J2000.0} & \multicolumn{1}{c}{J2000.0} & \multicolumn{1}{c}{(km~s$^{-1}$)} & \multicolumn{1}{c}{(pc)} & \multicolumn{1}{c}{(K)} & \multicolumn{1}{c}{(L$_{\odot}$)}  & \multicolumn{1}{c}{(M$_{\odot}$ yr$^{-1}$)} & \multicolumn{1}{c}{(km~s$^{-1}$)} &\multicolumn{1}{c}{(K)} & \multicolumn {1}{c}{(cm)} &   \\
\hline                                                               
IK\,Tau       & 03:53:28.87 & +11:24:21.7    &  +34.5 $^e$ &  285 $^a$    & 2100 $^b$     & 9250  $^{b*}$  &  $2.4\times10^{-5}$ $^{b*}$  &  17.5 $^e$  &  1000 $^b$       &    $1.8\times10^{14}$  $^b$     &  435 $^i$  \\
KU\,And       & 00:06:52.94 & +43:05:00.0    &  $-$22  &  680 $^c$    & 2000 $^c$     & 11800 $^c$     &  $9.4\times10^{-6}$ $^d$     &  19.5   &  1100 $^c$       &    $1.5\times10^{14}$  $^c$     &  200 $^m$   \\
RX\,Boo       & 14:24:11.63 & +25:42:13.4    &  +1.5   &  128 $^a$    & 1800 $^c$     & 4550  $^{c*}$  &  $6.1\times10^{-7}$ $^{b*}$  &  7.5    &  900  $^c$       &    $1.5\times10^{14}$  $^c$     &  144 $^x$ \\
RT\,Vir       & 13:02:37.98 & +05:11:08.4    & +18.5   &  226 $^l$    & 2000 $^b$     & 4500 $^b$      &  $4.5\times10^{-7}$ $^b$     &   7     &  1000 $^b$       &    $1.6\times10^{14}$  $^b$     & 2000 $^q$ \\
R\,Leo        & 09:47:33.49 & +11:25:43.7    &  +0.1   &  71  $^a$    & 2000 $^b$     & 2500  $^b$     &  $1.0\times10^{-7}$ $^b$     &   5     &  1200 $^b$       &    $1.3\times10^{14}$  $^b$     & 167 $^m$  \\
WX\,Psc       & 01:06:25.98 & +12:35:53.1    &  +9.5   &  700 $^b$    & 1800 $^b$     & 10300 $^b$     &  $4.0\times10^{-5}$ $^b$     &   19    &  800  $^b$       &    $3.2\times10^{14}$  $^b$     & 250 $^m$  \\
GX\,Mon       & 06:52:46.91 & +08:25:19.0    & $-$9.5  &  416 $^a$    & 2600 $^c$     & 4700  $^{c*}$  &  $4.9\times10^{-6}$ $^{d*}$  &   18    &  900  $^c$       &    $1.1\times10^{14}$  $^c$     & 200 $^m$ \\
NV\,Aur       & 05:11:19.44 & +52:52:33.2    & +3      & 1200 $^c$    & 2000 $^c$     & 9800  $^c$     &  $2.5\times10^{-5}$ $^d$     &   17.5  &  1100 $^c$       &    $1.7\times10^{14}$  $^c$     & 1000 $^m$   \\
V1111\,Oph    & 18:37:19.26 & +10:25:42.2    & $-$31   &  357 $^a$    & 1800 $^b$     & 2300  $^{b*}$  &  $2.7\times10^{-6}$ $^{d*}$  &   15.5  &  800  $^b$       &    $2.7\times10^{14}$  $^b$     & 200 $^m$ \\
RR\,Aql       & 19:57:36.06 & $-$01:53:11.3  &  +28   &  318 $^a$    & 2000 $^c$     & 2800  $^{c*}$  &  $8.6\times10^{-7}$ $^{d*}$  &   8.5   &  1500 $^c$       &    $5.9\times10^{13}$  $^c$     & 185 $^t$  \\
R\,LMi        & 09:45:34.28 & +34:30:42.8    & +0.9    &  330 $^d$    & 2400 $^d$     & 5500  $^d$     &  $2.6\times10^{-7}$ $^d$     &   5.5   &  1000 $^d$       &    $1.7\times10^{14}$  $^d$     & 115 $^t$   \\
BX\,Cam       & 05:46:44.10 & +69:58:25.2    &  $-$1   &  244 $^a$    & 2800 $^c$     & 1800  $^{c*}$  &  $1.0\times10^{-6}$ $^{d*}$  &   17    &  1500 $^c$       &    $7.1\times10^{13}$  $^c$     &  300 $^z$    \\
V1300\,Aql    & 20:10:27.87 & $-$06:16:13.6  & $-$17.5 &  620 $^c$    & 2000 $^c$     & 10600 $^c$     &  $1.0\times10^{-5}$  $^d$    &   15    &  1100 $^c$       &    $1.8\times10^{14}$  $^c$     & 1000 $^m$  \\
R\,Cas        & 23:58:24.87 & +51:23:19.7    &  +26.5  &  188 $^a$    & 1800 $^e$     & 10400 $^{e*}$  &  $9.5\times10^{-7}$ $^{e*}$  &   7.5   &  1050 $^e$       &    $2.5\times10^{14}$  $^e$     & 91 $^m$  \\
IRC\,$-$30398 & 18:59:13.85 & $-$29:50:20.4  & $-$7.5  &  390 $^m$    & 2000 $^m$     & 8700  $^m$     &  $6.0\times10^{-6}$ $^m$     &   14.5  &  800  $^b$       &    $2.6\times10^{14}$  $^b$     & 200 $^m$  \\
TX\,Cam       & 05:00:50.40 & +56:10:52.6    &  +11.5  &  334 $^a$    & 2600 $^c$     & 6600  $^{c*}$  &  $7.7\times10^{-6}$ $^{c*}$  &   17.5  &  1300 $^c$       &    $1.0\times10^{14}$  $^c$     &  500 $^x$  \\
S\,CrB        & 15:21:23.96 & +31:22:02.6    &  +1.5   &  431 $^a$    & 2400 $^d$     & 6300  $^{d*}$  &  $2.7\times10^{-7}$ $^{d*}$  &   5     &  1000 $^d$       &    $1.7\times10^{14}$  $^d$     &  300 $^z$   \\
IRC\,+60169   & 06:34:34.88 & +60:56:33.2    &  $-$22  &  510 $^a$    & 2200 $^c$     & 5900  $^{c*}$  &  $9.6\times10^{-6}$ $^{c*}$  &   15    &  1000 $^{c}$     &    $1.1\times10^{14}$  $^c$     &  300 $^z$    \\
R\,Hya        & 13:29:42.78 & $-$23:16:52.8  & $-$10 $^o$   &  224 $^a$    & 2600 $^c$     & 17200 $^{c*}$  &  $4.7\times10^{-7}$ $^{c*}$  &   5 $^o$    &  1500 $^{c}$     &    $6.1\times10^{13}$  $^c$     & 200 $^m$ \\
R\,Crt        & 11:00:33.85 & -18:19:29.6    & +11.5   &  236 $^a$    & 2800 $^c$     & 7700  $^{c*}$  &  $1.0\times10^{-6}$ $^{c*}$  &   11    &  600  $^{c}$     &    $3.5\times10^{14}$  $^c$     & 333 $^q$  \\
\textit{o}\,Ceti & 02:19:20.79 & $-$02:58:39.5  & +47  &  107 $^f$    & 3000 $^g$     & 9000  $^g$     &  $2.0\times10^{-7}$ $^h$     &   3     &  1000 $^{z}$     &    $9.7\times10^{13}$  $^y$     &  195 $^t$ \\
W\,Hya        & 13:49:02.00 & $-$28:22:03.5  &  +40.5  &  164 $^a$    & 2600 $^b$     & 16800 $^{b*}$  &  $4.2\times10^{-7}$ $^{b*}$  &   6     &  1200  $^b$      &   $6.3\times10^{13}$  $^b$      & 500 $^v$ \\
T\,Cep        & 21:09:31.78 & +68:29:27.2    &  $-$2.5 &  176 $^a$    & 2400 $^d$     & 4900 $^{d*}$   &  $7.8\times10^{-8}$ $^{d*}$  &   4     &  1000  $^d$      &   $1.8\times10^{14}$  $^d$      &  300 $^z$    \\
V1943\,Sgr    & 20:06:55.24 & $-$27:13:29.8  & $-$14.5 &  666 $^a$    & 2200 $^d$     & 55400 $^{d*}$  &  $1.0\times10^{-6}$ $^{d*}$  &   4.5   &  1000  $^d$      &   $1.6\times10^{14}$  $^d$      &  300 $^z$     \\ 
SW\,Vir       & 13:14:04.39 & $-$02:48:25.2  & $-$10.5 &  300 $^a$    & 2400 $^b$     & 17600 $^{b*}$  &  $2.2\times10^{-6}$ $^{b*}$  &   7.5   &  800   $^b$      &   $2.9\times10^{14}$  $^b$      & 1000  $^q$ \\
AFGL\,292     & 02:02:38.63 & +07:40:36.5    &  +23.7  &  253 $^a$    & 2200 $^d$     &  6000 $^d$     &  $1.3\times10^{-7}$ $^{d*}$  &   7     &  1000  $^d$      &   $1.8\times10^{14}$  $^d$      & 300 $^z$   \\ 
BK\,Vir       & 12:30:21.01 & +04:24:59.2    &  +17.5  &  234 $^a$    & 3000 $^n$     & 4500 $^{n*}$   & $2.3\times10^{-7}$ $^{m*}$   &   4     &  1000 $^{z}$     &   $8.6\times10^{13}$  $^z$      & 2000 $^q$   \\ 

\hline
\end{tabular}                                                                                               
}
\tablenote{\\
References and notes: The coordinates of the O-rich stars are taken from the literature. An asterisk in the value of the luminosity ($L_{\star}$) or mass loss rate ($\dot{M}$) indicates that the value has been scaled according to the updated value of the distance. $^a$~\cite{gaia2018}, $^b$~\cite{ram2014}, $^c$~\cite{sch2013}, $^d$~\cite{dan2015}, $^e$~\cite{mae2016}, $^f$~\cite{kna2003}, $^g$~\cite{woo2004},$^h$~\cite{ryd2001}, $^i$~\cite{gob2016}, $^k$~\cite{jus1996}, $^l$~\cite{zha2017}, $^m$~\cite{gon2003}, $^n$~\cite{ohn2011}, $^o$~\cite{kna1998}, $^p$~\cite{deb2010}, $^q$~\cite{olo2002}, $^r$~\cite{dyc1996}, $^s$~\cite{win2007}, $^t$~\cite{gro1999}, $^v$~\cite{kho2014}, $^x$~\cite{dha2018}, $^w$~\cite{gar2006} , $^y$~\cite{kam2016} $^{[z]}$ Assumed value for the condensation radius $r_{c}$ is 5 $R_{\star}$, for the dust temperature at the condensation radius $T_{\rm d}(r_{\rm c})$ is 1000 K, and for the gas-to-dust mass ratio $\Psi$ is 300.}
\end{table*}

\begin{table*}
\caption{Peculiar sources}\label{table:peculiar}
\centering
\resizebox{\linewidth}{!}{
\begin{tabular}{lrrccrcrccccc}
\hline \hline
\multicolumn{1}{l}{Name} & \multicolumn{1}{c}{R.A.} & \multicolumn{1}{c}{Decl.} & comp. & \multicolumn{1}{c}{$V_{\rm LSR}$} & \multicolumn{1}{c}{D} & \multicolumn{1}{c}{$T_{\star}$} & \multicolumn{1}{c}{$L_{\star}$} & \multicolumn{1}{c}{$\dot{M}$} & \multicolumn{1}{c}{$V_{\rm exp}$} &   \multicolumn{1}{c}{$T_{\rm d}(r_c)$} & \multicolumn{1}{c}{$r_c$} & $\Psi$ \\
\multicolumn{1}{c}{}          & \multicolumn{1}{c}{J2000.0} & \multicolumn{1}{c}{J2000.0} & & \multicolumn{1}{c}{(km~s$^{-1}$)} & \multicolumn{1}{c}{(pc)} & \multicolumn{1}{c}{(K)} & \multicolumn{1}{c}{(L$_{\odot}$)}  & \multicolumn{1}{c}{(M$_{\odot}$ yr$^{-1}$)} & \multicolumn{1}{c}{(km~s$^{-1}$)} &\multicolumn{1}{c}{(K)} & \multicolumn {1}{c}{(cm)} &   \\
\hline                                                                                                                             
Ep\,Aqr & 21:46:31.85 & $-$02:12:45.9 & Narrow  & $-$33.5 &  124 $^a$    & 3200 $^s$     & 4100$^{s*}$    & $1.7\times10^{-8}$ $^{s*}$   & 1 $^q$   &  1000 $^{z}$   &   $7.2\times10^{13}$ $^z$   &  860 $^x$ \\
        & &  & Broad   &  &     &      &   & $5.0\times10^{-7}$ $^{s*}$   & 9.2 $^q$ &     &     &  \\
\hline
X\,Her  & 16:02:39.17 & +47:14:25.3  & Narrow  &  $-$73  &  145 $^a$    & 3300 $^r$     & 5100 $^{w*}$    & $4.3\times10^{-8}$ $^{m*}$  &   2.2 $^m$ &  1000 $^{z}$  &  $6.7\times10^{13}$ $^z$    &  500 $^q$\\
        &  &  & Broad   &    &     &     &     & $1.6\times10^{-7}$ $^{m*}$  &   6.5 $^m$ &   &   &  \\
\hline
OH\,26.5+0.6  & 18:37:32.51 & $-$05:23:59.2  & AGB wind & +27     &  1370 $^k$   & 2200 $^k$     & 14000 $^k$     & $1.0\times10^{-6}$ $^k$      &   15.4 $^k$    &  1000 $^k$       &  $4.5\times10^{14}$ $^k$    & 278  $^k$    \\
              &  &   & Superwind &    &     &     &  & $5.5\times10^{-4}$ $^k$      &      &        &     &     \\ 
              \hline
\end{tabular}                                                                                               
}
\tablenote{\\
References in Table~\ref{table:parameters}}
\end{table*}

The sample contains 30 O-rich AGB stars, among which there are Mira variables (M), characterized by regular variations with a large amplitude ($>$ 2.5 mag in the V band), and semiregular variables (SR), characterized by a small amplitude ($<$ 2.5 mag in the V band). We selected sources from samples in the literature (e.g., \citealt{sch2013, ram2014, dan2015}) mainly based on strong line emission of molecules like CO, SiO, and SO. The sample was also chosen to cover a wide range of mass loss rates (10$^{-8}-10^{-5}$ M$_{\odot}$ yr$^{-1}$). The list of AGB stars are presented in Tables~\ref{table:parameters} and \ref{table:peculiar} for regular and peculiar sources respectively along with their coordinates, systemic velocity with respect to the Local Standard of Rest ($V_{\rm LSR}$), distance ($D$), effective temperature of the star ($T_*$), stellar luminosity ($L_*$), mass loss rate ($\dot{M}$),  terminal expansion velocity of the envelope ($V_{\rm exp}$), dust condensation radius ($r_c$), dust temperature at the condensation radius ($T_{\rm d}(r_{\rm c})$), gas-to-dust mass ratio ($\Psi$), and the corresponding references for each parameter.

Coordinates were taken from the literature and checked using the SIMBAD astronomical database\footnote{\tiny \texttt{http://simbad.u-strasbg.fr/Simbad}}. The parameters $V_{\rm LSR}$ and $V_{\rm exp}$ are determined from various strong molecular lines available in this study. These two parameters are reported in the literature mainly from CO and SiO lines with varying degrees of accuracy (e.g., \citealt{gro1999,gon2003,tey2006}). We carried out an evaluation of the values of $V_{\rm LSR}$ and $V_{\rm exp}$ derived from our data and compared with those in the literature. In cases where our lines have a well-defined shape, the values from our dataset were preferred, whereas when lines show a less clear shape, the values from literature were favored (as denoted in Table~\ref{table:parameters} where the lack of reference means that the values are derived from this work). The final values of $V_{\rm LSR}$ and $V_{\rm exp}$ adopted in this work are given in Tables~\ref{table:parameters} and \ref{table:peculiar}. We adopted the values of $T_*$ from studies where this parameter is derived by modeling the SED of each star. Stellar luminosities were adopted from the literature, where they are mostly estimated using the period-luminosity relation for Mira variables. Mass loss rates were taken from the literature, where they are determined by modeling observations of multiple CO lines. Distances were adopted from Gaia\footnote{\texttt{\tiny https://gea.esac.esa.int/archive/} \label{note:gaia}} for the stars that have available Gaia data. Although Gaia distances are known to be problematic for AGB stars due to the variability of the photocenter position (which may introduce an error of up to 20 \% in the parallax; \citealt{chi2018}), here we decided to favor distances from Gaia over those from Hipparcos or from the period-luminosity relation (see, e.g., \citealt{mcd2018,dia2019}). Mass loss rates and luminosities are two quantities that follow the inverse-square law as $\propto D^2$, so we consistently scaled them taking into account the newly adopted Gaia distance and mark the new values in Table~\ref{table:parameters} with an asterisk. Note however that empirical mass loss rates derived from CO lines may scale with distance in a slightly different way according to Appendix A of \cite{ram2008}, where scaling laws of the type $\propto D^{1.4-1.9}$ are found, depending on the CO line used. In any case, we evaluated the impact of adopting a scaling law $\propto D^{1.4}$ instead of $\propto D^{2}$ would have on the scaled mass loss rates and it is at most a factor of two.

\section{The observations} \label{sec:observations}

The observations were carried out in the period February to October 2018 with the IRAM 30m telescope, located at Pico Veleta (Spain). Table \ref{table:observed_lines} includes some basic information about the targeted lines: the rest frequency, the Einstein coefficient, A$_{ul}$, the upper level energy, E$_{u}$, and the beam size of the telescope, $\theta_{mb}$. We used the E150 receiver in dual sideband mode, with image rejections $>$10 dB, and observed the frequency ranges $128.5-136.2$ GHz and $144.1-151.9$ GHz (in the lower and upper side bands, respectively). The beam size of the telescope at these frequencies is in the range 16.2-19.0$''$. The observations were done in the wobbler-switching mode with a throw of 180$''$ in azimuth. This technique implies that the target source is measured (ON), followed by a measurement of the sky (OFF) with similar atmospheric conditions. The OFF measurement is then subtracted from the ON measurement to obtain a spectra of the source from which the contribution of the atmosphere to the signal has been removed. The focus was regularly checked on a planet and the pointing of the telescope was systematically checked on a nearby quasar before the observation of each AGB star. The error in the pointing is estimated to be 2-3$''$. The E150 receiver was connected to a fast Fourier transform spectrometer providing a spectral resolution of 0.2 MHz which corresponds to velocity resolutions 0.46 km\,s$^{-1}$ at 129 GHz and 0.39 km\,s$^{-1}$ at 151 GHz. The weather was good and stable during most of the observations, with typical amounts of precipitable water vapor of 2-4 mm and average system temperatures of 115 K. The observations were calibrated by observing the sky and two absorbers at different temperatures, a hot (ambient) and a cold (liquid nitrogen) load using the atmospheric transmission model ATM \citep{cer1985,par2001} adopted by the IRAM 30m telescope. The intensity scale of the output spectra obtained from the antenna is calibrated in antenna temperature ($T^*_{\rm A}$). To express the latter in terms of the main beam brightness temperature ($T_{\rm mb}$), we used the recommended values of $B_{\rm eff}$ and $F_{\rm eff}$ for EMIR\footnote{\texttt{\tiny Eight MIxer Receiver}} at the frequencies of the observed lines\footnote{\texttt{\tiny http://www.iram.es/IRAMES/mainWiki/Iram30mEfficiencies}}, where $B_{\rm eff} = 0.863\,\rm exp[-(\nu(GHz)/361)^2]$ and $F_{\rm eff} = 0.93$. The error in the intensities due to calibration is estimated to be $\sim$20 \%. Typical on source integration times, after averaging horizontal and vertical polarizations, were $\sim$1-2 hrs for each source, resulting in $T_{\rm mb}$ rms noise levels per 0.2 MHz channel of 3-7 mK. 

The data were reduced using the software CLASS\footnote{\texttt{\tiny Continuum and Line Analysis Single-dish Software}}\label{note:class} within the package GILDAS\footnote{\texttt{\tiny GILDAS is a software to reduce and analyze mainly (sub-)mm observations from single-dish and interferometric telescopes. See http://www.iram.fr/IRAMFR/GILDAS} \label{note:gildas}}. To obtain the final spectra for each source, we followed the standard procedure of data reduction that consists of removal of bad channels and low-quality scans, averaging the spectra corresponding to the horizontal and vertical polarizations, and subtracting a baseline of a first order polynomial. In the case of weak lines, the spectra were smoothed to a spectral resolution of 0.4 MHz to increase the signal-to-noise ratio. This corresponds to a velocity resolution of 0.8-1 km s$^{-1}$. When a line was undetected, we smoothed the spectrum to a spectral resolution of 0.8 MHz, corresponding to 1.6-1.9 km s$^{-1}$. 

\begin{table}
\caption{Targeted molecular lines} \label{table:observed_lines}
\centering
\begin{tabular}{lccrr}
\hline \hline
\multicolumn{1}{l}{Transition}  & \multicolumn{1}{c}{Frequency} & \multicolumn{1}{c}{$A_{ul}$} & \multicolumn{1}{c}{$E_u$} & $\theta_{mb}$\\
\multicolumn{1}{l}{}                 & \multicolumn{1}{c}{(MHz)}  & \multicolumn{1}{c}{(s$^{-1}$)} & \multicolumn{1}{c}{(K)} & ($''$)\\
\hline
SiO $J=3-2$ & 130268.665 & $1.06\times10^{-4}$ & 12.5 & 18.8 \\
CS  $J=3-2$ & 146969.025 & $6.07\times10^{-5}$ & 14.1 & 16.7 \\
SiS $J=8-7$ & 145227.052 & $5.05\times10^{-5}$ & 31.4 & 16.9 \\
SO  $3_{3}-2_{2}$          & 129138.983  & $2.21\times10^{-5}$ & 25.5 & 19.0 \\
SO$_{2}$ $8_{2-6}-8_{1-7}$ & 134004.811  & $2.50\times10^{-5}$ & 43.1 & 18.3 \\
SO$_{2}$ $5_{1-5}-4_{0,4}$ & 135696.016  & $2.21\times10^{-5}$  & 15.7 & 18.1 \\
SO$_{2}$ $4_{2-2}-4_{1,3}$ & 146605.519  & $2.47\times10^{-5}$ & 19.0 & 16.7 \\
SO$_{2}$ $2_{2-0}-2_{1,1}$& 151378.662  & $1.88\times10^{-5}$ & 12.6 & 16.2 \\
\hline
\end{tabular}
\end{table}

\section{Observational results} \label{sec:observation_results}

 A total of 30 O-rich CSEs were observed. The spectra obtained is shown in Fig. \ref{fig:models}. We clearly detected SiO $J=3-2$ in all the 30 sources, CS $J=3-2$ in 18 sources, SiS $J=8-7$ in 13 sources, SO $3_{3}-2_{2}$ in 26 sources, while SO$_2$ was detected in at least one of the targeted lines in 19 sources. The detection rates are therefore 100 \% for SiO, 60 \% for CS, 43 \% for SiS, 86 \% for SO, and 63 \% for SO$_2$.
 
 \begin{figure*}
\centering
\vspace{0.3cm}
\includegraphics[width=\textwidth]{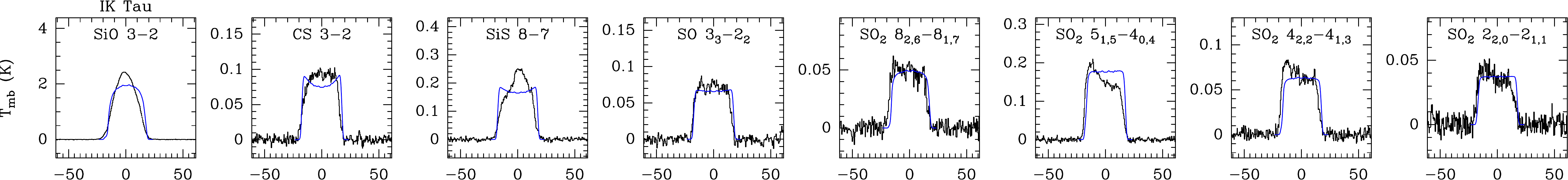}\vspace{0.1cm}
\includegraphics[width=\textwidth]{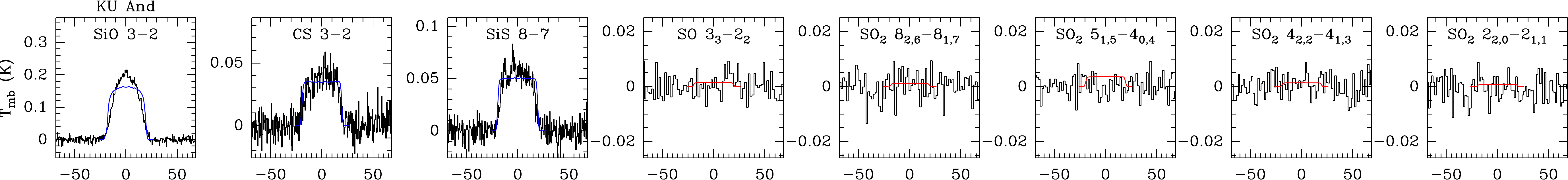}\vspace{0.1cm}
\includegraphics[width=\textwidth]{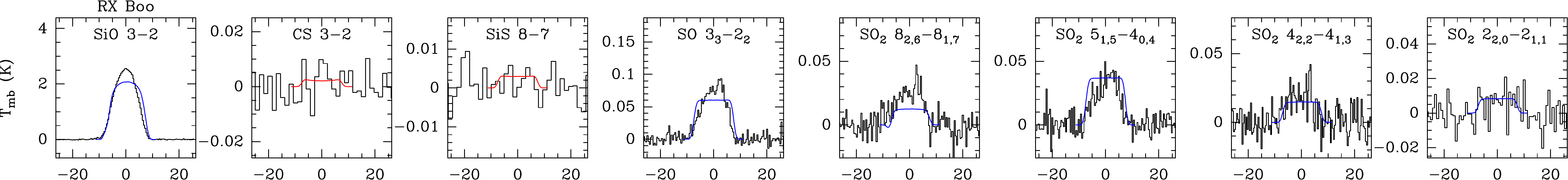}\vspace{0.1cm}
\includegraphics[width=\textwidth]{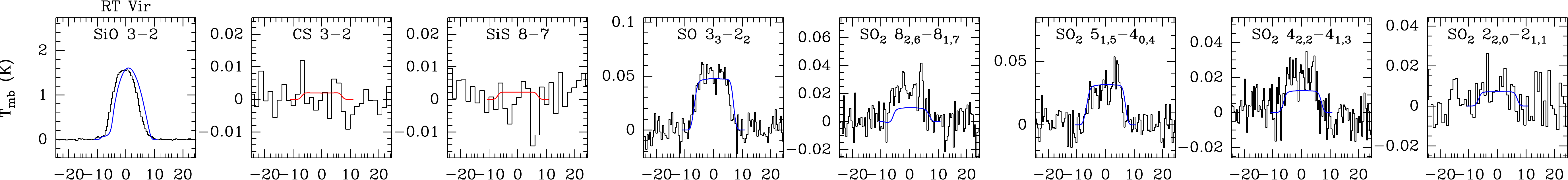}\vspace{0.1cm}
\includegraphics[width=\textwidth]{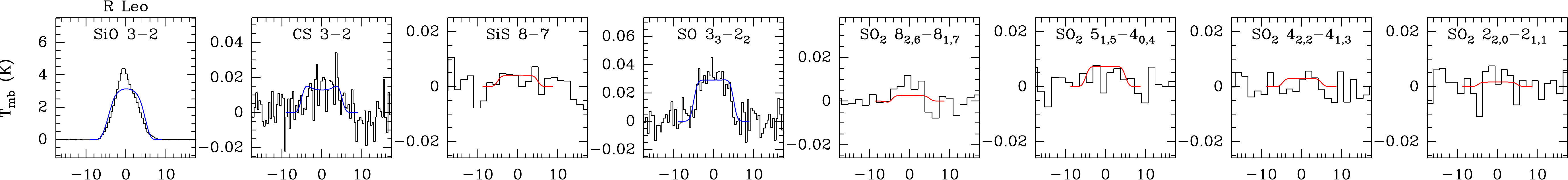}\vspace{0.1cm}
\includegraphics[width=\textwidth]{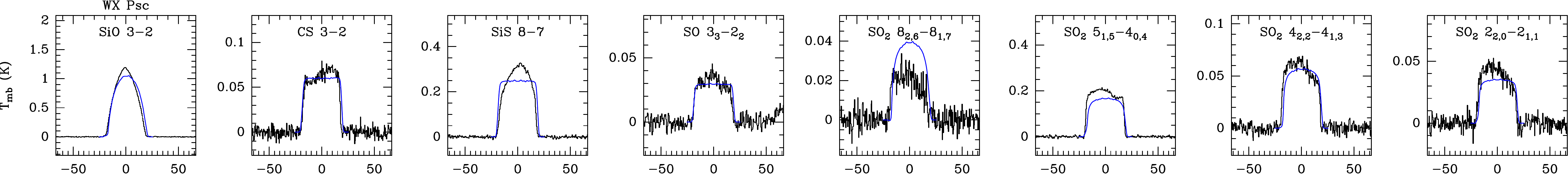}\vspace{0.1cm}
\includegraphics[width=\textwidth]{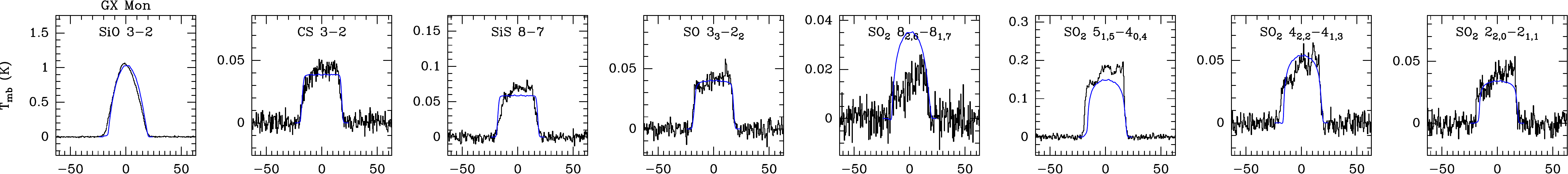}\vspace{0.1cm}
\includegraphics[width=\textwidth]{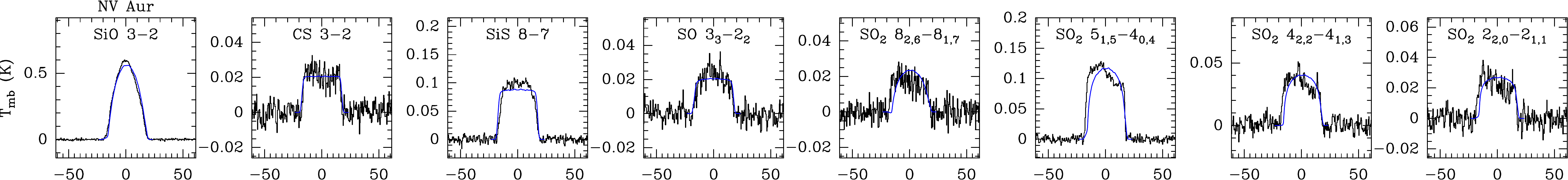}\vspace{0.1cm}
\includegraphics[width=\textwidth]{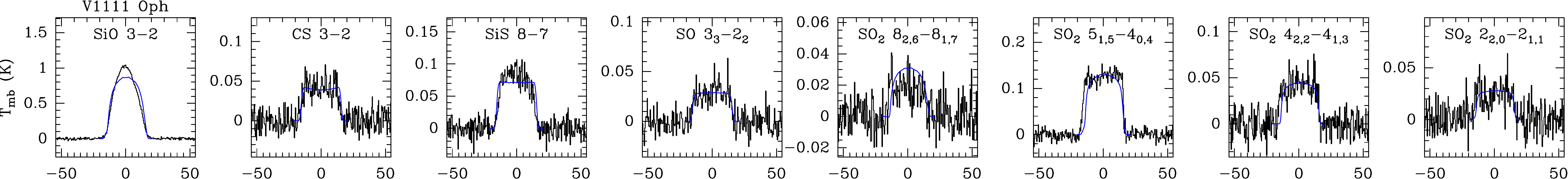}\vspace{0.1cm}
\includegraphics[width=\textwidth]{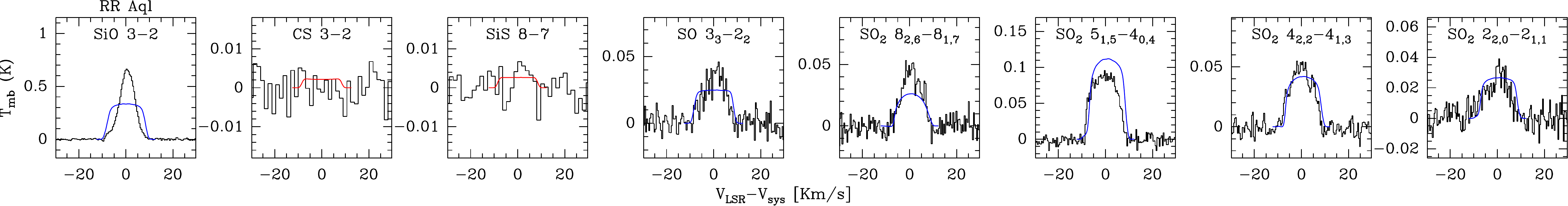}\vspace{0.1cm}
\caption{Rotational lines observed with the IRAM 30m telescope in the 30 O-rich CSEs (black histograms). The blue lines indicate the calculated line profiles from the best-fit LVG model. The red lines correspond to the calculated line profiles with the maximum intensity compatible with the nondetection.}
\label{fig:models}
\end{figure*}

\begin{figure*}
\ContinuedFloat
\captionsetup{list=off,format=cont}
\centering
\vspace{0.3cm}
\includegraphics[width=\textwidth]{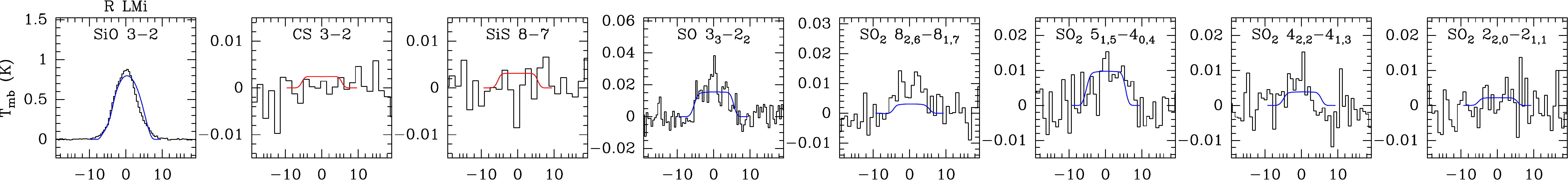}\vspace{0.1cm}
\includegraphics[width=\textwidth]{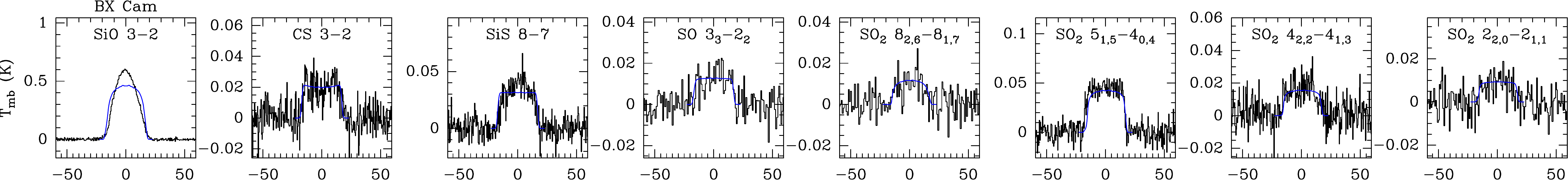}\vspace{0.1cm}
\includegraphics[width=\textwidth]{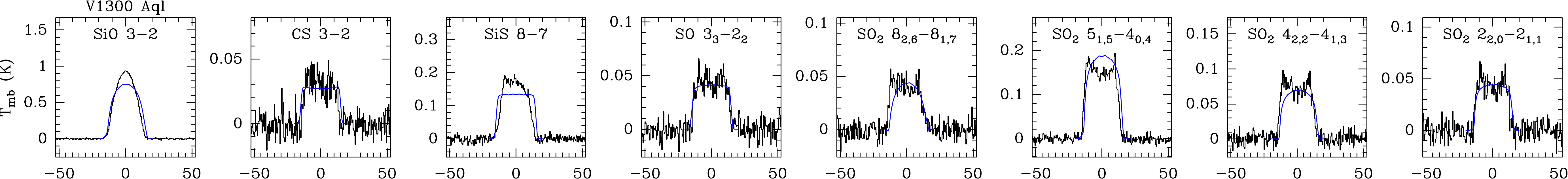}\vspace{0.1cm}
\includegraphics[width=\textwidth]{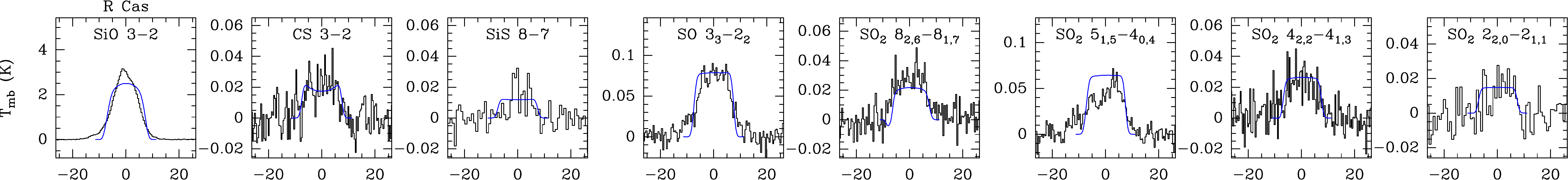}\vspace{0.1cm}
\includegraphics[width=\textwidth]{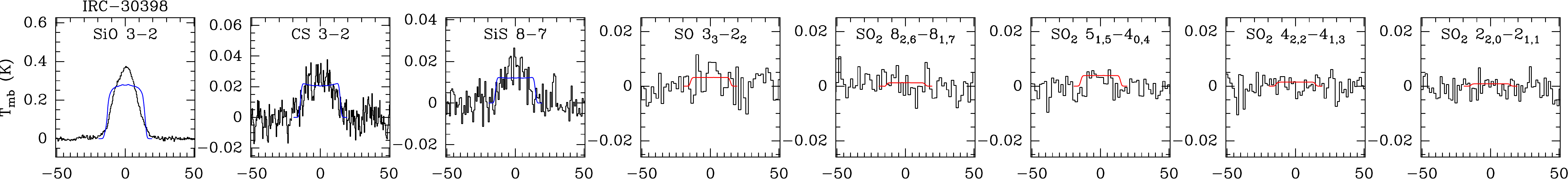}\vspace{0.1cm}
\includegraphics[width=\textwidth]{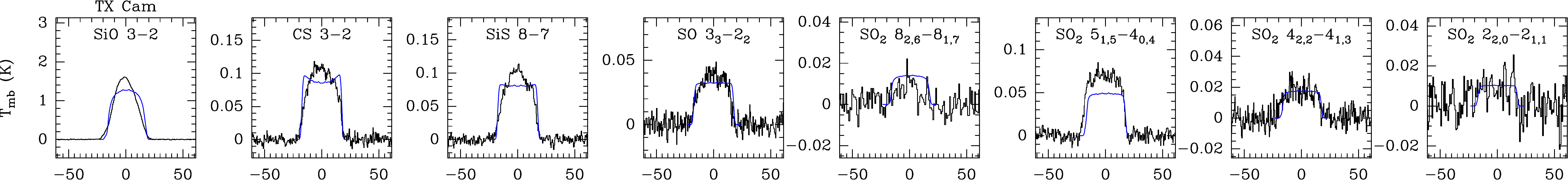}\vspace{0.1cm}
\includegraphics[width=\textwidth]{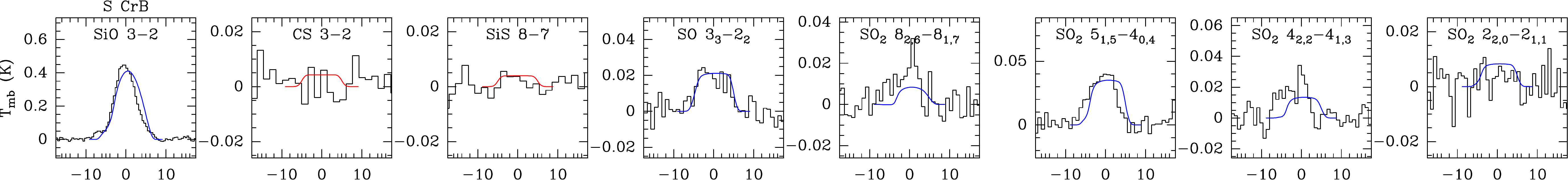}\vspace{0.1cm}
\includegraphics[width=\textwidth]{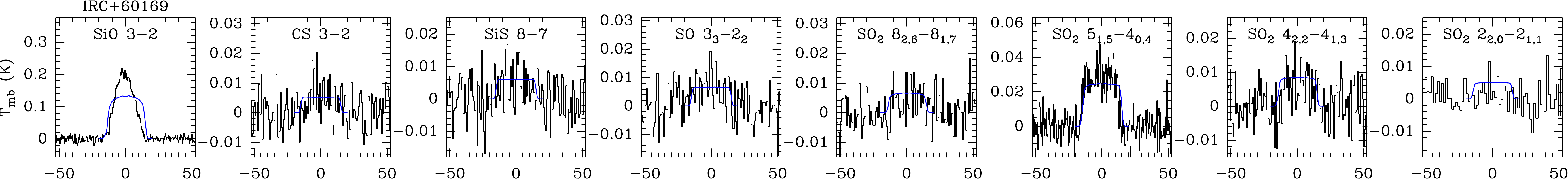}\vspace{0.1cm}
\includegraphics[width=\textwidth]{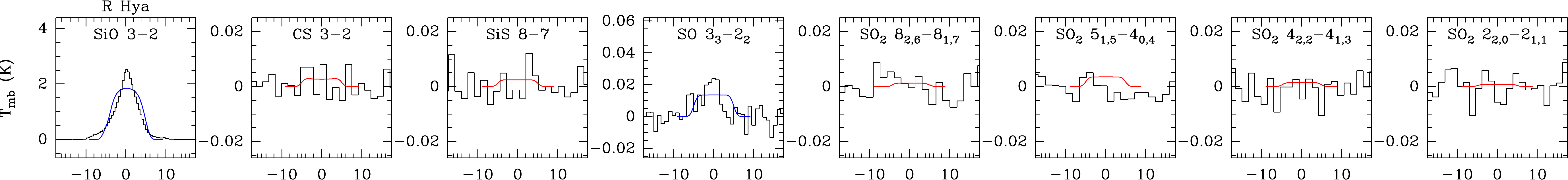}\vspace{0.1cm}
\includegraphics[width=\textwidth]{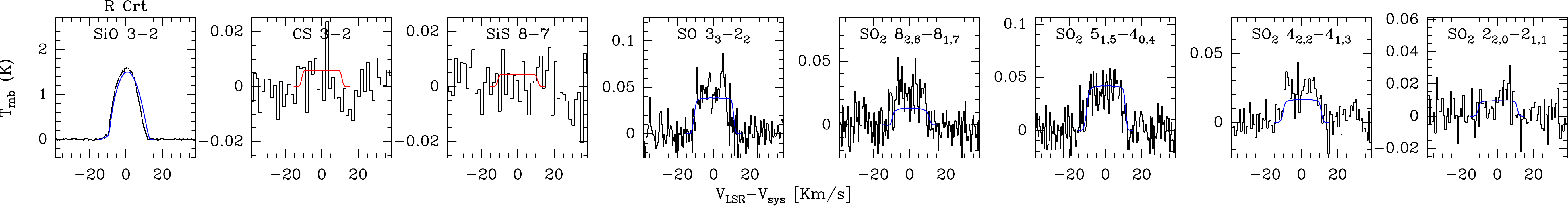}\vspace{0.1cm}
\caption{}
\end{figure*}

\begin{figure*}
\ContinuedFloat
\captionsetup{list=off,format=cont}
\centering
\vspace{0.3cm}
\includegraphics[width=\textwidth]{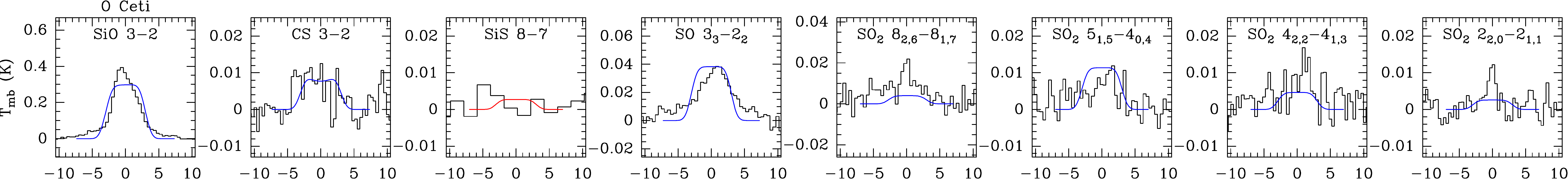}\vspace{0.1cm}
\includegraphics[width=\textwidth]{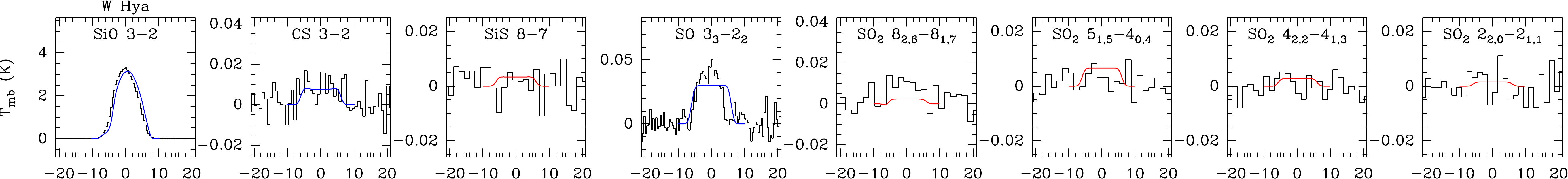}\vspace{0.1cm}
\includegraphics[width=\textwidth]{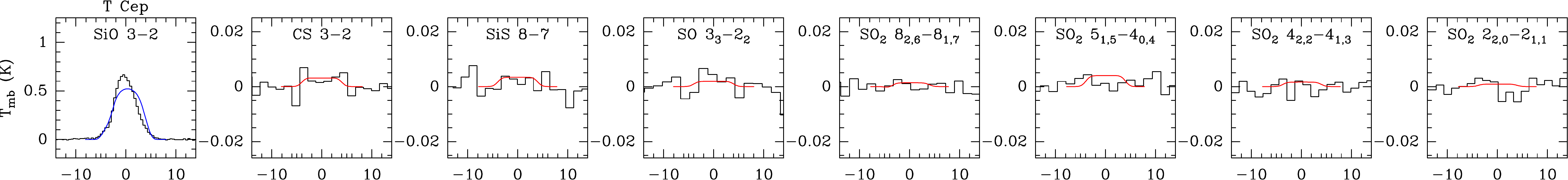}\vspace{0.1cm}
\includegraphics[width=\textwidth]{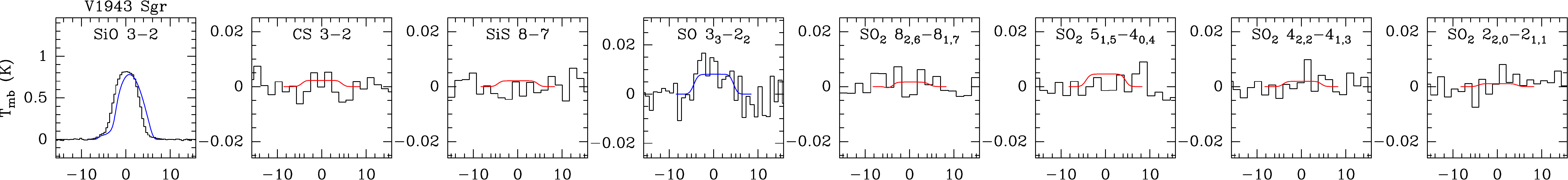}\vspace{0.1cm}
\includegraphics[width=\textwidth]{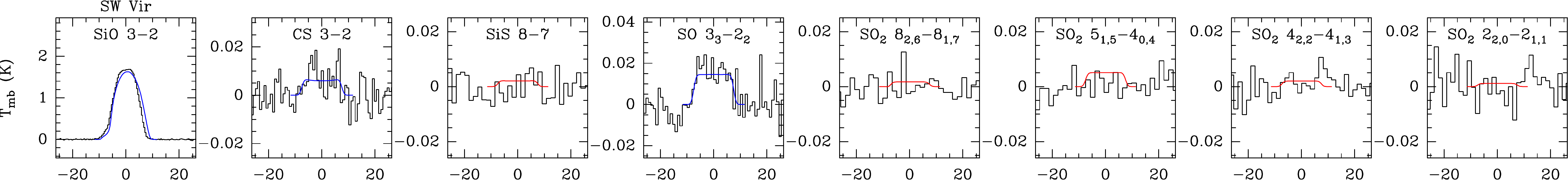}\vspace{0.1cm}
\includegraphics[width=\textwidth]{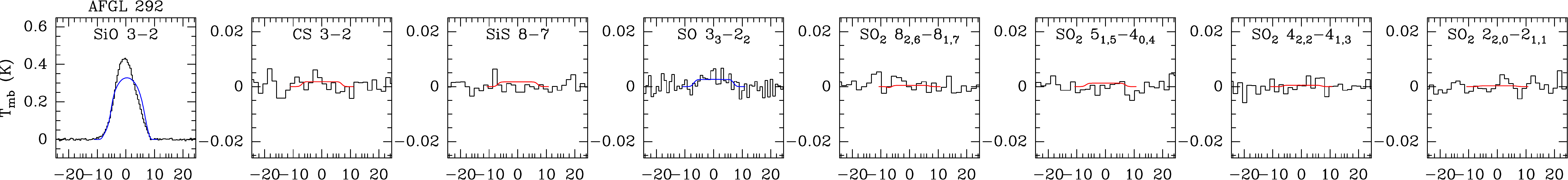}\vspace{0.1cm}
\includegraphics[width=\textwidth]{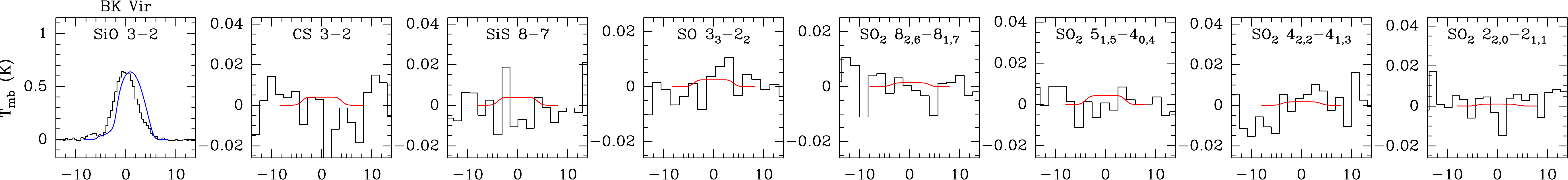}\vspace{0.1cm}
\includegraphics[width=\textwidth]{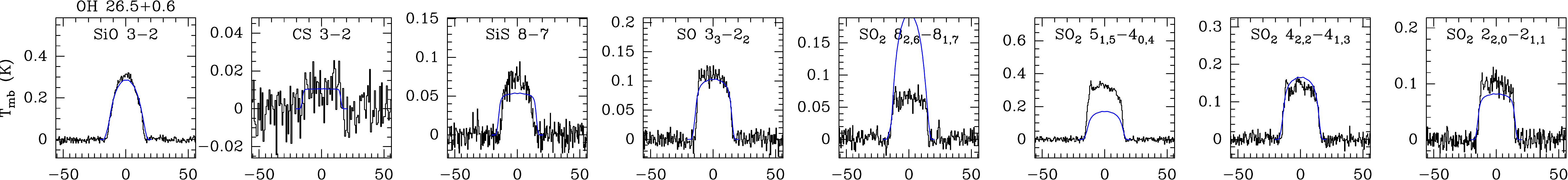}\vspace{0.1cm}
\includegraphics[width=\textwidth]{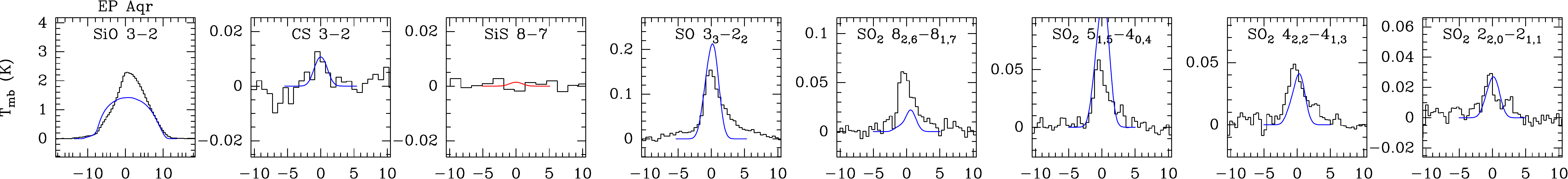}\vspace{0.1cm}
\includegraphics[width=\textwidth]{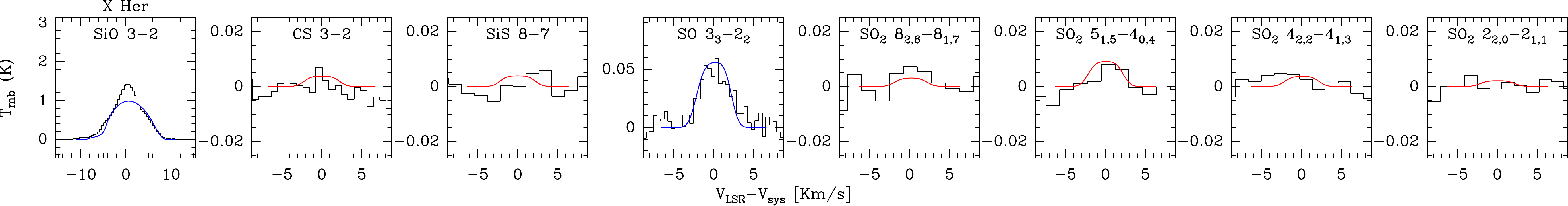}\vspace{0.1cm}
\caption{}
\end{figure*}

The lines were fit using the \texttt{shell} method of CLASS as described in \citet{mas2019}. By performing the fit, we aim to derive for each target lines in every source the centroid frequency in MHz, the expansion velocity in km s$^{-1}$, and the line area, that is, the velocity-integrated intensity in K km s$^{-1}$. These line parameters are given in Table \ref{table:lines}.

The shapes of the emission lines arising from spherically expanding envelopes are essentially determined by the angular size of the emitting source relative to the size of the telescope beam and the line opacity. Most of the line shapes observed here are typical of spherically expanding envelopes, that is, parabolic (optically thick spatially unresolved emission; e.g., SiO $J=3-2$ in KU\,And), flat-topped (optically thin spatially unresolved emission; e.g., CS $J=3-2$ in V1111\,Oph), or double-peaked (optically thin  spatially resolved emission; e.g., SO$_2$ lines in V1300\,Aql). However, there is a number of lines that show profiles with varying kinds of asymmetries. The triangular profile shown in SiO $J=3-2$ in R\,Leo and RR\,Aql is said to indicate that the emission is mainly originating from a region close to the star where the gas is still accelerating. Some striking lines show one side of the profile brighter than the other, sometimes in the blue-shifted side and sometimes in the red-shifted side. An example of these are the SO$_2$ lines in IK\,Tau (blue-shifted emission) and GX\,Mon (red-shifted emission). This indicates an asymmetry in the distribution of the gas emission. Another explanation could be due to self absorption in the line of sight, however, this effect is rather unlikely because the lines are optically thin. Regardless, the \texttt{shell} method of CLASS cannot deal with these kind of asymmetries, but we nevertheless use it on the account that the line area and the expansion velocity resulting from the fit should be trustworthy.

\section{Excitation and radiative transfer modeling} \label{sec:model}

We aim to derive the abundances of SiO, CS, SiS, SO, and SO$_2$ in each source of our sample to provide a statistically meaningful view of how abundant these molecules are in envelopes around O-rich stars. The five molecules studied here are not excited according to local thermodynamic equilibrium (LTE) in the regions of the envelope which contribute mostly to the observed emission (see Sec. \ref{sec:results_model}). Determining the level populations then requires detailed knowledge of collisional excitation data. In Section~\ref{sec:spectro_data} we describe the spectroscopic and collisional excitation data of the five molecules that were input into our calculations and in Section~\ref{sec:procedure} we briefly describe the CSE model and how information on the abundances can be derived from the observed lines using non-LTE radiative transfer modeling. 

\subsection{Molecular data}\label{sec:spectro_data}

In the excitation analysis of SiO, we considered the first 50 rotational levels within the $v=0$ and $v=1$ vibrational states (i.e., a total number of 100 energy levels). The level energies and transition frequencies were calculated from the Dunham coefficients given by \cite{sanz2003}. The dipole moments for pure rotational transitions within the $v=0$ and $v=1$ vibrational states, 3.0982 D and 3.1178 D, respectively, were taken from \cite{ray1970} and the Einstein coefficient for the ro-vibrational transition $\nu=1\rightarrow0$ P(1) of 6.61 s$^{-1}$ from \cite{dri1997}. As collisional rate coefficients for pure rotational transitions we adopted those calculated by \cite{bal2018} for H$_2$ as collider and by \cite{day2006} for He as collider, while for ro-vibrational transitions we used the values computed by \cite{bal2017} scaling from He to H$_2$ as collider (by multiplying by the squared ratio of the reduced masses of the SiO-H$_2$ and SiO-He colliding systems) when needed.

For CS, we included the first 50 rotational levels within the $v=0$ and $v=1$ vibrational states (i.e., a total number of 100 energy levels). The level energies and transition frequencies were calculated from the Dunham coefficients given by \cite{mul2005a}. The line strengths of pure rotational transitions were computed from the dipole moments for each vibrational state, $\mu_{v=0}$ =1.958 D and $\mu_{v=1}$ = 1.936 D \citep{win1968}, while for ro-vibrational transitions we used the Einstein coefficient of 15.8 s$^{-1}$ given for the $v=1\rightarrow0$ P(1) transition by \cite{cha1995}. We adopted the H$_2$ collision rate coefficients recently calculated by \cite{den2018} for pure rotational transitions and up to temperatures of 300 K. At higher temperatures and for ro-vibrational transitions we used the rate coefficients calculated by \cite{liq2007} scaling from He to H$_2$ as collider. Rate coefficients for collisions with He were taken from \cite{liq2006_cs} and \cite{liq2007}.
 
For SiS, we considered the first 70 rotational levels within the $v=0$ and $v=1$ vibrational states (i.e., a total number of 140 energy levels). Level energies were computed from the Dunham coefficients given by \cite{mul2007}. Line strengths were computed from the dipole moments $\mu_{v=0}$ =1.735 D, $\mu_{v=1}$ = 1.770 D, and $\mu_{v=1\rightarrow0}$ = 0.13 D \citep{mul2007,pin1987}. The rate coefficients for inelastic collisions with H$_2$ were taken from the calculations of \cite{klos2008}, while for temperatures higher than 300 K and for ro-vibrational transitions we adopted the rate coefficients computed by \cite{tob2008} scaled from He to H$_2$. Rate coefficients for He as collider were taken from \cite{tob2008}.
 
For SO, we considered rotational levels up to $J=30$ within the ground vibrational state $v=0$ (i.e., a total number of 91 energy levels). Level energies and transition frequencies were calculated from the rotational constants reported by \cite{bog1997}, and line strengths for rotational transitions were computed from the dipole moment, 1.52$\pm$0.02 D, measured by \cite{lov1992}. The rate coefficients for excitation through inelastic collisions were taken from \cite{liq2005} for temperatures up to 50 K and from \cite{liq2006} for higher temperatures, scaling from He to H$_2$ when needed. 

For SO$_2$, we included the first 31 energy levels within the ground vibrational state. We used the rotational constants reported by \cite{mul2005b}. Line strengths for rotational transitions were computed from the dipole moment measured by \cite{pat1979}. The rate coefficients for excitation through inelastic collisions with H$_2$ were taken from \cite{cer2011} for temperatures up to 30 K, and from \cite{bal2016} for higher temperatures, while for collisions with He rate coefficients were taken from \cite{gre1995}.

\subsection{Modeling procedure}\label{sec:procedure}

Here, we give a brief description of the model used to perform the non-LTE excitation and radiative transfer calculations (for details see \citealt{mas2018}). To derive accurate molecular abundances in each object, we take into account the specific physical properties for each envelope and source presented in Table~\ref{table:parameters} and Table~\ref{table:peculiar}.

The main assumption in the model is that the CSE that surrounds the central AGB star has a smooth and spherically symmetric geometry that is produced by an isotropic mass loss with a constant mass loss rate $\dot{M}$ and a constant expansion velocity $V_{\rm exp}$. We assume that the hydrogen in the CSE is molecular and its density structure (as a function of a distance $r$ from the star) follows an $r^{-2}$ law. The various physical quantities relating to the envelope such as the radial profiles of the gas density, gas temperature, and dust temperature, as well as the properties of the dust grains are described in \cite{mas2018}. The only difference in this study is that we consider spherical grains of silicate with a radius of 0.1 $\mu$m, a mass density of 3.3 g cm$^{-3}$, and optical properties for warm silicate from \cite{suh1999}.

We chose to model the molecular line emission using the multishell large velocity gradient (LVG) method explained in more detail in \cite{agu2009} and \cite{agu2012}. The LVG formalism, first developed by \cite{sob1960}, has been widely used to solve the molecular excitation and radiative transfer problem in environments with large velocity gradients. This approach is valid for molecular lines in circumstellar envelopes as long as they are not too optically thick. \citet{buj2013} showed that this formalism yields quite accurate excitation conditions even when the approximations of the method are marginally satisfied. These authors investigated the validity of the LVG formalism by studying the CO molecular excitation for different conditions and conclude that although the LVG approximation still produces good behavior in cases where the velocity gradient is low, the behavior is not as accurate for the very outer regions with high opacities. The LVG method provides a good compromise with respect to other methodologies such as Monte Carlo, which are more computationally expensive and exhibit problems of convergence when including a high number of energy levels.

Briefly, the CSE is divided into several concentric shells. The statistical equilibrium equations are then solved in each shell to determine the level populations. The radiation field which is needed to solve the statistical equilibrium equations is evaluated solving the radiative transfer under the LVG approximation. We assume that the molecules are excited by collisions with H$_2$ molecules and He atoms and through radiation from three sources: the cosmic microwave background, the stellar radiation, and the thermal emission from dust. We also include infrared (IR) pumping to excited vibrational states for SiO, CS, and SiS. In the cases of SO and SO$_2$, we only consider rotational levels within the ground vibrational state for simplicity and because of the less reliable collisional rate coefficients for ro-vibrational transitions.

\subsection{Adopted abundance distribution}\label{sec:distribution}

The abundance distribution is important in the radiative transfer modeling. For parent molecules that are injected from the inner parts of the envelope, the abundance may vary with radius due to two processes: condensation onto grains and photodissociation by ultraviolet (UV) photons. In reality, the abundance is expected to decrease from the thermochemical equilibrium (TE) value at the stellar surface in the dust formation region. The molecules are further depleted by the ambient radiation field which eventually determines the size of the emission envelope. Relating to this scenario, a few studies have reported on an abundance distribution to be consisting of two components, a compact high abundance in the inner regions of the CSE, and a lower abundance in the extended outer regions of the CSE (e.g., \citealt{sch2007}, \citealt{dec2010}). For example, \cite{sch2004} modeled the SiO emission of the $J=6-5$, $J=5-4$, $J=3-2$, and $J=2-1$ lines in the CSE of the M-type star R\,Dor and found the need for a two-component abundance distribution which was characterized by a high abundance of $4\times10^{-5}$ up to $1.2\times10^{15}$ cm and a lower abundance of $3\times10^{-6}$ at larger radii. This initial high abundance followed by a decrease was interpreted as adsorption of SiO onto dust. However, when \cite{van2018} modeled the SiO emission with several low- and high-$J$ transitions (up to $J=38-37$) in the same object, R\,Dor, they found no indication of a two-component abundance distribution. Similarly, in the case of SiS, \citet{sch2007} modeled the emission in IK\,Tau and found a better fit to their observations when they included an inner component out to $1\times10^{15}$ cm with a high SiS abundance of $2\times10^{-5}$. However, \citet{dan2019} performed sensitive ALMA observations with an angular resolution of $\sim$ 150 mas that correponds to $\sim$ $6\times10^{14}$ cm and did not find evidence for such a jump in the abundance of SiS in the same source. Other studies reported on the molecular abundance distribution in CSEs as well. \citet{agu2012} modeled lines of CS in the $\nu=0-3$ states in addition to several transitions of the isotopologues $^{13}$CS, C$^{34}$S, and C$^{33}$S in IRC\,+10216 and derived an abundance of $4\times10^{-6}$ in the inner regions that decreased to $7\times10^{-7}$ in the mid envelope at a radius of $2\times10^{15}$ cm. Their result is in good agreement with that derived by \citet{vel2019} that modeled the $J=2-1$ lines of CS, and $^{13}$CS, C$^{34}$S, and C$^{33}$S using ALMA which also showed a decline in the abundance toward the intermediate envelope of IRC\,+10216 supporting a depletion scenario. However, no such abundance distribution is reported for the sulfur oxides thusfar that evidence any depletion (e.g., \citealt{dan2016,dan2020}).

Regardless whether or not these molecules experience a first abundance depletion due to dust condensation, they maintain a significant abundance in the extended outer envelope where photodissociation further removes the molecules from the gas phase. The lines observed in this study probe intermediate/outer regions of the envelope (see Sec. \ref{sec:results_model}). That is, we are not sensitive to abundance gradients occurring in the inner envelope and the abundances derived are valid for the post-condensation region. We therefore adopt a simple scenario in which the fractional abundance remains constant throughout the envelope up to some region where it drops due to photodissociation. The adopted abundance radial distribution $f(r)$ is described by a Gaussian as:

\begin{equation}\label{eq:abundance_equation}
f(r) = f_0\,\exp{\Big(-(r/r_e)^{2}\Big)},
\end{equation}

where $f$ is the fractional abundance of the molecule relative to H$_2$, $f_0$ is the initial abundance, and $r_e$ is the $e$-folding radius at which the abundance has dropped by a factor $e$. \citet{dan2016} observed SO and SO$_2$ emission in a sample of five stars O-rich CSEs, and found that SO$_2$ has a Gaussian abundance distribution, whereas the SO abundance distribution differs between Gaussian for high mass-loss rate envelopes, and shell-like for low mass-loss rate envelopes where the abundance peaks at a distance from the central star. In their recent study on these two molecules using ALMA in two CSEs R\,Dor and IK\,Tau, \citet{dan2020} found that SO and SO$_2$ in R\,Dor have Gaussian distribution, while in IK\,Tau, SO and probably SO$_2$ have a shell-like abundance distribution. Here, we adopt a Gaussian abundance distribution for all the molecules.\\

In their study on M-type stars, \cite{gon2003} estimated the SiO emission size by using a scaling law that correlated $r_e$ and the envelope density evaluated through the quantity $\dot{M}/V_{\rm exp}$,

\begin{equation}\label{eq:scaling_law1}
\log r_{e}{\rm (SiO)} = 19.2 + 0.48 \log \left(\frac{\dot{M}}{V_{\rm exp}}\right), 
\end{equation}

where $r_{e}$ is given in cm, $\dot{M}$ in M$_{\odot}$ yr$^{-1}$, and $V_{\rm exp}$ in km s$^{-1}$. We use Eq.~(\ref{eq:scaling_law1}) to determine the emission size of SiO, SiS, SO, and SO$_2$ in our sample. The assumption of similar radial extents for SiO and SiS is discussed in our previous study of these molecules in C-rich CSEs in \citet{mas2019}. \citet{dan2018} derived empirical relations between $r_e$ and $\dot{M}/V_{\rm exp}$ for SiS and CS from a limited sample of M-, C-, and S-type stars which we noticed are unreliable outside the relatively narrow range of $\dot{M}/V_{\rm exp}$ over which they were derived (for details see \citealt{mas2019}). In their recent study using ALMA data, \citet{dan2020} found that SO and SO$_2$ are colocated around the O-rich R\,Dor, with SO being slightly more extended than SO$_2$. However, statistically robust empirical relations for the emission size of SO and SO$_2$ have not been derived. Since the photodissociation rates of SO and SO$_2$ under the interstellar radiation field are of the same order of that of SiO, a few times 10$^{-9}$ s$^{-1}$ \citep{hea2017,agu2018}, in the lack of better constraints on the emission size, we adopt the same radial extent for SiO, SiS, SO, and SO$_2$. For CS, which has a significantly lower photodissociation rate than SiO, SO, and SO$_2$, a few times 10$^{-10}$ s$^{-1}$ \citep{pat2018}, we use a larger emission size as suggested by \cite{mas2019} for C-rich AGB stars and described by the following empirical relation: 

\begin{equation}\label{eq:scaling_law2}
\log r_{e}{\rm (CS)} = 19.65 + 0.48 \log \left(\frac{\dot{M}}{V_{\rm exp}}\right),
\end{equation}
 
Briefly, we construct a physical model of the envelope for each source with the parameters given in Table~\ref{table:parameters} and \ref{table:peculiar}. Adopting the fractional abundance distributions given in the previous section, we perform excitation and radiative transfer calculations by varying the initial fractional abundance, $f_0$, until the calculated line profiles match the observed ones. The criteria we adopted to determine how well the model fits the data is by matching the area of the calculated line to the observed one. The agreement between observed and calculated line area was better than 3\% for the molecules for which we have only one line, SiO, CS, SiS, and SO, and better than 30\% for SO$_2$ because for this molecule we have four observed lines. When the line is undetected, we derive upper limits of the fractional abundance by choosing the maximum abundance that results in line intensities compatible with the noise level of the observations.

In our sample there are three sources which deserve to be considered separately, EP\,Aqr, X\,Her, and OH\,26.5+0.6. We discuss the reasons and the procedure adopted for the line modeling of these sources in Sec.~\ref{sec:peculiar}.

\subsection{Peculiar sources} \label{sec:peculiar}

Some AGB stars are known to exhibit a double-component profile in some molecular lines like CO, with a narrow spectral feature centered on a much broader plateau, with both components having the same LSR velocity \citep{kah1996,kna1998,ker1999,olo2002}. The origin of these double-component profiles is still not clear. \cite{kna1998} suggested that it is an effect of episodic mass loss with highly varying gas expansion velocities that produces multiple shells where each shell has a different expansion velocity and different mass loss rate. Other studies argued that complicated geometries and kinematics play a role in the rise of the effect \citep{ner1998,nak2005,cas2010,kim2012,hom2015,kim2019}. Two of the sources that are known to exhibit this type of profile, X\,Her and Ep\,Aqr \citep{hom2018}, are in our sample and their spectra are shown in Fig.\ref{fig:models}. The two stars are M-type semiregular variable AGBs. From the rotational transitions observed here there is no sign of the superimposition of the narrow profile on the broader one in both sources. However, based on the observed line widths and expansion velocities it appears that the SiO line emission of X\,Her and EP\,Aqr are coming from the broad component, while the CS, SO and SO$_2$ line emissions arise from the narrow component. The SiS rotational line is not detected in any of the sources, but in deriving abundance upper limits we assume that it behaves as CS, SO, and SO$_2$ and arises from the narrow component. We then consider two different winds for the narrow and the broad component each with different values of the expansion velocity and the mass loss rate, as given in Table~\ref{table:peculiar}, and perform the radiative transfer and excitation analysis independently. 

Another source we encountered problems modeling is the extreme OH/IR AGB star OH\,26.5+0.6. This star is thought to have gone through a superwind phase characterized by a dramatic increase of the mass loss rate (by a factor of ten at least) toward the end of the AGB \citep{ibe1983} which ejects most of the remaining envelope and the initial mass of the star allowing the latter to evolve toward the planetary nebula phase. \cite{jus1996} show evidence of two mass loss regimes: a higher density superwind that started recently ($<200$ yr), and a lower density AGB wind that started earlier. From \cite{jus1996}, the reported gas mass loss rate for the superwind is $\dot{M} = 5.5\times10^{-4}$ M$_{\odot}$ yr$^{-1}$ at $r\textless$ $8.0\times10^{15}$ cm and for the outer AGB wind is $\dot{M} = 1\times10^{-6}$ M$_{\odot}$ yr$^{-1}$ at greater radii. Moreover, we adopt the gas kinetic temperature profile for the object reported by the authors as well (solid line in Fig.~7b of \citealt{jus1996}). After setting the density and temperature structure in the envelope, we model the rotational line emission of the molecules.

\section{Results from line modeling} \label{sec:results_model}

The calculated line profiles from our best-fit LVG model for each of the sources are shown in blue in Fig.~\ref{fig:models}, where they are compared with the observed line profiles (black histograms). In those cases in which the lines are not detected, the calculated line profile is plotted in red instead. 

In general, the calculated line shapes agree well with the observed ones. However, there are profiles that exhibit complexities which our model is not able to reproduce, for example, the SO$_2$ emission in IK\,Tau. This is probably due to the simplicity of our approximated physical model which assumes smooth and spherically symmetric envelopes and does not take into consideration any deviations from that. In the case of the SiO $J=3-2$ line, the observed profiles are mostly parabolic or triangular, although for some sources, such as RR\,Aql, the model produces flat-topped shapes. The assumption of a constant expansion velocity in our model could be playing a factor in the discrepency between the calculated and the observed line profile since as mentioned previously triangular line profiles probably indicate emission from accelerating regions close to the star. Another explanation could be related to the line opacity $\tau$. In these cases, the line opacity is probably around one, with the modeled line having $\tau$ slightly below one and the observed line having $\tau$ slightly above one. In any case, the overall agreement between calculated and observed line profiles is good.

The excitation and radiative transfer calculations give us information about the excitation and emission region of the molecules in the envelope. In regards to the emission region, the model indicates that most of the contribution to the line emission of the five molecules is coming from regions $10^{15}-10^{16}$ cm from the star, i.e, most of the emission arises from intermediate and outer regions of the envelope, rather than from the inner regions. To evaluate the role of IR pumping to vibrationally excited states for SiO, CS, and SiS, we ran models excluding IR pumping. Our calculations indicate that in the absence of IR pumping, the emission is more compact than when IR pumping is included. The calculations also show that IR pumping has an effect on the intensities of the observed lines of SiO, CS and SiS. Neglecting IR pumping results in a systematic decrease in the integrated line intensities of $\sim 40 \%$ for SiO $J=3-2$, $\sim 50\%$ for CS $J=3-2$, and $\sim 35\%$ for SiS $J=8-7$.

In regards to the excitation of the rotational levels, the model indicates that the rotational levels of the five studied molecules are thermalized in the warm and dense inner layers of the envelopes. However, as the radial distance from the star increases and the gas density decreases, the rotational levels become suprathermally populated (with the ratio of the excitation temperature, $T_{ex}$, to the kinetic temperature, $T_k$, greater than unity, $T_{ex}$/$T_k$>1) for SiS, CS, SO and SO$_2$ in the regions where most of the emission is coming from. For SiO, the population of rotational levels vary. Mostly, they are subthermally populated ($T_{ex}$/$T_k$<1), however for a few envelopes (e.g., RR\,Aql, NV\,Aur, WX\,PSc), the levels are suprathermally populated . The behavior for SiO, CS, and SiS, is largely caused by IR pumping. In comparison, in C-rich CSEs, IR pumping causes the rotational transitions of these molecules to be all suprathermally excited \citep{mas2019}. We can conclude that IR pumping plays an important role in the excitation of the rotational emission of SiO, CS, and SiS whether in C-rich or O-rich envelopes. 

\section{Discussion}\label{sec:discussion}

\FloatBarrier
\begin{table*}
\caption{Fractional abundances of SiO, CS, SiS, SO, and SO$_2$ derived}\label{table:abundances}
\centering
\resizebox{1.8\columnwidth}{!}{
\begin{tabular}{lcccccccc}
\hline \hline
\multicolumn{1}{l}{Name}  & Star &  \multicolumn{1}{c}{$\dot{M}$} & \multicolumn{1}{c}{$V_{\rm exp}$}  & \multicolumn{1}{c}{$f_0$(SiO)} & \multicolumn{1}{c}{$f_0$(CS)} & \multicolumn{1}{c}{$f_0$(SiS)} & \multicolumn{1}{c}{$f_0$(SO)}  & \multicolumn{1}{c}{$f_0$(SO$_2$)} \\
\multicolumn{1}{c}{}  & Var. & \multicolumn{1}{c}{(M$_{\odot}$ yr$^{-1}$)} & \multicolumn{1}{c}{(km~s$^{-1}$)} & & & & & \\
\hline
IK\,Tau         & M & $2.4\times10^{-5}$    &  17.5    &  $3.1\times10^{-7}$      &  $1.0\times10^{-8}$      &  $1.0\times10^{-7}$     &  $1.7\times10^{-7}$     &  $3.2\times10^{-7}$  \\
KU\,And         & M & $9.4\times10^{-6}$   &  19.5    &  $6.2\times10^{-7}$      &  $8.2\times10^{-8}$      &  $9.9\times10^{-7}$     & < $9.8\times10^{-8}$    & < $1.8\times10^{-7}$         \\
RX\,Boo         & SRb & $6.1\times10^{-7}$    &  7.5     &  $1.7\times10^{-6}$      & < $1.0\times10^{-9}$     & < $1.2\times10^{-8}$    &  $8.3\times10^{-7}$     &  $3.5\times10^{-7}$     \\
RT\,Vir         &  SRb & $4.5\times10^{-7}$   &   7      &  $4.5\times10^{-5}$      & < $3.0\times10^{-9}$     & < $2.5\times10^{-8}$    &  $2.7\times10^{-6}$     &  $1.2\times10^{-6}$  \\
R\,Leo          & M & $1.0\times10^{-7}$    &   5      &  $5.7\times10^{-6}$      &  $1.2\times10^{-8}$      & < $3.0\times10^{-8}$    & $6.6\times10^{-7}$      & < $1.1\times10^{-7}$          \\
WX\,Psc         & M & $4.0\times10^{-5}$    &   19     &  $8.9\times10^{-7}$      &  $2.7\times10^{-8}$      &  $4.5\times10^{-7}$     &  $2.5\times10^{-7}$     &   $1.0\times10^{-6}$   \\
GX\,Mon         & M & $4.8\times10^{-6}$    &   18     &  $1.0\times10^{-5}$      &  $8.9\times10^{-8}$      &  $7.6\times10^{-7}$     & $2.5\times10^{-6}$      &  $7.4\times10^{-6}$         \\
NV\,Aur         & M & $2.5\times10^{-5}$    &   17.5   &  $2.9\times10^{-6}$      &  $2.8\times10^{-8}$      &  $7.8\times10^{-7}$     &  $8.5\times10^{-7}$     & $4.1\times10^{-6}$    \\
V1111\,Oph      & M & $2.7\times10^{-6}$    &   15.5   &  $5.8\times10^{-6}$      &  $7.7\times10^{-8}$      &  $1.3\times10^{-6}$     &  $2.0\times10^{-6}$     & $7.8\times10^{-6}$  \\
RR\,Aql         & M & $8.6\times10^{-7}$    &   8.5    &  $1.3\times10^{-6}$      & <  $4.5\times10^{-9}$    & < $6.3\times10^{-8}$    &  $1.7\times10^{-6}$     &  $6.5\times10^{-6}$       \\
R\,LMi          &  M & $2.6\times10^{-7}$   &   5.5    &  $2.6\times10^{-5}$      & <  $1.0\times10^{-8}$    & <  $1.5\times10^{-7}$   &  $2.3\times10^{-6}$     &  $9.8\times10^{-7}$   \\
BX\,Cam         & M & $1.0\times10^{-6}$    &   17     &  $5.5\times10^{-6}$      &  $1.1\times10^{-7}$      &  $1.9\times10^{-6}$     &  $2.3\times10^{-6}$     &  $5.8\times10^{-6}$    \\
V1300\,Aql      & M & $1.0\times10^{-5}$    &   15     &  $1.5\times10^{-6}$      &  $2.2\times10^{-8}$      &  $9.3\times10^{-7}$     &  $1.2\times10^{-6}$     &  $4.5\times10^{-6}$      \\
R\,Cas          & M &$9.5\times10^{-7}$   &   7.5    &  $3.1\times10^{-6}$      &  $1.2\times10^{-8}$      &  $7.4\times10^{-8}$     &  $1.2\times10^{-6}$     & $6.8\times10^{-7}$   \\
IRC\,-30398   & M & $6.0\times10^{-6}$   &   14.5   &  $2.9\times10^{-7}$      &  $1.5\times10^{-8}$      &  $6.5\times10^{-8}$     &  <  $6.8\times10^{-8}$     &  < $6.0\times10^{-8}$   \\
TX\,Cam         & M & $7.7\times10^{-6}$    &   17.5   &  $1.4\times10^{-6}$      &  $5.1\times10^{-8}$      &  $3.9\times10^{-7}$     &  $5.8\times10^{-7}$     &   $6.3\times10^{-7}$    \\
S\,CrB          & M & $2.7\times10^{-7}$    &   5      &  $1.8\times10^{-5}$      & <  $2.0\times10^{-8}$    & <  $2.0\times10^{-7}$   &  $4.2\times10^{-6}$     &   $4.9\times10^{-6}$      \\
IRC\,+60169     & M & $9.6\times10^{-6}$    &   15     &  $1.5\times10^{-7}$      &  $5.0\times10^{-9}$      &  $2.6\times10^{-8}$     &  $1.3\times10^{-7}$     &   $3.5\times10^{-7}$       \\
R\,Crt          & SRb & $1.0\times10^{-6}$    &   11     &  $2.3\times10^{-5}$      &  <  $1.0\times10^{-8}$      & <  $5.0\times10^{-8}$   &  $2.3\times10^{-6}$     &  $1.7\times10^{-6}$         \\
R\,Hya          & M & $4.7\times10^{-7}$   &   5      &  $3.3\times10^{-6}$      & <  $2.0\times10^{-9}$    & < $2.0\times10^{-8}$    &  $3.1\times10^{-7}$     & <  $5.2\times10^{-8}$         \\
\textit{o}\,Ceti& M & $2.0\times10^{-7}$    &   3      &  $7.0\times10^{-8}$      &  $1.4\times10^{-9}$      & <  $4.0\times10^{-9}$     & $2.0\times10^{-7}$      & $4.0\times10^{-8}$           \\
W\,Hya          & SRa & $4.2\times10^{-7}$    &   6      &  $1.4\times10^{-5}$      &  $5.0\times10^{-9}$      & <   $2.0\times10^{-8}$    &   $6.9\times10^{-7}$    & < $9.8\times10^{-8}$  \\
T\,Cep          & M & $7.8\times10^{-8}$    &   4      &  $4.4\times10^{-6}$      &  < $9.0\times10^{-9}$    & < $1.0\times10^{-7}$      & < $2.2\times10^{-7}$     &< $3.0\times10^{-7}$         \\
V1943\,Sgr      &  SRb & $1.0\times10^{-6}$   &   4.5    &  $1.5\times10^{-5}$      & <  $3.0\times10^{-9}$    & <  $3.0\times10^{-8}$     &   $4.1\times10^{-7}$    & <  $1.5\times10^{-7}$        \\
SW\,Vir         & SRb & $2.2\times10^{-6}$   &   7.5    &  $2.0\times10^{-6}$      &  $2.1\times10^{-9}$      & <  $5.0\times10^{-9}$      & $1.6\times10^{-7}$      &< $3.9\times10^{-8}$  \\
AFGL\,292       & $...$ & $1.3\times10^{-7}$    &   7      &  $1.2\times10^{-5}$      & <  $2.0\times10^{-8}$    & <  $2.0\times10^{-7}$      &  $1.2\times10^{-6}$     &< $3.6\times10^{-7}$     \\
BK\,Vir         & SRb & $2.3\times10^{-7}$    &   4      &  $7.0\times10^{-6}$      &  < $5.0\times10^{-9}$    & <  $3.0\times10^{-8}$     & < $1.0\times10^{-7}$     &<  $1.2\times10^{-7}$          \\
OH\,26.5+0.6    & M & $1.0\times10^{-6}$   &   15.4   &  $2.2\times10^{-6}$  &  $3.1\times10^{-8}$     & $5.2\times10^{-7}$      &  $6.0\times10^{-6}$     & $2.1\times10^{-6}$   \\
Ep\,Aqr         & SRb & $1.7\times10^{-8}$    &   1      &  -                       & $6.0\times10^{-9}$     &  < $6.0\times10^{-9}$     & $5.4\times10^{-6}$      &  $1.9\times10^{-6}$         \\
                & & $5.0\times10^{-7}$    &   9.2    &  $3.6\times10^{-6}$      &  - &   -  &   -  &   -       \\
X\,Her          & SRb & $4.3\times10^{-8}$    &   2.2    &  -                       &  < $4.2\times10^{-9}$      & < $3.6\times10^{-8}$     &   $2.5\times10^{-6}$      &  < $2.5\times10^{-7}$        \\
                &  &  $1.6\times10^{-7}$    &   6.5    &  $1.4\times10^{-5}$      &   -   &   -  &   -    &    -      \\              
\hline
\end{tabular}
}
\tablenotea{\small }
\end{table*}

\begin{figure*}
\centering
\includegraphics[width=0.33\textwidth]{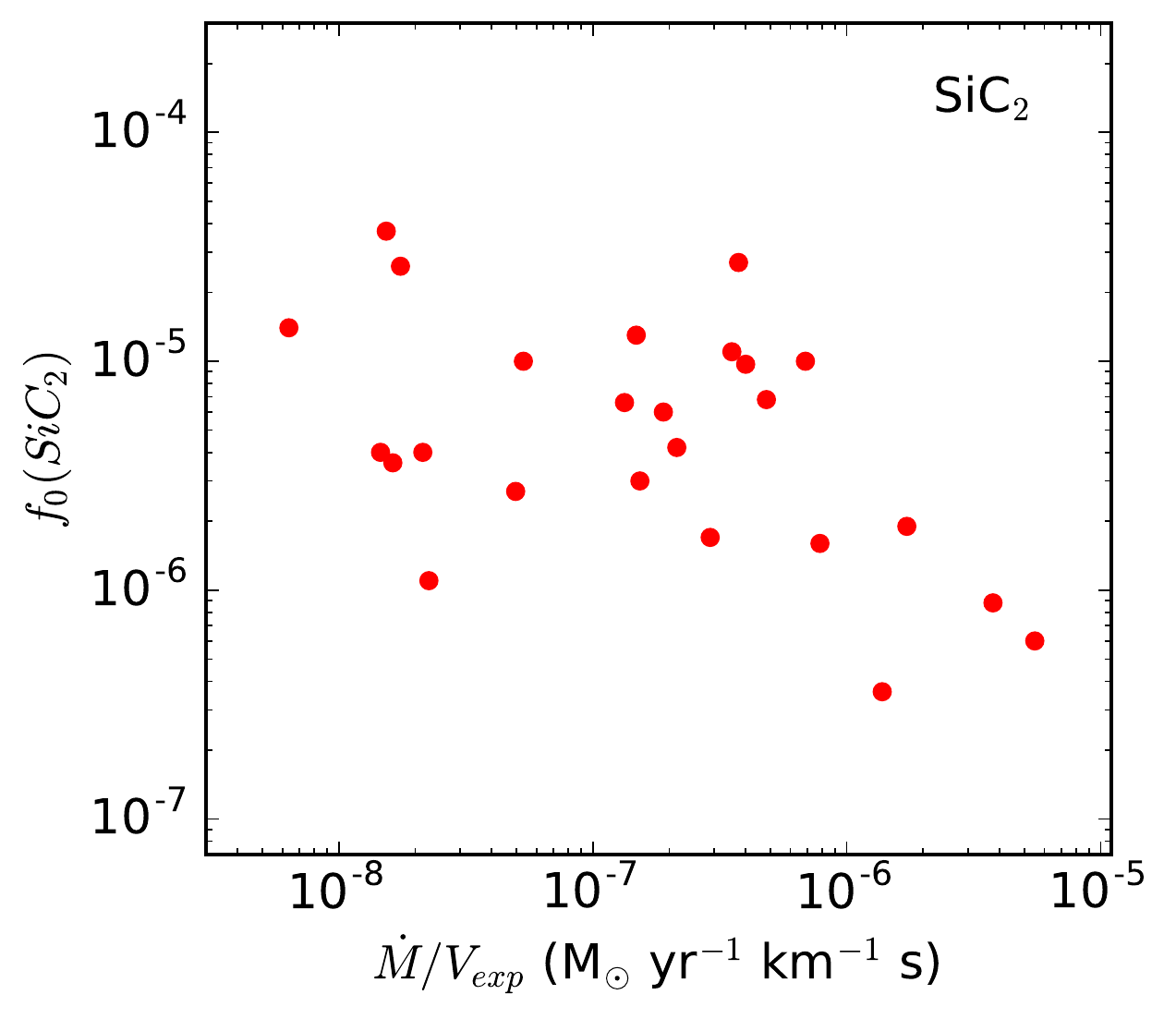}
\includegraphics[width=0.33\textwidth]{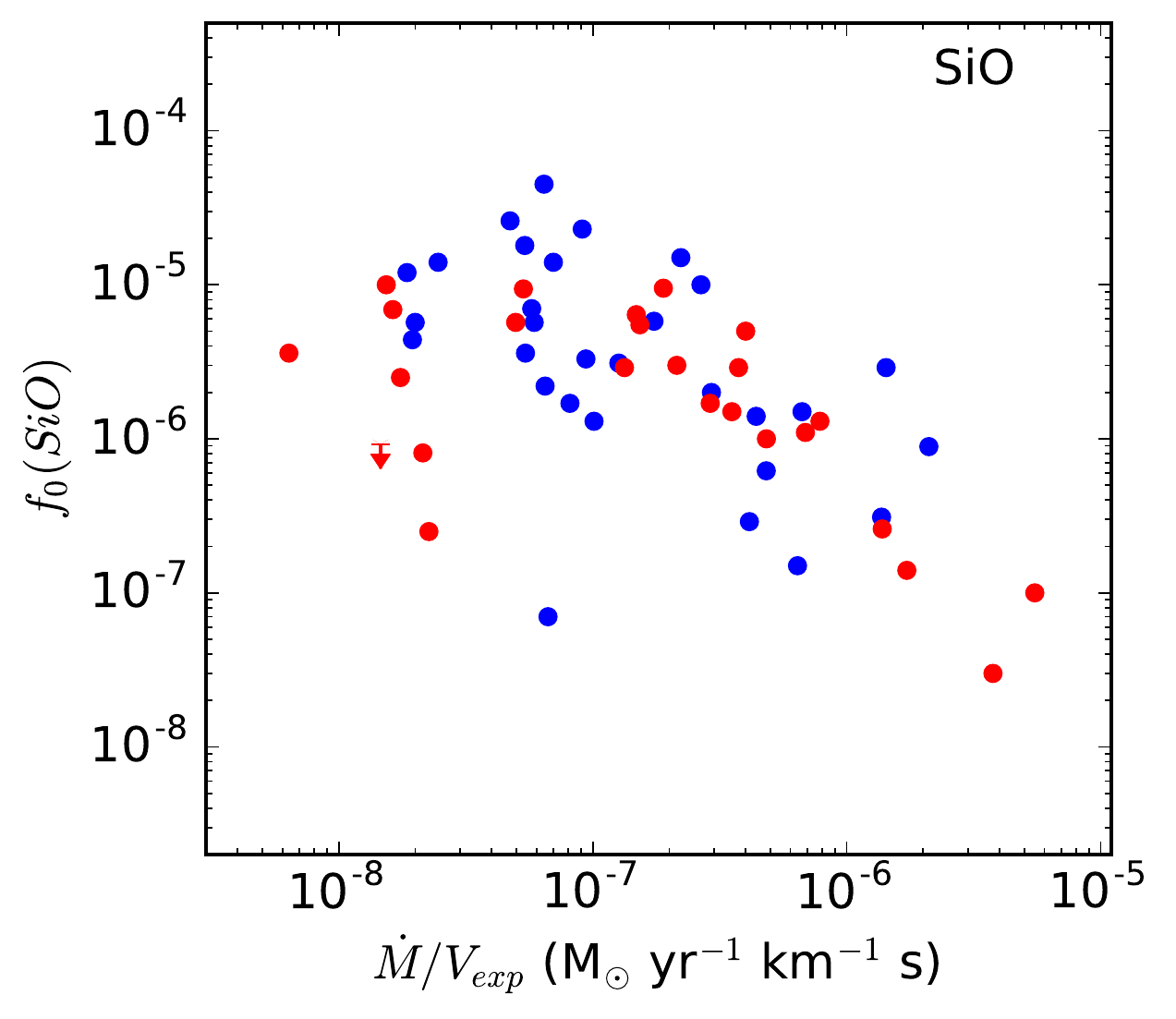}
\includegraphics[width=0.33\textwidth]{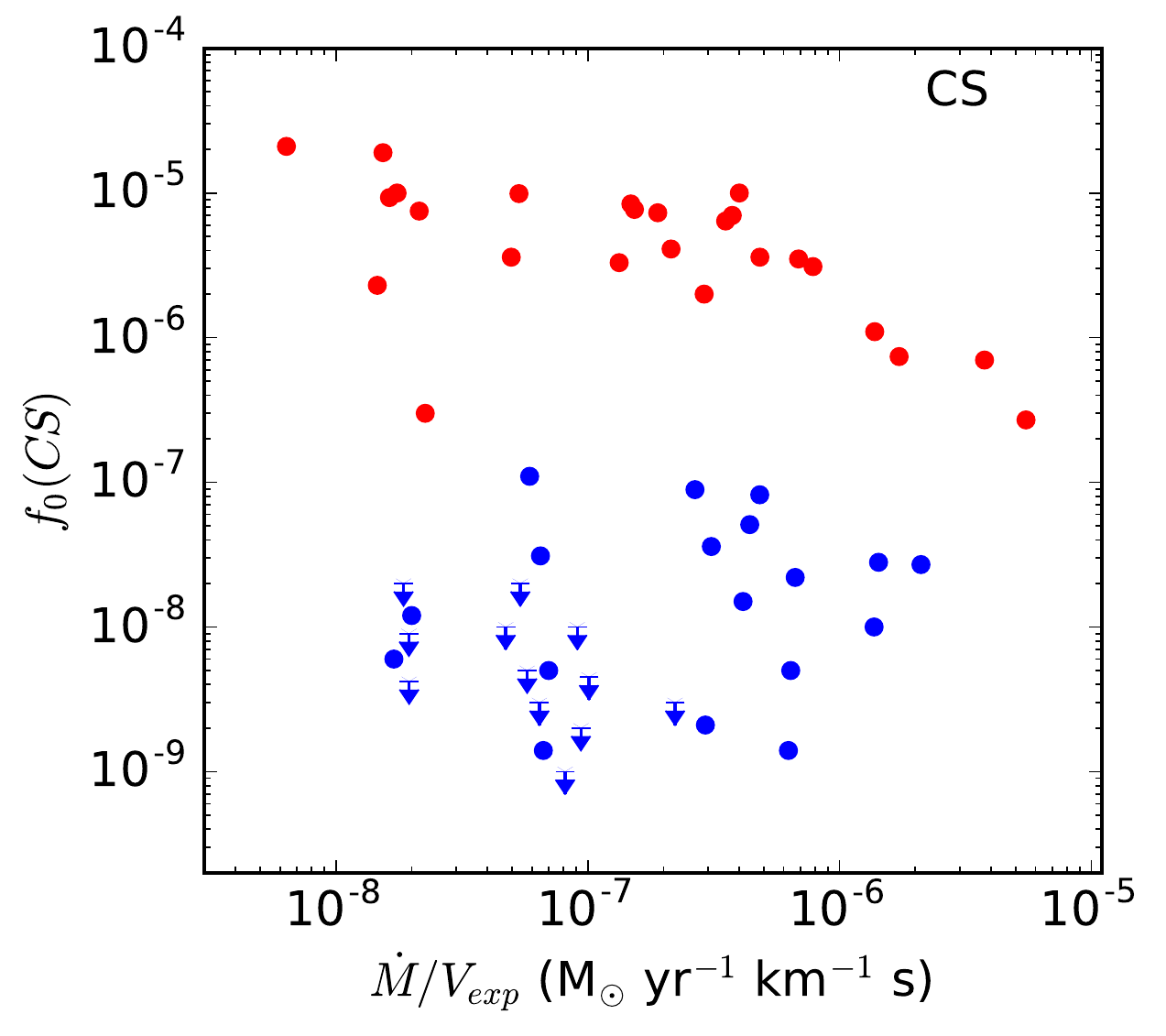}
\includegraphics[width=0.33\textwidth]{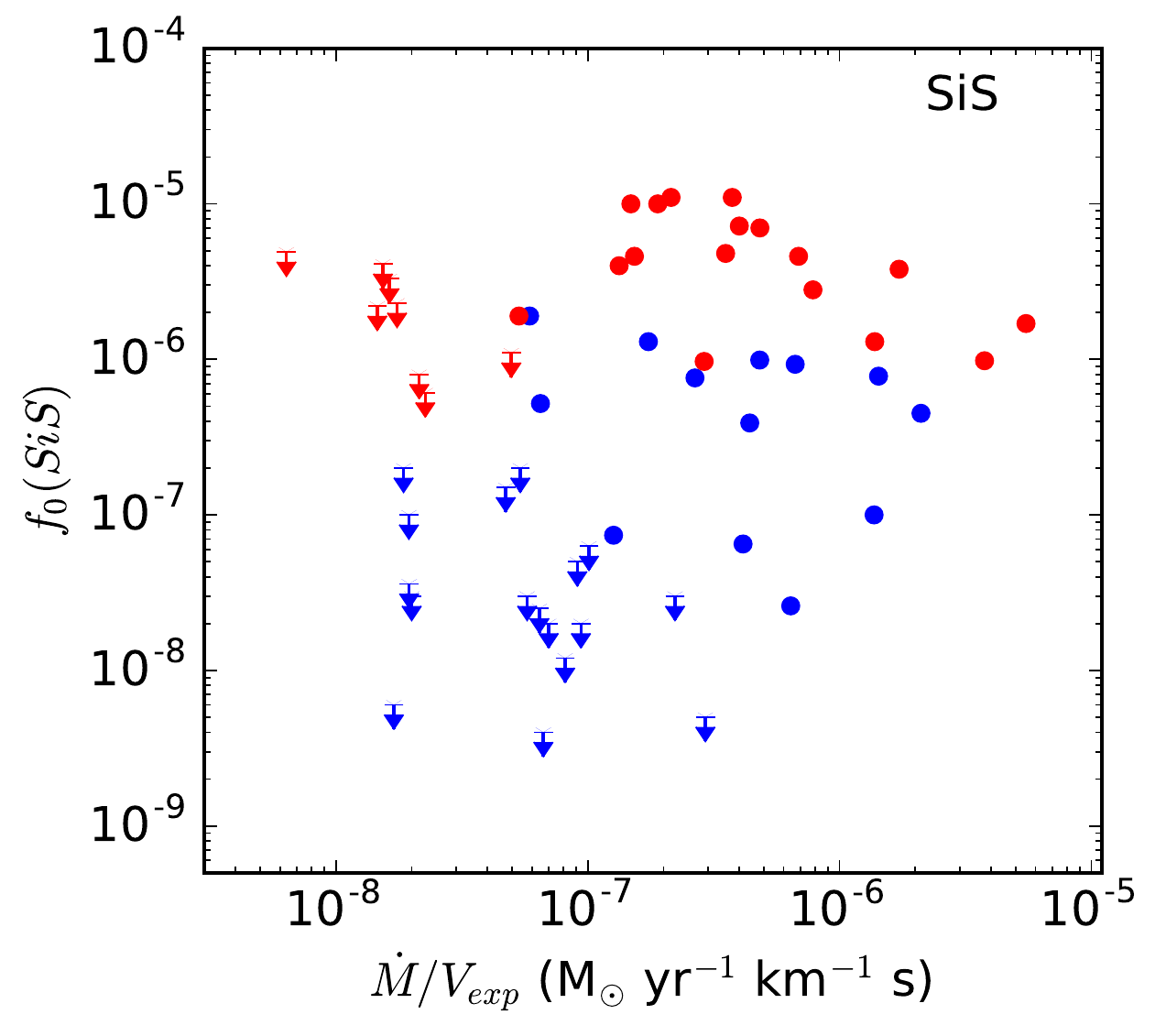}
\includegraphics[width=0.33\textwidth]{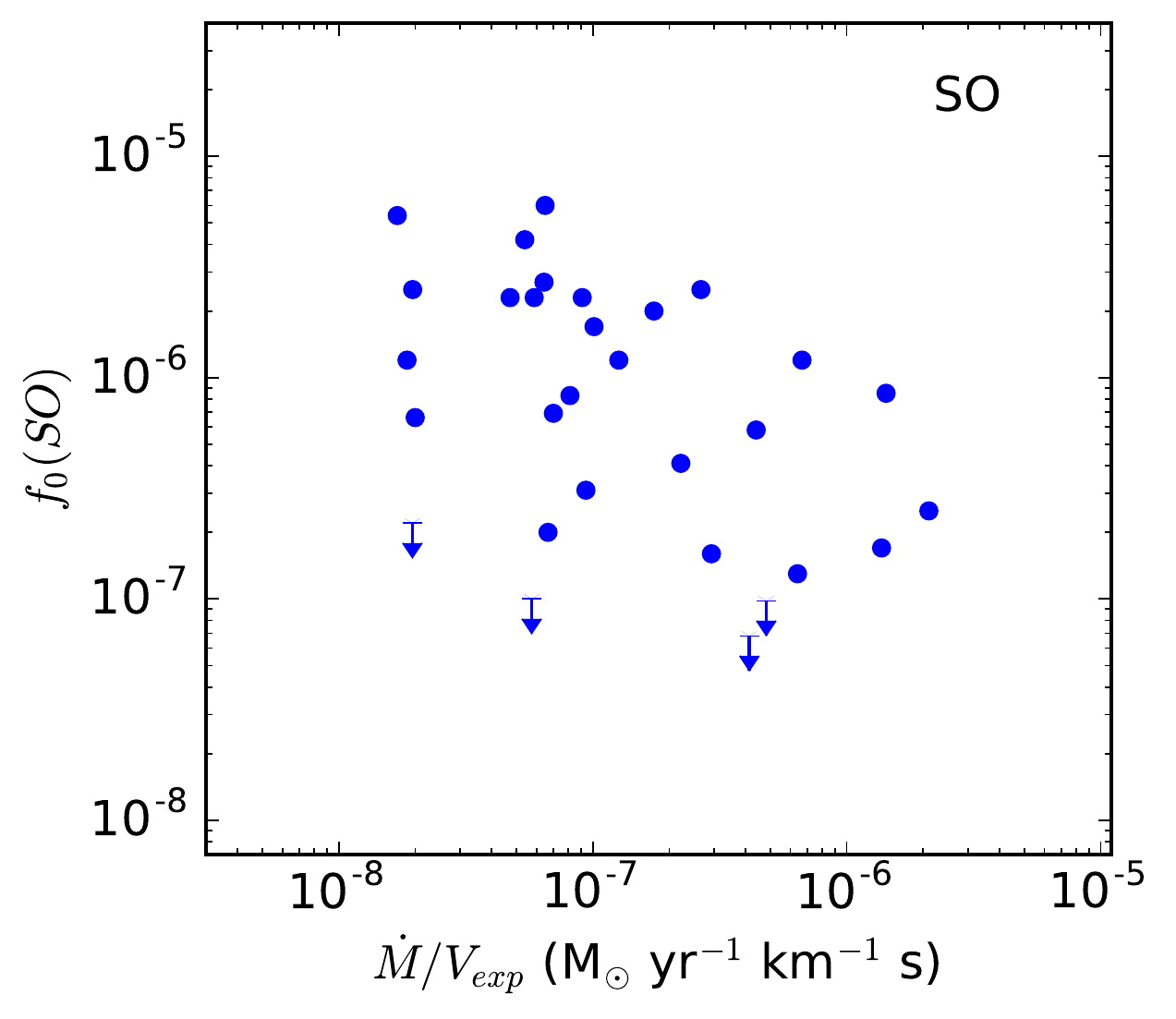}
\includegraphics[width=0.33\textwidth]{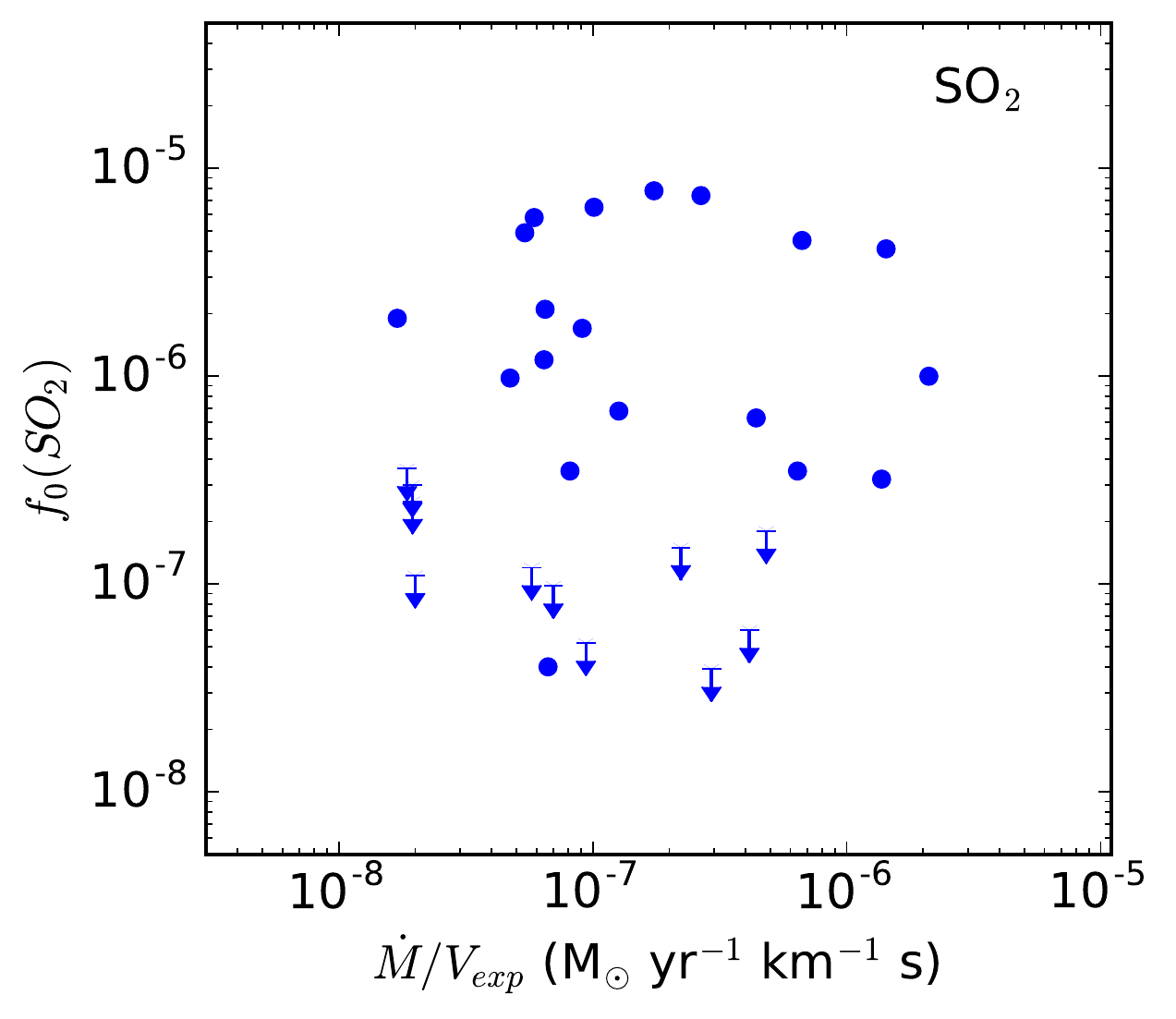}
\caption{Results for the radiative transfer and excitation analysis toward the sample of AGB stars. Fractional abundance $f_0$ derived for SiC$_2$ obtained by \citet{mas2018} for C-rich AGB stars (\textit{upper left}), fractional abundance for SiO (\textit{upper middle}), fractional abundance for CS (\textit{upper right}), and fractional abundance for SiS (\textit{lower left}) as a function of density measure $\dot{M}/V_{\rm exp}$ for oxygen stars (blue) and carbon stars (red) (\citealt{mas2019}). Fractional abundance for SO (\textit{lower middle}) and for SO$_2$ (\textit{lower right}) in O-rich envelopes as a function of density measure $\dot{M}/V_{\rm exp}$. Downward arrows represent upper limits to $f_0$.}\label{fig:trends}
\end{figure*}

The fractional abundances relative to H$_2$, $f_0$, derived for SiO, CS, SiS, SO, and SO$_2 $ in the 30 O-rich envelopes are presented in Table~\ref{table:abundances} and are shown as a function of the envelope density proxy, $\dot{M}/V_{\rm exp}$, in blue in Fig.~\ref{fig:trends}. In the panels of SiO, CS, and SiS we also include (plotted in red) the fractional abundances derived in a sample of 25 C-rich envelopes by \cite{mas2019}.

\subsection{Fractional abundances derived}

\subsubsection{SiO}

The SiO fractional abundances $f_0$(SiO) derived in this study for the 30 O-rich AGB stars are presented in blue in the upper middle panel of Fig.~\ref{fig:trends}. We overplot in red the SiO abundances derived by \citet{mas2019} for 25 C-rich AGB stars. 

One of the first comprehensive studies of molecular abundances in circumstellar envelopes was done by \cite{gon2003}, who focused on SiO in a large sample of about 40 O-rich CSEs. They used multiline data to determine the size of the emitting region and the SiO abundance simultaneously. We share 21 objects with these authors. In general, our derived SiO abundances are very similar to theirs. 

From Fig.~\ref{fig:trends}, we notice from the SiO fractional abundances we are not able to distinguish the chemical type, either O-rich (in blue) or C-rich (in red). \citet{sch2006} found the same result when comparing the distribution of their derived SiO abundances in C-rich stars to the distribution of SiO abundances in M-type stars derived by \citet{gon2003}. That is, observations indicate that the SiO abundance does not depend on the C/O ratio at the stellar surface. The mean fractional abundance of SiO we obtain is similar in both types of CSEs, with $\log f_0$(SiO) = -5.5 $\pm$ 0.7 in O-rich CSEs and $\log f_0$(SiO) = -5.8 $\pm$ 0.6 in C-rich CSEs\footnote{We consider upper limits as abundances to compute mean abundances and standard deviations. The same approach is adopted for the rest of molecules: CS, SiS, SO and SO$_2$.}.

The fact that the SiO abundance injected into the expanding wind is not sensitive to the C/O ratio is in line with theoretical expectations. Thermochemical equilibrium calculations predict that SiO maintains a uniform and high fractional abundance of several 10$^{-5}$ from the photosphere out to 10 $R_{\star}$ in O-rich CSEs, while in C-rich CSEs the predicted fractional abundance from 5 $R_{\star}$ is also on the order of 10$^{-5}$. The main difference occurs in the innermost region, from the photosphere to around 5 $R_{\star}$, where SiO has a low abundance in C-rich conditions while it maintains a high abundance in O-rich CSEs \citep{agu2006,agu2020}. In a scenario of chemical equilibrium, it seems that the low SiO abundance within 5 $R_{\star}$ in C-rich CSEs does not have an influence on the final SiO abundance that is injected into the expanding wind, and that chemical equilibrium holds for SiO in the high abundance region located beyond 5 $R_{\star}$. The low SiO abundance predicted by chemical equilibrium in the innermost envelope has been inferred by modeling multiwavelength observations of the carbon star IRC\,+10216 by \cite{sch2006_irc}. The nonequilibrium scenario of shocks induced by the stellar pulsation of \cite{che2006} also predicts a low sensitivity of the SiO abundance on the photospheric C/O ratio. In the model of \cite{che2006}, for C/O $<$ 1 shocks have a very limited effect on the SiO fractional abundance in the inner part of the wind as it stays around $\sim 10^{-5}$ from the photosphere out to 5 $R_{\star}$. For C/O $>$ 1, the authors find that the low chemical equilibrium abundance of $\sim 10^{-8}$ is enhanced rapidly to values around $10^{-5}$ in the 1-5 $R_{\star}$ region. Therefore, the SiO abundances calculated at 5 $R_{\star}$, which are supposed to be the ones injected into the expanding wind, are of the same order in O-rich and C-rich stars. In summary, theoretical studies show that the SiO abundance has no apparent dependence on the C/O ratio in the outer wind which is in agreement with the findings from our observational study. The different behavior of the SiO abundance in the inner wind as predicted by theoretical studies probably explains why strong SiO maser emission is detected toward O-rich stars and not toward carbon stars (e.g., \citealt{par2004,cot2004}). In view of the high SiO abundance observed in C-rich stars, this could suggest that the SiO molecules are formed further out in the wind in C-rich envelopes where the physical conditions are not likely to allow the pumping by IR photons, and thus the inversion of SiO level populations.

The fractional abundance of SiO in the O-rich sample varies substantially from as low as $7.0\times10^{-8}$ up to $4.5\times10^{-5}$. This variation in the SiO abundance as illustrated in Fig.~\ref{fig:trends}, whether C-rich or O-rich, shows a clear trend in which SiO becomes less abundant as the density in the wind, $\dot{M}/V_{\rm exp}$, increases. \citet{sch2006} and \citet{mas2019} presented analysis of circumstellar SiO abundances for carbon stars, and \citet{gon2003} for M-type stars and likewise they find a similar behavior of a strong anticorrelation between the abundance and the wind density which was interpreted as an effect of increased adsorption of SiO onto dust grains at high densities. Here, we confirm the results found for O-rich stars by \citet{gon2003}. We found a similar trend when we investigated SiC$_2$ in a sample of 25 carbon-rich AGB stars (see upper left panel in Fig.~\ref{fig:trends} for comparison), which was interpreted as that SiC$_2$ is being efficiently incorporated into dust grains and playing an important role in the formation of silicon carbide dust in C-rich envelopes \citep{mas2018}. Adsorption of SiO onto dust grains in O-rich envelopes is predicted theoretically by chemical kinetics models \citep{van2019}. We note that the median fractional abundance of SiO of the Mira-type variables, $2.5\times10^{-6}$, is lower by a factor of 6 with respect to the median value of the semiregular variables, $1.5\times10^{-5}$, which may be related to the mass loss rate rather than the stellar variability type. \citet{gon2003} found a similar result where the high mass-loss rate Miras in their sample have a median abundance that is more than six times lower than that of the irregular and semiregular variables.

Silicates are known to be one of the most important types of dust in oxygen-rich envelopes and SiO has long been discussed to be the gas-phase precursor of silicate dust, mainly because of its high abundance in O-rich envelopes. The trend that we see here between the fractional abundance and the wind density supports this hypothesis.

\subsubsection{CS}

The fractional abundances derived for CS in the O-rich envelopes are shown in blue as a function of $\dot{M}/V_{\rm exp}$ in the upper right panel of Fig.~\ref{fig:trends}. We also show in red the CS fractional abundances derived for the 25 C-rich AGB stars studied in \citet{mas2019}. 

\citet{buj1994} searched for CS $J=3-2$ and $J=5-4$ transitions in a large sample of evolved stars. Their sample contains 17 O-rich stars, 10 of which are in our sample. These authors derived abundances using a simple analytical expression based on the integrated intensities of the observed lines and assumes a constant fractional abundance inside a given radius. In general, their CS abundances are higher than ours with varying degrees, for example, ranging from a factor of two for some sources, like V1300\,Aql, to one order of magnitude for other sources, like IK\,Tau, to a highest factor of 47 for RX\,Boo, where our value is an upper limit. These authors remark that their approach holds for optically thin lines and estimated only a lower limit if the line was optically thick. \citet{dan2018} surveyed a diverse sample of AGB stars. They detected CS in only the highest mass loss rate O-rich stars and derived CS abundances in agreement with ours for some sources, such as GX\,Mon and V1300\,Aql, while for other sources their derived values were approximately an order of magnitude higher than ours, such as IK\,Tau and RR\,Aql, the latter being an upper limit in both studies.

Comparing the values of $f_0$(CS) in oxygen-rich and carbon-rich envelopes in Fig.~\ref{fig:trends}, the derived abundances show substantial variations between the two chemical types where the mean fractional abundance for O-rich CSEs is $\log f_0$(CS) = -8.0 $\pm$ 0.6,  more than two orders of magnitude lower than for C-rich CSEs, $\log f_0$(CS) = -5.4 $\pm$ 0.5. It is clear that the formation of CS is dependent on the photospheric C/O ratio of the star. We also notice that CS is mostly detected in O-rich CSEs with high mass-loss rates, while in C-rich CSEs, CS is detected in all the sources of the sample, regardless of the mass loss rate. Carbon monosulfide forms more readily in C-rich environments since there is available carbon, that is, not trapped by CO, to form C-bearing molecules. On the other hand, the formation of CS in O-rich CSEs is more surprising as all the available carbon is expected to be locked up in CO.

Chemical equilibrium calculations predict negligible abundances for CS in O-rich CSEs, more than 3-4 orders of magnitude below the observed values \citep{agu2020}. It is clear that some nonequilibrium process is enhancing the abundance of CS in O-rich envelopes. A possible explanation for the synthesis of this molecule could be related to photochemistry in a clumpy CSE, as investigated by \cite{agu2010}. In this scenario, interstellar UV photons penetrate into the inner regions of the envelope, break the CO bond and induce changes in the chemical composition which ultimately allow for the formation of CS and other C-bearing molecules. Their calculations predict abundances of $\sim 10^{-9}-10^{-8}$ for mass-loss rates in the 10$^{-7}$-10$^{-5}$ M$_{\odot}$ yr$^{-1}$, in agreement with the abundances we find here. Similar models by Van de Sande et al. (2018) examined the effects that clumping and porosity have on the chemistry in the AGB outflow and found slightly higher peak abundances of $\sim10^{-7}$. However, in their recently published corrigendum these authors no longer find this peak, instead the fractional abundance of CS drop to $\sim10^{-10}$ \citep{van2020}. Another explanation for the formation of CS in O-rich environments could be related to the variable nature of AGB stars. Periodic shock waves caused by stellar pulsations propagate through the photosphere and alter the gas chemistry where the collisional destruction of CO in the shocks could release free atomic carbon and trigger the formation of CS in O-rich environments \citep{dua1999,che2006,gob2016}. The shock-induced chemistry model of \citet{che2006} predicts that CS reaches abundances of a few times 10$^{-6}$ in envelopes with C/O $<$ 1, that is to say, significantly above the mean abundance derived here from observations.

To investigate the reason for the nondetection of CS in the low mass loss rate objects, we consider the variability type (Mira variable and semiregular variable, see Table~\ref{table:abundances}) of the O-rich stars. The type of variability is generally attributed to the pulsation of the star and therefore could influence the shock conditions and provide an explanation of the abundance differentiation between high- and low-mass loss rate O-rich stars. However, we see no indication of a dependence between the abundance of CS and the variability type. On one hand, the nondetection of CS in these envelopes could be due to a low fractional abundance of the molecule, on the other hand, it could be due to a lack of sensitivity.

The CS fractional abundance in the O-rich sample varies by about two orders of magnitude, ranging from as low as $1\times10^{-9}$ to as high as $1.1\times10^{-7}$, yet unlike the case of SiO, this variation shows no apparent trend that the CS abundance decreases as the density in the wind ($\dot{M}/V_{\rm exp}$) increases for O-rich envelopes. Such a trend is however evident for carbon-rich envelopes. This suggests that CS molecules are more likely to adsorb onto dust grains in C-rich CSEs than in O-rich ones. While CS is thought to play a role in the formation of MgS dust in C-rich envelopes \citep{mas2019}, in the case of O-rich envelopes CS does not seem to be affected by adsorption onto dust grains and to be playing a role in the formation of dust.

\subsubsection{SiS}

In the lower left panel of Fig.~\ref{fig:trends} we show as a function of $\dot{M}/V_{\rm exp}$ the fractional abundances of SiS derived in the 30 O-rich envelopes studied here (in blue) and in the 25 C-rich envelopes studied by \citet{mas2019} (in red). 

\citet{sch2007} reported on the detection of SiS line emission in 8 oxygen-rich envelopes, all of which are included in our sample. They performed radiative transfer calculations to derive abundances adopting, similarly to us, an abundance distribution based on the scaling law established by \citet{gon2003} for SiO in M-type stars. Our SiS abundances are in good agreement with theirs for all of the sources. 

By looking to the fractional abundances $f_0$(SiS) derived in the O-rich sample (see Table~\ref{table:abundances}) we notice that they vary considerably among different sources, between $<4.0\times10^{-9}$ and $1.9\times10^{-6}$. The mean fractional abundance in the O-rich sample is $\log f_0$(SiS) = -7.0 $\pm$ 0.7, while in C-rich AGB stars is $\log f_0$(SiS) = -5.5 $\pm$ 0.4, that is, an order of magnitude higher than in O-rich envelopes. Similarly, in their study of SiS in a small sample of oxygen- and carbon-rich envelopes, \citet{sch2007} found SiS abundances in carbon-rich envelopes about an order of magnitude higher than in oxygen-rich envelopes. This indicates that SiS has a marked chemical differentiation based on the photospheric C/O ratio, being more preferentially formed in C-rich environments than in O-rich ones.

According to chemical equilibrium, in C-rich envelopes SiS reaches a high fractional abundance of about 10$^{-5}$ from 2 $R_{\star}$, while in O-rich envelopes the abundance of SiS is low in the very inner regions but rises to values slightly below 10$^{-5}$ from 2 $R_{\star}$ beyond 5 $R_{\star}$ \citep{agu2020}. That is, the differentiation between C-rich and O-rich is restricted to the very inner regions, but beyond 5 $R_{\star}$ SiS is predicted to reach high abundances, on the order of 10$^{-5}$, in both C- and O-rich envelopes. The low SiS abundances observed here in some O-rich envelopes are thus not expected according to chemical equilibrium. On the other hand, nonequilibrium chemical models based on shocks induced by the stellar pulsation predict a dependence of the SiS abundance on the C/O ratio, with abundances on the order of 10$^{-5}$ and 10$^{-8}$ at 5 $R_{\star}$ in the inner regions of C- and O-rich, respectively, envelopes \citep{che2006}. The SiS abundances derived from observations here agree with the predictions of these models in terms of differentiation based on the C/O ratio, although there is a discrepancy because some of our observed abundances are significantly above those predicted by the model. For example, \citet{che2006} predicts an SiS fractional abundance for TX\,Cam of $\sim10^{-8}$ at 5 $R_{\star}$, while for this source we derive an abundance of $3.9\times10^{-7}$. Therefore, chemical equilibrium overestimates the SiS abundances in O-rich envelopes while nonequilibrium models underestimate them.
 
Strikingly, we noticed that SiS is not detected in envelopes with low mass loss rates $<10^{-6}$, while it is detected in all sources above this threshold. This fact has been reported by some previous observational studies \citep{buj1994,dan2015,dan2018,mas2019}. \cite{mas2019} surveyed a sample of 25 C-rich AGB stars in SiS $J=8-7$ and $J=7-6$ emission and did not detect emission below the same threshold as well. They speculated that the nondetection of SiS in the low mass-loss rate C-rich envelopes could be either caused by a lack of the constituent elements, which would be trapped in other S- and Si-bearing molecules like SiO, SiC$_2$, and CS, or could be due to sensitivity which might be the case here as well. \citet{dan2019} discussed based on previous studies that SiS does not form readily in low-mass loss rate semi-regular variables, where the molecule otherwise reaches higher abundunces in Mira variable type CSEs. However, these authors detect faint SiS $J=19-18$ emission toward the low mass-loss rate semiregular variable, R\,Dor, using ALMA and derive an abundance of $1.5\times10^{-8}$ which indicates that the nondetection of SiS in low mass loss rate semiregular variables could be due to low sensitivity. In this study, we do not detect SiS in low-mass loss rate objects of both, semiregular and Mira variables, that is to say, we do not see a dependence of the SiS fractional abundance on the variability type, similar to the case of CS. 

While SiS shows a tentative trend of a decreasing abundance with increasing envelope density for C-rich CSEs \citep{mas2019}, which was interpreted in terms of adsorption onto dust grains, here we do not see any similar trend, implying that SiS is probably not an important gas-phase precursor of dust in O-rich CSEs. However, chemical kinetics models by \citet{van2019} predict adsorption of SiS onto dust grains in O-rich outflows. 

\subsubsection{SO}

The resulting SO fractional abundances are shown as a function of $\dot{M}/V_{\rm exp}$ in the lower middle panel of Fig.~\ref{fig:trends}. Systematic studies of the abundance of SO on large samples of AGB stars are scarce. In fact, for some of the sources in our sample, SO abundances are reported for the first time. One of those studies was made by \citet{buj1994}, who surveyed a large sample of evolved stars in several molecular lines, including SO 6$_5-$5$_4$. Their sample contains 18 O-rich objects, 11 of which are in our sample. For some sources, our derived abundances are higher than theirs, and for other sources, the opposite is found. But in general the difference is within a factor of a few. The modeling performed by these authors is based on a somewhat simple method as mentioned previously, in which an analytical expression is used to derive the abundance of the molecules. A more complex abundance derivation based on radiative transfer modeling was done by \citet{dan2016} using high- and low-$E_u$ lines of SO (and SO$_2$; see below) in a small sample of five M-type AGB stars, four of which are in our sample. The authors find that the spatial distribution of SO differs between the low mass loss rate objects (R\,Dor, and W\,Hya) and the high mass loss rate ones (IK\,Tau, R\,Cas, and TX\,Cam), where the former were best reproduced by a Gaussian disribution whereas the latter by a shell-like one. For the four sources we have in common, their derived abundances are higher than ours by a factor of a few. In their recent study of two O-rich envelopes using ALMA, \citet{dan2020} confirmed their previous findings in that the SO abundance distribution in IK\,Tau is shell-like with a constant inner abundance of $4.1\times10^{-7}$, not very different from the value derived in this study ($1.7\times10^{-7}$), that increases to $2.2 \times 10^{-6}$ at $5 \times 10^{15}$ cm followed by a decline at $e-$folding radius $1.3 \times 10^{16}$ cm. \citet{vel2017} also surveyed IK\,Tau and derived $f_0$(SO) $\geq 8\times10^{-6}$, which is significantly higher than the value derived here ($1.7\times10^{-7}$). These authors discuss that their derived SO abundance for this source may be overestimated since it is higher than previous observational studies and higher than abundances predicted by chemical equilibrium models, and that the reason behind this discrepancy is the uncertainty in the adopted SO emitting region. 

The values of $f_0$(SO) range between $<6.8\times10^{-8}$ and $6\times10^{-6}$ and have a mean fractional abundance of $\log f_0$(SO) = -6.1 $\pm$ 0.6. Chemical equilibrium calculations predict a peak SO abundance in the 1-10 $R_{\star}$ region of $\sim10^{-7}$ \citep{agu2020}, while nonequilibrium chemical models considering shocks induced by the pulsation of the star predict similar abundances at 5 $R_{\star}$ \citep{che2006}. Therefore, on average, our observed SO abundances are higher than theoretical predictions of the inner wind. For SO, we see no dependence of the fractional abundance on the stellar variability type.

The distribution of the fractional abundances derived in the O-rich sample show hints of decreasing SO abundance with increasing density. This is however tentative as it is not as evident as in the case of SiO. \citet{dan2016} found a similar trend of SO being less abundant with wind density, although this result was based on a reduced sample of only 3 objects (TX\,Cam, IK\,Tau, and R\,Cas) with a limited range of mass loss rates. If the tentative decrease in the abundance of SO with increasing envelope density that we see here is interpreted in terms of adsorption onto dust grains, SO could emerge as a candidate to gas-phase precursor of dust. To date, no sulfur-containing condensate has been identified in the spectra of O-rich envelopes, although CaS and FeS are expected to be important solid carriers of sulfur in these environments \citep{lod1999,agu2020}.

\subsubsection{SO$_2$}

The fractional abundances derived for SO$_2$ are shown as a function of $\dot{M}/V_{\rm exp}$ in the lower right panel of Fig.~\ref{fig:trends}. For some of the sources in our sample, SO$_2$ abundances are reported for the first time.

Infrared observations, in particular, the ISO/SWS detection of the \SI{7.3}{\micro\meter} $\nu_{3}$ band in a few AGB stars by \citet{yam1999} and observations of high energy rotational lines by, e.g., \cite{dan2016} and \cite{vel2017} indicate that SO$_2$ is formed in the inner layers of the CSE. \citet{omo1993} surveyed a diverse sample of evolved stars in sulfur-bearing molecules and derived SO$_2$ abundances for 7 of the objects in our sample. For some sources, the abundances derived by these authors are similar to the values derived in this work. While for other sources, the derived abundances are different, like RX\,Boo, their value is 50 times higher than ours. These authors used a relatively simple method for estimating the molecular abundances, which is based on an analytical expression in which they assumed a constant excitation temperature for simplicity. A study was conducted by \citet{dan2016} to investigate the SO$_2$ rotational lines observed with Herschel/HIFI in addition to further archival data toward a small sample M-type AGB stars. They performed radiative transfer modeling and derived SO$_2$ abundance toward three of the objects that are in our sample, IK\,Tau, W\,Hya, and R\,Cas. They find SO$_2$ fractional abundance in IK\,Tau similar to ours assuming a Gaussian distribution. They note that their SO$_2$ model for IK\,Tau is uncertain due to the difficulty of determining an abundance distribution. In fact, in their recent study using ALMA, these authors suspect that the SO$_2$ abundance in IK\,Tau is consistent with a shell-like distribution and not a Gaussian distribution \citep{dan2020}. For W\,Hya and R\,Cas, their derived abundances are higher than ours, by a factor of 50 (with ours being an upper limit) and an order of magnitude, respectively. They also note that their SO$_2$ model for R\,Cas is very uncertain due to the fact that they had only two detected lines toward that source. The mean fractional abundance of SO$_2$ in the 30 O-rich envelopes studied here is  $\log f_0$(SO$_2$) = -6.2 $\pm$ 0.7, that is, very similar to that of SO.   

Sulfur dioxide is predicted to have low abundances ($<10^{-10}$), well below the observed values, in the inner regions of O-rich envelopes according to chemical equilibrium \citep{agu2020}. There must be a nonequilibrium process that enhances the formation of SO$_2$ in the inner envelope. The shock-induced chemistry scenario of \cite{che2006} also predicts very low abundances (10$^{-13}$-10$^{-12}$) for SO$_2$ in the inner winds of O-rich AGB stars. Clearly, observations and theory maintain a severe discrepancy with respect to the abundance of SO$_2$ in the inner envelope of M-type stars. Similar to SO, we see no dependence of the SO$_2$ fractional abundance on the stellar variability type.

In their study on sulfur molecules in M-type AGB stars, \citet{dan2016} reported that SO and SO$_2$ are the main reservoirs of sulfur in the inner regions of the CSE of W\,Hya and R\,Cas, with more uncertainties for the latter. For W\,Hya, they derived a combined fractional abundance of SO and SO$_2$ of $\sim10^{-5}$ within the inner layers of the wind, thus accounting for most of the sulfur. Here in this work, the combined fractional abundance of CS, SiS, SO, and SO$_2$ in the intermediate regions of the W\,Hya envelope is just $\sim8\times10^{-7}$, well below the elemental abundance of S. This could point to depletion of sulfur through dust condensation in this object. A large fraction of the sulfur could also be trapped as gaseous H$_2$S, which is abundant in O-rich CSEs \citep{dan2017}. In any case, for SO$_2$ we do not see any clear trend of decreasing $f_0$(SO$_2$) with increasing envelope density that could point to this molecule as a gas-phase precursor of dust in O-rich envelopes.

\begin{figure}
\centering
\includegraphics[width=0.85\columnwidth]{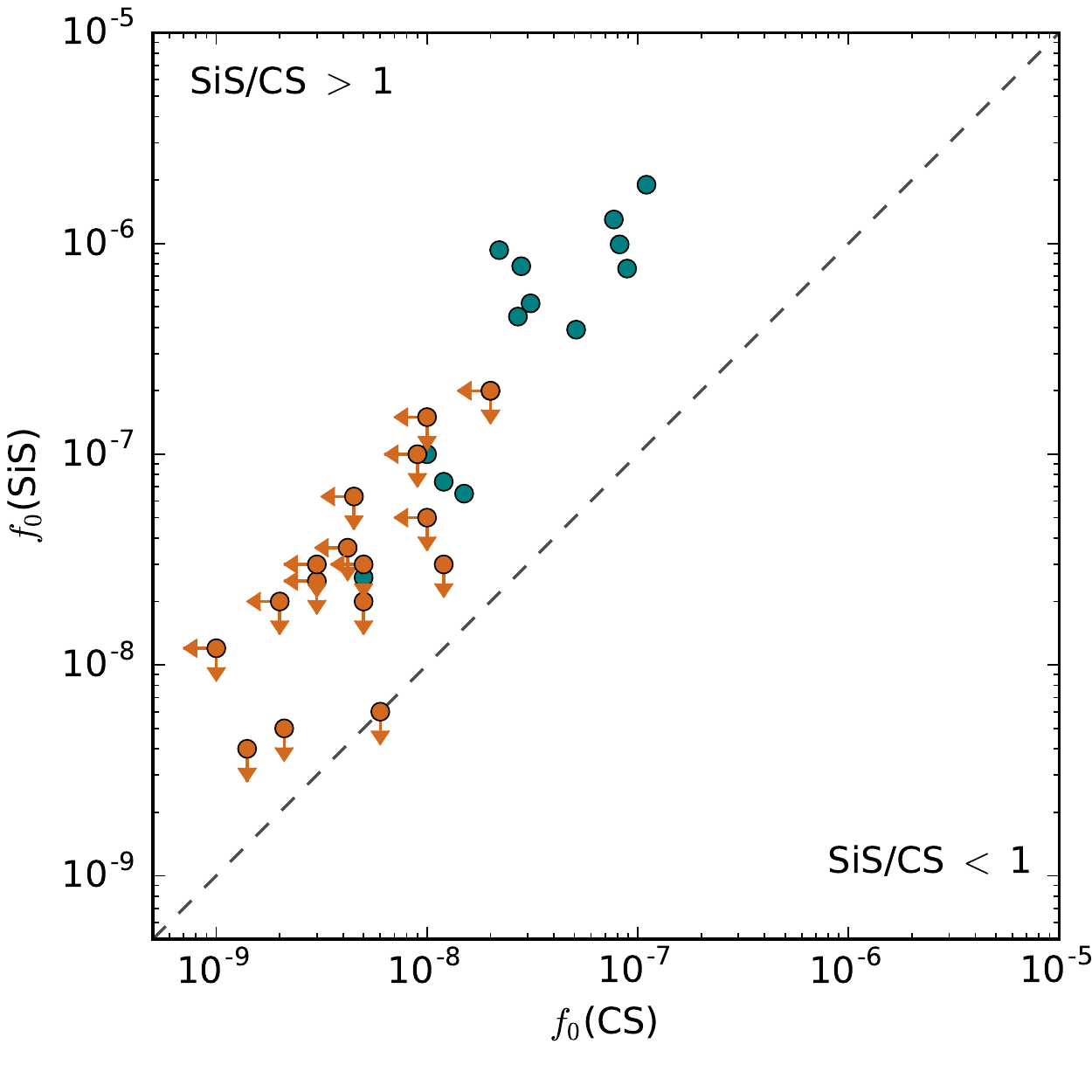}
\includegraphics[width=0.85\columnwidth]{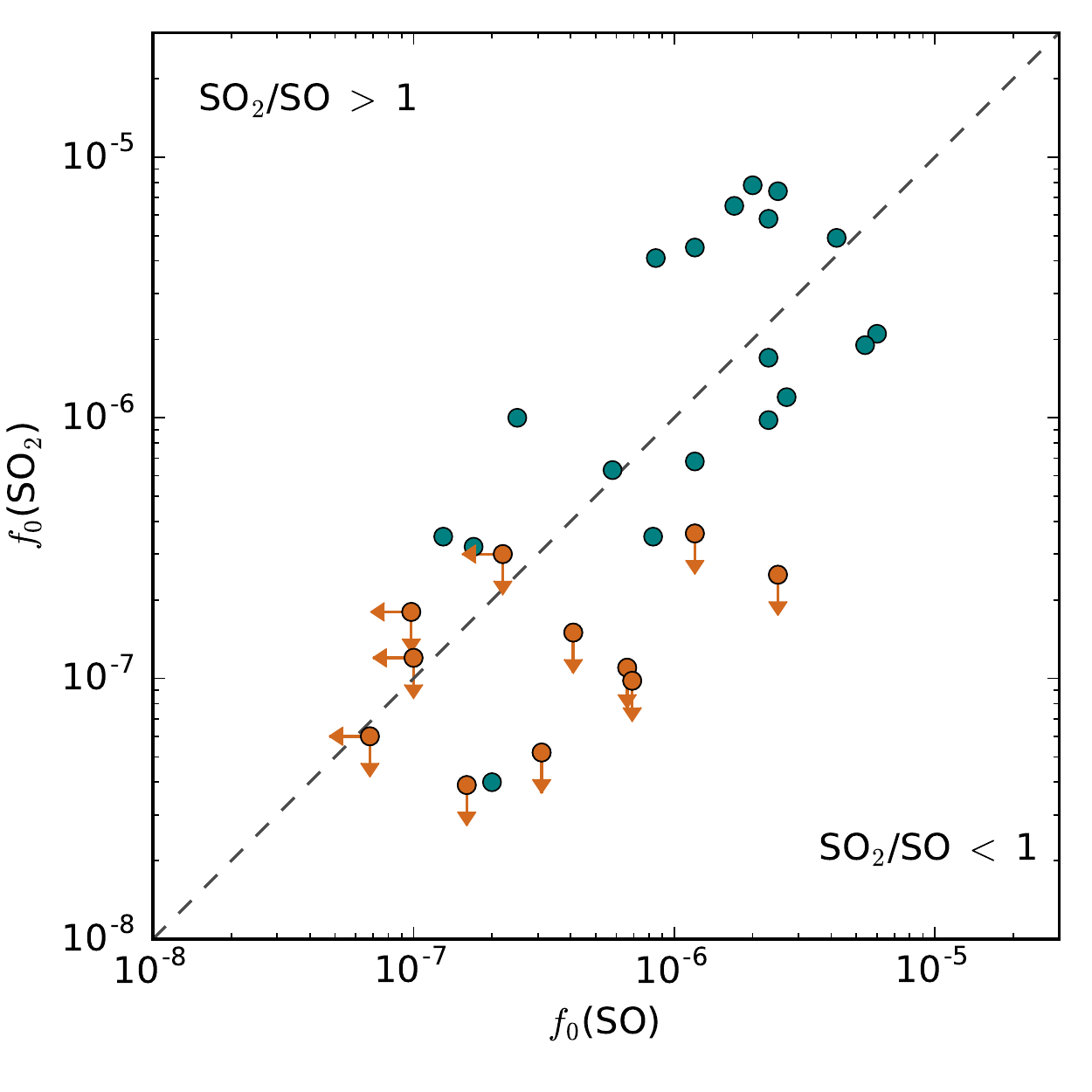}
\caption{Comparison of abundances between different pairs of molecules. The derived fractional abundances relative to H$_2$ of CS vs SiS (\textit{upper panel}) and SO vs. SO$_2$ (\textit{lower panel}). Those sources with nondetections are denoted with arrows. The diagonal dashed lines indicate where the abundances of the two molecules become equal.The orange circles indicate fractional abundances with an upper limit.}
\label{fig:ratios}
\end{figure}

\subsection{Correlations between abundances of different molecules}

In Fig.~\ref{fig:ratios} we plot the derived fractional abundance of CS against that of SiS (upper panel) and the fractional abundance of SO against that of SO$_2$ (lower panel). We find that SiS is systematically more abundant than CS in the 30 O-rich envelopes studied, as indicated by the fact that all sources lie in the region of SiS/CS $>$ 1, apart from EP\,Aqr that falls on the dashed line representing equal amounts of SiS and CS. Similarly, \citet{dan2018} determined the CS and SiS abundances in a sample of AGB stars, and found SiS to be systematically more abundant than CS in their O-rich sample. Therefore, SiS seems to be a more abundant gas-phase reservoir of sulfur than CS in oxygen star envelopes. The behavior is thus different to that of C-rich envelopes, where CS and SiS have comparable abundances \citep{mas2019}. Moving on to the lower panel in Fig.~\ref{fig:ratios}, the comparison of SO and SO$_2$ shows that in some sources SO is more abundant, like in R\,LMi and BK\,Vir, while in other sources SO$_2$ is more abundant, like in NV\,Aur and V1111\,Oph. In general, the data points fall along the line defined by $f_0$(SO) = $f_0$(SO$_2$) and there is no clear preference for either the SO/SO$_2$ $>$ 1 or the SO/SO$_2$ $<$ 1 sides. By looking to those sources where both SO and SO$_2$ are detected, it seems that in oxygen-rich envelopes, SO and SO$_2$ have abundances of the same order, carrying similar amounts of sulfur.

Regardless of which pair of molecules from those shown in Fig.~\ref{fig:ratios}, in both cases there is a trend in which the higher the abundance of one molecule the more abundant the other is, ie., the abundances of SO and SO$_2$, and of SiS and CS, seem to scale with each other which suggests a chemical connection between the members of each couple of molecules. \citet{dan2018} found this type of correlation for CS and SiS in a sample including C-, M-, and S-type stars, although in that study the trend is considered tentative because of the small number of sources included. \cite{mas2019} also found a similar correlation between the three molecules CS, SiO, and SiS in their large sample of C-rich CSEs. We remark, however, that the trends in Fig.~\ref{fig:ratios} become less robust given the upper limits on some of the fractional abundances in the sources where these molecules are not detected.

\section{Conclusion}\label{sec:conclusion}

In this study we observed SiO, CS, SiS, SO, and SO$_2$ using the IRAM 30m telescope in a statistically meaningful sample of 30 O-rich AGB stars covering a wide range of mass-loss rates and circumstellar properties. We performed an extensive radiative transfer and excitation analysis based on the LVG method to derive the fractional abundance of these molecules in the circumstellar envelopes. 

We found that the derived circumstellar abundances of SiS and CS have a clear dependence on the photospheric C/O ratio of the star, while SiO is not sensitive to it. Moreover, the fractional abundance of CS and SiS in carbon-rich CSEs are about two and one orders of magnitude, respectively, higher than in oxygen-rich envelopes, whereas the fractional abundance of SiO in both chemical types is of the same order of magnitude. Chemical equilibrium correctly predicts that SiO is abundant and that SiS and SO can reach high abundances in O-rich stars. However, the observed abundances of CS and SO$_2$ are higher than predicted by several orders of magnitude. Nonequilibrium chemical models succeed to different extents in reproducing the observed abundances. A scenario of photochemistry in a clumpy envelope accounts for the abundance enhancement of CS. On the other hand, a scenario of shocks induced by the stellar pulsation results in abundances that are 1-3 orders of magnitude too high for CS, somewhat lower than observed for SiS and SO, and well below the observed values for SO$_2$.

We find that the abundances of SiS and CS, on one hand, and SO and SO$_2$, on the other, are positively correlated which suggests a chemical connection between the members of each couple. Moreover, as already found for C-rich envelopes, we find a clear trend of decreasing SiO abundance with increasing envelope density in O-rich envelopes, which points to adsorption of SiO onto dust grains. A similar trend is observed for SO, although not as clear as for SiO. Therefore, SiO and SO are likely candidates to act as gas-phase precursors of dust in O-rich envelopes. In the cases of CS, SiS, and SO$_2$, abundances span over 2-3 orders of magnitude with no obvious correlation with the envelope density. These three molecules are thus less attractive candidates to be precursors of dust.

Our conclusions on the role of these molecules as gas-phase precursors of dust are based on low energy rotational lines, which probe post-condensation regions. More observations, in particular high-J lines and interferometric observations probing the inner regions of the envelopes, are needed to affirm the conclusions obtained in this study.

\begin{acknowledgements}

We thank the IRAM 30m staff for their help during the observations. This research has made use of the SIMBAD database, operated at CDS, Strasbourg, France. We acknowledge funding support from the European Research Council (ERC Grant 610256: NANOCOSMOS) and from Spanish MINECO through grant AYA2016-75066-C2-1-P. M.A. thanks Spanish MINECO for funding support through the Ram\'on y Cajal programme (RyC-2014-16277). LVP acknowledges funding support from the Swedish Research Council and the European Research Council (ERC Consolidator Grant 614264).

\end{acknowledgements}

\FloatBarrier
\bibliographystyle{aa} 
\bibliography{mybib} 

\begin{thebibliography}{110}
\expandafter\ifx\csname natexlab\endcsname\relax\def\natexlab#1{#1}\fi

\bibitem[{{Ag\'undez}(2009)}]{agu2009}
{Ag\'undez}, M. 2009, PhD thesis, Universidad Aut\'onoma de Madrid

\bibitem[{{Ag{\'u}ndez} \& {Cernicharo}(2006)}]{agu2006}
{Ag{\'u}ndez}, M. \& {Cernicharo}, J. 2006, \apj, 650, 374

\bibitem[{{Ag{\'u}ndez} {et~al.}(2010){Ag{\'u}ndez}, {Cernicharo}, \&
  {Gu{\'e}lin}}]{agu2010}
{Ag{\'u}ndez}, M., {Cernicharo}, J., \& {Gu{\'e}lin}, M. 2010, \apjl, 724, L133

\bibitem[{{Ag{\'u}ndez} {et~al.}(2012){Ag{\'u}ndez}, {Fonfr{\'{\i}}a},
  {Cernicharo}, {Kahane}, {Daniel}, \& {Gu{\'e}lin}}]{agu2012}
{Ag{\'u}ndez}, M., {Fonfr{\'{\i}}a}, J.~P., {Cernicharo}, J., {et~al.} 2012,
  \aap, 543, A48

\bibitem[{{Ag{\'u}ndez} {et~al.}(2020){Ag{\'u}ndez}, {Mart{\'\i}nez}, {de
  Andres}, {Cernicharo}, \& {Mart{\'\i}n-Gago}}]{agu2020}
{Ag{\'u}ndez}, M., {Mart{\'\i}nez}, J.~I., {de Andres}, P.~L., {Cernicharo},
  J., \& {Mart{\'\i}n-Gago}, J.~A. 2020, \aap, 637, A59

\bibitem[{{Ag{\'u}ndez} {et~al.}(2018){Ag{\'u}ndez}, {Roueff}, {Le Petit}, \&
  {Le Bourlot}}]{agu2018}
{Ag{\'u}ndez}, M., {Roueff}, E., {Le Petit}, F., \& {Le Bourlot}, J. 2018,
  \aap, 616, A19

\bibitem[{{Balan{\c c}a} \& {Dayou}(2017)}]{bal2017}
{Balan{\c c}a}, C. \& {Dayou}, F. 2017, \mnras, 469, 1673

\bibitem[{{Balan{\c{c}}a} {et~al.}(2018){Balan{\c{c}}a}, {Dayou}, {Faure},
  {Wiesenfeld}, \& {Feautrier}}]{bal2018}
{Balan{\c{c}}a}, C., {Dayou}, F., {Faure}, A., {Wiesenfeld}, L., \&
  {Feautrier}, N. 2018, \mnras, 479, 2692

\bibitem[{{Balan{\c{c}}a} {et~al.}(2016){Balan{\c{c}}a}, {Spielfiedel}, \&
  {Feautrier}}]{bal2016}
{Balan{\c{c}}a}, C., {Spielfiedel}, A., \& {Feautrier}, N. 2016, \mnras, 460,
  3766

\bibitem[{{Banerjee} {et~al.}(2012){Banerjee}, {Varricatt}, {Mathew},
  {Launila}, \& {Ashok}}]{ban2012}
{Banerjee}, D.~P.~K., {Varricatt}, W.~P., {Mathew}, B., {Launila}, O., \&
  {Ashok}, N.~M. 2012, \apjl, 753, L20

\bibitem[{Bogey {et~al.}(1997)Bogey, Civi{\v{s}}, Delcroix, Demuynck, Krupnov,
  Quiguer, Tretyakov, \& Walters}]{bog1997}
Bogey, M., Civi{\v{s}}, S., Delcroix, B., {et~al.} 1997, Journal of Molecular
  Spectroscopy, 182, 85

\bibitem[{{Bujarrabal} \& {Alcolea}(2013)}]{buj2013}
{Bujarrabal}, V. \& {Alcolea}, J. 2013, \aap, 552, A116

\bibitem[{{Bujarrabal} {et~al.}(1994){Bujarrabal}, {Fuente}, \&
  {Omont}}]{buj1994}
{Bujarrabal}, V., {Fuente}, A., \& {Omont}, A. 1994, \aap, 285, 247

\bibitem[{{Castro-Carrizo} {et~al.}(2010){Castro-Carrizo}, {Quintana-Lacaci},
  {Neri}, {Bujarrabal}, {Sch{\"o}ier}, {Winters}, {Olofsson}, {Lindqvist},
  {Alcolea}, {Lucas}, \& {Grewing}}]{cas2010}
{Castro-Carrizo}, A., {Quintana-Lacaci}, G., {Neri}, R., {et~al.} 2010, \aap,
  523, A59

\bibitem[{Cernicharo(1985)}]{cer1985}
Cernicharo, J. 1985, IRAM Internal Report 52

\bibitem[{{Cernicharo} {et~al.}(2011){Cernicharo}, {Spielfiedel},
  {Balan{\c{c}}a}, {Dayou}, {Senent}, {Feautrier}, {Faure}, {Cressiot-Vincent},
  {Wiesenfeld}, \& {Pardo}}]{cer2011}
{Cernicharo}, J., {Spielfiedel}, A., {Balan{\c{c}}a}, C., {et~al.} 2011, \aap,
  531, A103

\bibitem[{{Chandra} {et~al.}(1995){Chandra}, {Kegel}, {Le Roy}, \&
  {Hertenstein}}]{cha1995}
{Chandra}, S., {Kegel}, W.~H., {Le Roy}, R.~J., \& {Hertenstein}, T. 1995,
  \aaps, 114, 175

\bibitem[{{Cherchneff}(2006)}]{che2006}
{Cherchneff}, I. 2006, \aap, 456, 1001

\bibitem[{{Chiavassa} {et~al.}(2018){Chiavassa}, {Freytag}, \&
  {Schultheis}}]{chi2018}
{Chiavassa}, A., {Freytag}, B., \& {Schultheis}, M. 2018, \aap, 617, L1

\bibitem[{{Cotton} {et~al.}(2004){Cotton}, {Mennesson}, {Diamond}, {Perrin},
  {Coud{\'e} du Foresto}, {Chagnon}, {van Langevelde}, {Ridgway}, {Waters},
  {Vlemmings}, {Morel}, {Traub}, {Carleton}, \& {Lacasse}}]{cot2004}
{Cotton}, W.~D., {Mennesson}, B., {Diamond}, P.~J., {et~al.} 2004, \aap, 414,
  275

\bibitem[{{Danilovich} {et~al.}(2016){Danilovich}, {De Beck}, {Black},
  {Olofsson}, \& {Justtanont}}]{dan2016}
{Danilovich}, T., {De Beck}, E., {Black}, J.~H., {Olofsson}, H., \&
  {Justtanont}, K. 2016, \aap, 588, A119

\bibitem[{{Danilovich} {et~al.}(2018){Danilovich}, {Ramstedt}, {Gobrecht},
  {Decin}, {De Beck}, \& {Olofsson}}]{dan2018}
{Danilovich}, T., {Ramstedt}, S., {Gobrecht}, D., {et~al.} 2018, \aap, 617,
  A132

\bibitem[{{Danilovich} {et~al.}(2020){Danilovich}, {Richards}, {Decin}, {Van de
  Sand e}, \& {Gottlieb}}]{dan2020}
{Danilovich}, T., {Richards}, A.~M.~S., {Decin}, L., {Van de Sand e}, M., \&
  {Gottlieb}, C.~A. 2020, \mnras, 494, 1323

\bibitem[{{Danilovich} {et~al.}(2019){Danilovich}, {Richards}, {Karakas}, {Van
  de Sande}, {Decin}, \& {De Ceuster}}]{dan2019}
{Danilovich}, T., {Richards}, A.~M.~S., {Karakas}, A.~I., {et~al.} 2019,
  \mnras, 484, 494

\bibitem[{{Danilovich} {et~al.}(2015){Danilovich}, {Teyssier}, {Justtanont},
  {Olofsson}, {Cerrigone}, {Bujarrabal}, {Alcolea}, {Cernicharo},
  {Castro-Carrizo}, {Garc{\'{\i}}a-Lario}, \& {Marston}}]{dan2015}
{Danilovich}, T., {Teyssier}, D., {Justtanont}, K., {et~al.} 2015, \aap, 581,
  A60

\bibitem[{{Danilovich} {et~al.}(2017){Danilovich}, {Van de Sande}, {De Beck},
  {Decin}, {Olofsson}, {Ramstedt}, \& {Millar}}]{dan2017}
{Danilovich}, T., {Van de Sande}, M., {De Beck}, E., {et~al.} 2017, \aap, 606,
  A124

\bibitem[{{Dayou} \& {Balan{\c c}a}(2006)}]{day2006}
{Dayou}, F. \& {Balan{\c c}a}, C. 2006, \aap, 459, 297

\bibitem[{{De Beck} {et~al.}(2010){De Beck}, {Decin}, {de Koter}, {Justtanont},
  {Verhoelst}, {Kemper}, \& {Menten}}]{deb2010}
{De Beck}, E., {Decin}, L., {de Koter}, A., {et~al.} 2010, \aap, 523, A18

\bibitem[{{De Beck} {et~al.}(2017){De Beck}, {Decin}, {Ramstedt}, {Olofsson},
  {Menten}, {Patel}, \& {Vlemmings}}]{deb2017}
{De Beck}, E., {Decin}, L., {Ramstedt}, S., {et~al.} 2017, \aap, 598, A53

\bibitem[{{Decin} {et~al.}(2010){Decin}, {De Beck}, {Br{\"u}nken},
  {M{\"u}ller}, {Menten}, {Kim}, {Willacy}, {de Koter}, \&
  {Wyrowski}}]{dec2010}
{Decin}, L., {De Beck}, E., {Br{\"u}nken}, S., {et~al.} 2010, \aap, 516, A69

\bibitem[{{Decin} {et~al.}(2017){Decin}, {Richards}, {Waters}, {Danilovich},
  {Gobrecht}, {Khouri}, {Homan}, {Bakker}, {Van de Sande}, {Nuth}, \& {De
  Beck}}]{dec2017}
{Decin}, L., {Richards}, A.~M.~S., {Waters}, L.~B.~F.~M., {et~al.} 2017, \aap,
  608, A55

\bibitem[{{Denis-Alpizar} {et~al.}(2018){Denis-Alpizar}, {Stoecklin},
  {Guilloteau}, \& {Dutrey}}]{den2018}
{Denis-Alpizar}, O., {Stoecklin}, T., {Guilloteau}, S., \& {Dutrey}, A. 2018,
  \mnras, 478, 1811

\bibitem[{{Dharmawardena} {et~al.}(2018){Dharmawardena}, {Kemper}, {Scicluna},
  {Wouterloot}, {Trejo}, {Srinivasan}, {Cami}, {Zijlstra}, \&
  {Marshall}}]{dha2018}
{Dharmawardena}, T.~E., {Kemper}, F., {Scicluna}, P., {et~al.} 2018, \mnras,
  479, 536

\bibitem[{{D{\'\i}az-Luis} {et~al.}(2019){D{\'\i}az-Luis}, {Alcolea},
  {Bujarrabal}, {Santand er-Garc{\'\i}a}, {Castro-Carrizo},
  {G{\'o}mez-Garrido}, \& {Desmurs}}]{dia2019}
{D{\'\i}az-Luis}, J.~J., {Alcolea}, J., {Bujarrabal}, V., {et~al.} 2019, \aap,
  629, A94

\bibitem[{{Drira} {et~al.}(1997){Drira}, {Hure}, {Spielfiedel}, {Feautrier}, \&
  {Roueff}}]{dri1997}
{Drira}, I., {Hure}, J.~M., {Spielfiedel}, A., {Feautrier}, N., \& {Roueff}, E.
  1997, \aap, 319, 720

\bibitem[{{Duari} {et~al.}(1999){Duari}, {Cherchneff}, \& {Willacy}}]{dua1999}
{Duari}, D., {Cherchneff}, I., \& {Willacy}, K. 1999, \aap, 341, L47

\bibitem[{{Dyck} {et~al.}(1996){Dyck}, {Benson}, {van Belle}, \&
  {Ridgway}}]{dyc1996}
{Dyck}, H.~M., {Benson}, J.~A., {van Belle}, G.~T., \& {Ridgway}, S.~T. 1996,
  \aj, 111, 1705

\bibitem[{{Engels}(1979)}]{eng1979}
{Engels}, D. 1979, \aaps, 36, 337

\bibitem[{{Gaia Collaboration} {et~al.}(2018){Gaia Collaboration}, {Brown},
  {Vallenari}, {Prusti}, {de Bruijne}, {Babusiaux}, {Bailer-Jones}, {Biermann},
  {Evans}, {Eyer}, \& et~al.}]{gaia2018}
{Gaia Collaboration}, {Brown}, A.~G.~A., {Vallenari}, A., {et~al.} 2018, \aap,
  616, A1

\bibitem[{{Gail} \& {Sedlmayr}(1998)}]{gai1998}
{Gail}, H.~P. \& {Sedlmayr}, E. 1998, The Molecular Astrophysics of Stars and
  Galaxies, 4, 285

\bibitem[{{Gardan} {et~al.}(2006){Gardan}, {G{\'e}rard}, \& {Le
  Bertre}}]{gar2006}
{Gardan}, E., {G{\'e}rard}, E., \& {Le Bertre}, T. 2006, \mnras, 365, 245

\bibitem[{{Gehrz}(1989)}]{geh1989}
{Gehrz}, R. 1989, in IAU Symposium, Vol. 135, Interstellar Dust, ed. L.~J.
  {Allamandola} \& A.~G.~G.~M. {Tielens}, 445

\bibitem[{{Gobrecht} {et~al.}(2016){Gobrecht}, {Cherchneff}, {Sarangi},
  {Plane}, \& {Bromley}}]{gob2016}
{Gobrecht}, D., {Cherchneff}, I., {Sarangi}, A., {Plane}, J.~M.~C., \&
  {Bromley}, S.~T. 2016, \aap, 585, A6

\bibitem[{{Gonz{\'a}lez Delgado} {et~al.}(2003){Gonz{\'a}lez Delgado},
  {Olofsson}, {Kerschbaum}, {Sch{\"o}ier}, {Lindqvist}, \&
  {Groenewegen}}]{gon2003}
{Gonz{\'a}lez Delgado}, D., {Olofsson}, H., {Kerschbaum}, F., {et~al.} 2003,
  \aap, 411, 123

\bibitem[{{Green}(1995)}]{gre1995}
{Green}, S. 1995, \apjs, 100, 213

\bibitem[{{Groenewegen} {et~al.}(1999){Groenewegen}, {Baas}, {Blommaert},
  {Stehle}, {Josselin}, \& {Tilanus}}]{gro1999}
{Groenewegen}, M.~A.~T., {Baas}, F., {Blommaert}, J.~A.~D.~L., {et~al.} 1999,
  \aaps, 140, 197

\bibitem[{{Heays} {et~al.}(2017){Heays}, {Bosman}, \& {van Dishoeck}}]{hea2017}
{Heays}, A.~N., {Bosman}, A.~D., \& {van Dishoeck}, E.~F. 2017, \aap, 602, A105

\bibitem[{{Homan} {et~al.}(2015){Homan}, {Decin}, {de Koter}, {van Marle},
  {Lombaert}, \& {Vlemmings}}]{hom2015}
{Homan}, W., {Decin}, L., {de Koter}, A., {et~al.} 2015, \aap, 579, A118

\bibitem[{{Homan} {et~al.}(2018){Homan}, {Richards}, {Decin}, {de Koter}, \&
  {Kervella}}]{hom2018}
{Homan}, W., {Richards}, A., {Decin}, L., {de Koter}, A., \& {Kervella}, P.
  2018, \aap, 616, A34

\bibitem[{{Iben} \& {Renzini}(1983)}]{ibe1983}
{Iben}, I., J. \& {Renzini}, A. 1983, \araa, 21, 271

\bibitem[{{Justtanont} {et~al.}(1996){Justtanont}, {Skinner}, {Tielens},
  {Meixner}, \& {Baas}}]{jus1996}
{Justtanont}, K., {Skinner}, C.~J., {Tielens}, A.~G.~G.~M., {Meixner}, M., \&
  {Baas}, F. 1996, \apj, 456, 337

\bibitem[{{Kahane} \& {Jura}(1996)}]{kah1996}
{Kahane}, C. \& {Jura}, M. 1996, \aap, 310, 952

\bibitem[{{Kami{\'n}ski} {et~al.}(2013){Kami{\'n}ski}, {Gottlieb}, {Menten},
  {Patel}, {Young}, {Br{\"u}nken}, {M{\"u}ller}, {McCarthy}, {Winters}, \&
  {Decin}}]{kam2013}
{Kami{\'n}ski}, T., {Gottlieb}, C.~A., {Menten}, K.~M., {et~al.} 2013, \aap,
  551, A113

\bibitem[{{Kami{\'n}ski} {et~al.}(2017){Kami{\'n}ski}, {M{\"u}ller}, {Schmidt},
  {Cherchneff}, {Wong}, {Br{\"u}nken}, {Menten}, {Winters}, {Gottlieb}, \&
  {Patel}}]{kam2017}
{Kami{\'n}ski}, T., {M{\"u}ller}, H.~S.~P., {Schmidt}, M.~R., {et~al.} 2017,
  \aap, 599, A59

\bibitem[{{Kami{\'n}ski} {et~al.}(2016){Kami{\'n}ski}, {Wong}, {Schmidt},
  {M{\"u}ller}, {Gottlieb}, {Cherchneff}, {Menten}, {Keller}, {Br{\"u}nken}, \&
  {Winters}}]{kam2016}
{Kami{\'n}ski}, T., {Wong}, K.~T., {Schmidt}, M.~R., {et~al.} 2016, \aap, 592,
  A42

\bibitem[{{Kerschbaum} \& {Olofsson}(1999)}]{ker1999}
{Kerschbaum}, F. \& {Olofsson}, H. 1999, \aaps, 138, 299

\bibitem[{{Khouri} {et~al.}(2014){Khouri}, {de Koter}, {Decin}, {Waters},
  {Lombaert}, {Royer}, {Swinyard}, {Barlow}, {Alcolea}, \&
  {Blommaert}}]{kho2014}
{Khouri}, T., {de Koter}, A., {Decin}, L., {et~al.} 2014, \aap, 561, A5

\bibitem[{{Kim} {et~al.}(2019){Kim}, {Liu}, \& {Taam}}]{kim2019}
{Kim}, H., {Liu}, S.-Y., \& {Taam}, R.~E. 2019, \apjs, 243, 35

\bibitem[{{Kim} \& {Taam}(2012)}]{kim2012}
{Kim}, H. \& {Taam}, R.~E. 2012, \apj, 759, 59

\bibitem[{{K{\l}os} \& {Lique}(2008)}]{klos2008}
{K{\l}os}, J. \& {Lique}, F. 2008, \mnras, 390, 239

\bibitem[{{Knapp} {et~al.}(2003){Knapp}, {Pourbaix}, {Platais}, \&
  {Jorissen}}]{kna2003}
{Knapp}, G.~R., {Pourbaix}, D., {Platais}, I., \& {Jorissen}, A. 2003, \aap,
  403, 993

\bibitem[{{Knapp} {et~al.}(1998){Knapp}, {Young}, {Lee}, \&
  {Jorissen}}]{kna1998}
{Knapp}, G.~R., {Young}, K., {Lee}, E., \& {Jorissen}, A. 1998, \apjs, 117, 209

\bibitem[{Lique {et~al.}(2006)Lique, Dubernet, Spielfiedel, \&
  Feautrier}]{liq2006}
Lique, F., Dubernet, M.-L., Spielfiedel, A., \& Feautrier, N. 2006, Astronomy
  \& Astrophysics, 450, 399

\bibitem[{{Lique} \& {Spielfiedel}(2007)}]{liq2007}
{Lique}, F. \& {Spielfiedel}, A. 2007, \aap, 462, 1179

\bibitem[{{Lique} {et~al.}(2006){Lique}, {Spielfiedel}, \&
  {Cernicharo}}]{liq2006_cs}
{Lique}, F., {Spielfiedel}, A., \& {Cernicharo}, J. 2006, \aap, 451, 1125

\bibitem[{Lique {et~al.}(2005)Lique, Spielfiedel, Dubernet, \&
  Feautrier}]{liq2005}
Lique, F., Spielfiedel, A., Dubernet, M.-L., \& Feautrier, N. 2005, The Journal
  of chemical physics, 123, 134316

\bibitem[{{Lodders} \& {Fegley}(1999)}]{lod1999}
{Lodders}, K. \& {Fegley}, B., J. 1999, in IAU Symposium, Vol. 191, Asymptotic
  Giant Branch Stars, ed. T.~{Le Bertre}, A.~{Lebre}, \& C.~{Waelkens}, 279

\bibitem[{Lovas {et~al.}(1992)Lovas, Suenram, Ogata, \& Yamamoto}]{lov1992}
Lovas, F.~J., Suenram, R., Ogata, T., \& Yamamoto, S. 1992, The Astrophysical
  Journal, 399, 325

\bibitem[{{Maercker} {et~al.}(2016){Maercker}, {Danilovich}, {Olofsson}, {De
  Beck}, {Justtanont}, {Lombaert}, \& {Royer}}]{mae2016}
{Maercker}, M., {Danilovich}, T., {Olofsson}, H., {et~al.} 2016, \aap, 591, A44

\bibitem[{{Massalkhi} {et~al.}(2019){Massalkhi}, {Ag{\'u}ndez}, \&
  {Cernicharo}}]{mas2019}
{Massalkhi}, S., {Ag{\'u}ndez}, M., \& {Cernicharo}, J. 2019, \aap, 628, A62

\bibitem[{{Massalkhi} {et~al.}(2018){Massalkhi}, {Ag{\'u}ndez}, {Cernicharo},
  {Velilla Prieto}, {Goicoechea}, {Quintana-Lacaci}, {Fonfr{\'{\i}}a},
  {Alcolea}, \& {Bujarrabal}}]{mas2018}
{Massalkhi}, S., {Ag{\'u}ndez}, M., {Cernicharo}, J., {et~al.} 2018, \aap, 611,
  A29

\bibitem[{{McDonald} {et~al.}(2018){McDonald}, {De Beck}, {Zijlstra}, \&
  {Lagadec}}]{mcd2018}
{McDonald}, I., {De Beck}, E., {Zijlstra}, A.~A., \& {Lagadec}, E. 2018,
  \mnras, 481, 4984

\bibitem[{M{\"u}ller {et~al.}(2007)M{\"u}ller, McCarthy, Bizzocchi, Gupta,
  Esser, Lichau, Caris, Lewen, Hahn, Degli~Esposti, {et~al.}}]{mul2007}
M{\"u}ller, H., McCarthy, M., Bizzocchi, L., {et~al.} 2007, Physical Chemistry
  Chemical Physics, 9, 1579

\bibitem[{M{\"u}ller \& Br{\"u}nken(2005)}]{mul2005b}
M{\"u}ller, H.~S. \& Br{\"u}nken, S. 2005, Journal of Molecular Spectroscopy,
  232, 213

\bibitem[{M{\"u}ller {et~al.}(2005)M{\"u}ller, Schl{\"o}der, Stutzki, \&
  Winnewisser}]{mul2005a}
M{\"u}ller, H.~S., Schl{\"o}der, F., Stutzki, J., \& Winnewisser, G. 2005,
  Journal of Molecular Structure, 742, 215

\bibitem[{{Nakashima}(2005)}]{nak2005}
{Nakashima}, J.-i. 2005, \apj, 620, 943

\bibitem[{{Neri} {et~al.}(1998){Neri}, {Kahane}, {Lucas}, {Bujarrabal}, \&
  {Loup}}]{ner1998}
{Neri}, R., {Kahane}, C., {Lucas}, R., {Bujarrabal}, V., \& {Loup}, C. 1998,
  \aaps, 130, 1

\bibitem[{{Ohnaka} {et~al.}(2011){Ohnaka}, {Weigelt}, {Millour}, {Hofmann},
  {Driebe}, {Schertl}, {Chelli}, {Massi}, {Petrov}, \& {Stee}}]{ohn2011}
{Ohnaka}, K., {Weigelt}, G., {Millour}, F., {et~al.} 2011, \aap, 529, A163

\bibitem[{{Olofsson} {et~al.}(2002){Olofsson}, {Gonz{\'a}lez Delgado},
  {Kerschbaum}, \& {Sch{\"o}ier}}]{olo2002}
{Olofsson}, H., {Gonz{\'a}lez Delgado}, D., {Kerschbaum}, F., \& {Sch{\"o}ier},
  F.~L. 2002, \aap, 391, 1053

\bibitem[{{Omont} {et~al.}(1993){Omont}, {Lucas}, {Morris}, \&
  {Guilloteau}}]{omo1993}
{Omont}, A., {Lucas}, R., {Morris}, M., \& {Guilloteau}, S. 1993, \aap, 267,
  490

\bibitem[{{Pardo} {et~al.}(2004){Pardo}, {Alcolea}, {Bujarrabal}, {Colomer},
  {del Romero}, \& {de Vicente}}]{par2004}
{Pardo}, J.~R., {Alcolea}, J., {Bujarrabal}, V., {et~al.} 2004, \aap, 424, 145

\bibitem[{Pardo {et~al.}(2001)Pardo, Cernicharo, \& Serabyn}]{par2001}
Pardo, J.~R., Cernicharo, J., \& Serabyn, E. 2001, IEEE Transactions on
  Antennas and Propagation, 49, 1683

\bibitem[{Patel {et~al.}(1979)Patel, Margolese, \& Dykea)}]{pat1979}
Patel, D., Margolese, D., \& Dykea), T. 1979, The Journal of Chemical Physics,
  70, 2740

\bibitem[{{Pattillo} {et~al.}(2018){Pattillo}, {Cieszewski}, {Stancil},
  {Forrey}, {Babb}, {McCann}, \& {McLaughlin}}]{pat2018}
{Pattillo}, R.~J., {Cieszewski}, R., {Stancil}, P.~C., {et~al.} 2018, \apj,
  858, 10

\bibitem[{{Pi{\~n}eiro} {et~al.}(1987){Pi{\~n}eiro}, {Tipping}, \&
  {Chackerian}}]{pin1987}
{Pi{\~n}eiro}, A.~L., {Tipping}, R.~H., \& {Chackerian}, C. 1987, Journal of
  Molecular Spectroscopy, 125, 184

\bibitem[{{Ramstedt} \& {Olofsson}(2014)}]{ram2014}
{Ramstedt}, S. \& {Olofsson}, H. 2014, \aap, 566, A145

\bibitem[{{Ramstedt} {et~al.}(2009){Ramstedt}, {Sch{\"o}ier}, \&
  {Olofsson}}]{ram2009}
{Ramstedt}, S., {Sch{\"o}ier}, F.~L., \& {Olofsson}, H. 2009, \aap, 499, 515

\bibitem[{{Ramstedt} {et~al.}(2008){Ramstedt}, {Sch{\"o}ier}, {Olofsson}, \&
  {Lundgren}}]{ram2008}
{Ramstedt}, S., {Sch{\"o}ier}, F.~L., {Olofsson}, H., \& {Lundgren}, A.~A.
  2008, \aap, 487, 645

\bibitem[{Raymonda {et~al.}(1970)Raymonda, Muenter, \& Klemperer}]{ray1970}
Raymonda, J.~W., Muenter, J.~S., \& Klemperer, W.~A. 1970, The Journal of
  Chemical Physics, 52, 3458

\bibitem[{{Ryde} \& {Sch{\"o}ier}(2001)}]{ryd2001}
{Ryde}, N. \& {Sch{\"o}ier}, F.~L. 2001, \apj, 547, 384

\bibitem[{Sanz {et~al.}(2003)Sanz, McCarthy, \& Thaddeus}]{sanz2003}
Sanz, M.~E., McCarthy, M.~C., \& Thaddeus, P. 2003, The Journal of chemical
  physics, 119, 11715

\bibitem[{{Sch{\"o}ier} {et~al.}(2007){Sch{\"o}ier}, {Bast}, {Olofsson}, \&
  {Lindqvist}}]{sch2007}
{Sch{\"o}ier}, F.~L., {Bast}, J., {Olofsson}, H., \& {Lindqvist}, M. 2007,
  \aap, 473, 871

\bibitem[{{Sch{\"o}ier} {et~al.}(2006{\natexlab{a}}){Sch{\"o}ier}, {Fong},
  {Olofsson}, {Zhang}, \& {Patel}}]{sch2006_irc}
{Sch{\"o}ier}, F.~L., {Fong}, D., {Olofsson}, H., {Zhang}, Q., \& {Patel}, N.
  2006{\natexlab{a}}, \apj, 649, 965

\bibitem[{{Sch{\"o}ier} {et~al.}(2006{\natexlab{b}}){Sch{\"o}ier}, {Olofsson},
  \& {Lundgren}}]{sch2006}
{Sch{\"o}ier}, F.~L., {Olofsson}, H., \& {Lundgren}, A.~A. 2006{\natexlab{b}},
  \aap, 454, 247

\bibitem[{{Sch{\"o}ier} {et~al.}(2004){Sch{\"o}ier}, {Olofsson}, {Wong},
  {Lindqvist}, \& {Kerschbaum}}]{sch2004}
{Sch{\"o}ier}, F.~L., {Olofsson}, H., {Wong}, T., {Lindqvist}, M., \&
  {Kerschbaum}, F. 2004, \aap, 422, 651

\bibitem[{{Sch{\"o}ier} {et~al.}(2013){Sch{\"o}ier}, {Ramstedt}, {Olofsson},
  {Lindqvist}, {Bieging}, \& {Marvel}}]{sch2013}
{Sch{\"o}ier}, F.~L., {Ramstedt}, S., {Olofsson}, H., {et~al.} 2013, \aap, 550,
  A78

\bibitem[{{Sobolev}(1960)}]{sob1960}
{Sobolev}, V.~V. 1960, {Moving envelopes of stars} (Harvard University Press)

\bibitem[{{Suh}(1999)}]{suh1999}
{Suh}, K.-W. 1999, \mnras, 304, 389

\bibitem[{{Teyssier} {et~al.}(2006){Teyssier}, {Hernandez}, {Bujarrabal},
  {Yoshida}, \& {Phillips}}]{tey2006}
{Teyssier}, D., {Hernandez}, R., {Bujarrabal}, V., {Yoshida}, H., \&
  {Phillips}, T.~G. 2006, \aap, 450, 167

\bibitem[{{Tobo{\l}a} {et~al.}(2008){Tobo{\l}a}, {Lique}, {K{\l}os}, \&
  {Cha{\l}asi{\'n}ski}}]{tob2008}
{Tobo{\l}a}, R., {Lique}, F., {K{\l}os}, J., \& {Cha{\l}asi{\'n}ski}, G. 2008,
  Journal of Physics B Atomic Molecular Physics, 41, 155702

\bibitem[{{Van de Sande} {et~al.}(2018){Van de Sande}, {Sundqvist}, {Millar},
  {Keller}, {Homan}, {de Koter}, {Decin}, \& {De Ceuster}}]{van2018}
{Van de Sande}, M., {Sundqvist}, J.~O., {Millar}, T.~J., {et~al.} 2018, \aap,
  616, A106

\bibitem[{{Van de Sande} {et~al.}(2020){Van de Sande}, {Sundqvist}, {Millar},
  {Keller}, {Homan}, {de Koter}, {Decin}, \& {De Ceuster}}]{van2020}
{Van de Sande}, M., {Sundqvist}, J.~O., {Millar}, T.~J., {et~al.} 2020, \aap,
  634, C1

\bibitem[{{Van de Sande} {et~al.}(2019){Van de Sande}, {Walsh}, {Mangan}, \&
  {Decin}}]{van2019}
{Van de Sande}, M., {Walsh}, C., {Mangan}, T.~P., \& {Decin}, L. 2019, \mnras,
  490, 2023

\bibitem[{{Velilla-Prieto} {et~al.}(2019){Velilla-Prieto}, {Cernicharo},
  {Ag{\'u}ndez}, {Fonfr{\'\i}a}, {Quintana-Lacaci}, {Marcelino}, \&
  {Castro-Carrizo}}]{vel2019}
{Velilla-Prieto}, L., {Cernicharo}, J., {Ag{\'u}ndez}, M., {et~al.} 2019, \aap,
  629, A146

\bibitem[{{Velilla Prieto} {et~al.}(2017){Velilla Prieto}, {S{\'a}nchez
  Contreras}, {Cernicharo}, {Ag{\'u}ndez}, {Quintana-Lacaci}, {Bujarrabal},
  {Alcolea}, {Balan{\c c}a}, {Herpin}, {Menten}, \& {Wyrowski}}]{vel2017}
{Velilla Prieto}, L., {S{\'a}nchez Contreras}, C., {Cernicharo}, J., {et~al.}
  2017, \aap, 597, A25

\bibitem[{Winnewisser \& Cook(1968)}]{win1968}
Winnewisser, G. \& Cook, R.~L. 1968, Journal of Molecular Spectroscopy, 28, 266

\bibitem[{{Winters} {et~al.}(2007){Winters}, {Le Bertre}, {Pety}, \&
  {Neri}}]{win2007}
{Winters}, J.~M., {Le Bertre}, T., {Pety}, J., \& {Neri}, R. 2007, \aap, 475,
  559

\bibitem[{{Woodruff} {et~al.}(2004){Woodruff}, {Eberhardt}, {Driebe},
  {Hofmann}, {Ohnaka}, {Richichi}, {Schertl}, {Sch{\"o}ller}, {Scholz}, \&
  {Weigelt}}]{woo2004}
{Woodruff}, H.~C., {Eberhardt}, M., {Driebe}, T., {et~al.} 2004, \aap, 421, 703

\bibitem[{{Yamamura} {et~al.}(1999){Yamamura}, {de Jong}, {Onaka}, {Cami}, \&
  {Waters}}]{yam1999}
{Yamamura}, I., {de Jong}, T., {Onaka}, T., {Cami}, J., \& {Waters},
  L.~B.~F.~M. 1999, \aap, 341, L9

\bibitem[{{Zhang} {et~al.}(2017){Zhang}, {Zheng}, {Reid}, {Honma}, {Menten},
  {Brunthaler}, \& {Kim}}]{zha2017}
{Zhang}, B., {Zheng}, X., {Reid}, M.~J., {et~al.} 2017, \apj, 849, 99

\end{thebibliography}

\appendix
\section{Observed lines}

\longtab[1]{
\begin{longtable}{lccccc}
\caption{Observed line parameters.}\label{table:lines} \\
\hline \hline
\multicolumn{1}{l}{Line} &  & \multicolumn{1}{c}{$\nu_{calc}$} & \multicolumn{1}{c}{$\nu_{obs}$} & \multicolumn{1}{c}{$V_{e}$}      & \multicolumn{1}{c}{$\int T_{\rm mb}\,dv$} \\
& & \multicolumn{1}{c}{(MHz)}           & \multicolumn{1}{c}{(MHz)}          & \multicolumn{1}{c}{(km s$^{-1}$)} & \multicolumn{1}{c}{(K km s$^{-1}$)} \\
\hline
\endfirsthead
\caption{Continued.} \\
\hline
\multicolumn{1}{l}{Line} & & \multicolumn{1}{c}{$\nu_{calc}$} & \multicolumn{1}{c}{$\nu_{obs}$} & \multicolumn{1}{c}{$V_{e}$}  & \multicolumn{1}{c}{$\int T_A^* dv$} \\
& &  \multicolumn{1}{c}{(MHz)}   & \multicolumn{1}{c}{(MHz)}  & \multicolumn{1}{c}{(km s$^{-1}$)} & \multicolumn{1}{c}{(K km s$^{-1}$)} \\
\hline
\endhead
\hline
\endfoot
\multicolumn{6}{c}{IK\,Tau} \\
\hline
SiO &$3-2$                  & 130268.665  & 130268.7(1)   &  17.7(2)    &  55.6(55)  \\
SiS &$8-7$                  & 145227.054  & 145226.9(5)   &  19.5(5)    &  6.1(6) \\
CS &$3-2$                   & 146969.025  & 146969.4(5)   &  17.3(8)    &  3.11(3) \\
SO & $3_{3}-2_{2}$          & 129138.983  & 129139.5(5)   &  16.8(6)    &  2.29(2) \\
SO$_{2}$ &$8_{2-6}-8_{1-7}$ & 134004.811  & 134005.3(5)   &  16.6(5)    &  3.80(4) \\
SO$_{2}$ &$5_{1-5}-4_{0,4}$ & 135696.016  & 135696.7(5)   &  16.8(8)    &  5.20(5) \\
SO$_{2}$ &$4_{2-2}-4_{1,3}$ & 146605.519  & 146606.3(6)   &  16.6(6)    &  2.19(2)   \\
SO $_{2}$&$2_{2-0}-2_{1,1}$ & 151378.662  & 151379.4(5)   &  17.0(6)    &  1.21(1)  \\
\hline
\multicolumn{6}{c}{KU\,And} \\
\hline
SiO &$3-2$                  & 130268.665  & 130268.3(1)  &  20.7(1)     & 5.62(6)   \\
SiS &$8-7$                  & 145227.054  &  145227.1(1) &  19.0(1)     & 1.88(2)    \\
CS &$3-2$                   & 146969.025  & 146968.5(1)  & 19.4(1)     & 1.4(4) \\
SO & $3_{3}-2_{2}$          & 129138.983  & -  & -   & -   \\
SO$_{2}$ &$8_{2-6}-8_{1-7}$ & 134004.811  & - &   -    &  - \\
SO$_{2}$ &$5_{1-5}-4_{0,4}$ & 135696.016  & - &  -   &  -   \\
SO$_{2}$ &$4_{2-2}-4_{1,3}$ & 146605.519  &   &   &  \\
SO $_{2}$&$2_{2-0}-2_{1,1}$ & 151378.662  & - &-  & - \\ 
\hline
\multicolumn{6}{c}{RX\,Boo} \\
\hline
SiO &$3-2$                  & 130268.665  & 130268.7(1)  &  7.8(1)     & 26.0(26)    \\
SiS &$8-7$                  & 145227.054  &  - &   -    & -    \\
CS &$3-2$                   & 146969.025  & - & - & - \\
SO & $3_{3}-2_{2}$          & 129138.983  & 129138.9(1)  &  7.8(1)   & 0.88(1)   \\
SO$_{2}$ &$8_{2-6}-8_{1-7}$ & 134004.811  & 134005.0(5)  & 7.9(5)    &  0.83(8)  \\
SO$_{2}$ &$5_{1-5}-4_{0,4}$ & 135696.016  & 135695.9(1)  &  6.5(1)   &  0.37(4)   \\
SO$_{2}$ &$4_{2-2}-4_{1,3}$ & 146605.519  & 146606.4(5)  & 6.3(4)    & 0.23(5)$^a$   \\
SO $_{2}$&$2_{2-0}-2_{1,1}$ & 151378.662  & 151377.8(10) & 9.0(10)   & 0.12(2)$^a$ \\ 
\hline
\multicolumn{6}{c}{RT\,Vir} \\
\hline
SiO &$3-2$                  & 130268.665  & 130268.7(1) & 7.1(1) & 6.56(6)\\
SiS &$8-7$                  & 145227.054  & - & -   & - \\
CS &$3-2$                   & 146969.025  & -  &  -  & -\\
SO & $3_{3}-2_{2}$          & 129138.983  & 129139.2(1)  & 6.9(1)   & 0.64(1) \\
SO$_{2}$ &$8_{2-6}-8_{1-7}$ & 134004.811  & 134004.9(5)  & 6.0(4)   &  0.67(3)\\
SO$_{2}$ &$5_{1-5}-4_{0,4}$ & 135696.016  & 135696.0(5)  & 5.8(6)  &  0.35(3)\\
SO$_{2}$ &$4_{2-2}-4_{1,3}$ & 146605.519  & 146605.5(5)  & 5.5(6)   & 0.24(2) \\
SO $_{2}$&$2_{2-0}-2_{1,1}$ & 151378.662  & 151379.3(10) & 4.5(10)  & 0.09(2)$^a$  \\
\hline
\multicolumn{6}{c}{R\,Leo} \\
\hline
SiO &$3-2$                  & 130268.665  &130268.6(5)  & 5.1(5) & 15.09(15)\\
SiS &$8-7$                  & 145227.054  & -  & - & -\\ 
CS &$3-2$                   & 146969.025  & 146968.5(5) & 4.8(5) & 0.10(1)$^a$ \\
SO & $3_{3}-2_{2}$          & 129138.983  & 129138.9(1) & 4.5(3) & 0.27(3) \\
SO$_{2}$ &$8_{2-6}-8_{1-7}$ & 134004.811  & - & - & - \\
SO$_{2}$ &$5_{1-5}-4_{0,4}$ & 135696.016  & - &- &- \\
SO$_{2}$ &$4_{2-2}-4_{1,3}$ & 146605.519  & -  & - & - \\
SO $_{2}$&$2_{2-0}-2_{1,1}$ & 151378.662  & - & - & - \\
\hline
\multicolumn{6}{c}{WX\,Psc} \\
\hline
SiO &$3-2$                  & 130268.665  & 130268.7(1)   & 18.8(4)   & 29.0(30) \\
SiS &$8-7$                  & 145227.054  & 145227.0(1)   & 19.4(4)   & 26.2(26) \\
CS &$3-2$                   & 146969.025  & 146969.0(1)   & 17.8(6)   & 2.41(2)\\
SO & $3_{3}-2_{2}$          & 129138.983  & 129139.2(1)   & 18.5(3)   & 1.11(1) \\
SO$_{2}$ &$8_{2-6}-8_{1-7}$ & 134004.811  & 134005.2(5)   & 17.7(5)   & 1.7(2) \\
SO$_{2}$ &$5_{1-5}-4_{0,4}$ & 135696.016  & 135696.2(1)   & 18.3(4)   & 6.81(7)\\
SO$_{2}$ &$4_{2-2}-4_{1,3}$ & 146605.519  & 146605.8(1)   & 18.4(2)   & 2.02(2) \\
SO $_{2}$&$2_{2-0}-2_{1,1}$ & 151378.662  & 151378.8(1)   & 18.0(6)   & 1.40(1) \\
\hline
\multicolumn{6}{c}{GX\,Mon} \\
\hline
SiO &$3-2$                  & 130268.665  & 130268.6(1)   & 18.4(1)    & 25.0(25)   \\
SiS &$8-7$                  & 145227.054  & 145226.8(4)   & 18.2(6)    & 2.10(20)\\
CS &$3-2$                   & 146969.025  & 146968.9(5)   & 17.4(8)    & 1.51(1) \\
SO & $3_{3}-2_{2}$          & 129138.983  & 129138.9(1)   &  17.6(6)   & 1.39(1)  \\
SO$_{2}$ &$8_{2-6}-8_{1-7}$ & 134004.811  & 134005.2(5)   & 16.6(6)    & 1.2(1)  \\
SO$_{2}$ &$5_{1-5}-4_{0,4}$ & 135696.016  &135696.2(5)    & 18.1(5)    & 5.79(8) \\
SO$_{2}$ &$4_{2-2}-4_{1,3}$ & 146605.519  & 146605.4(5)   &  18.5(6)   & 1.54(1)  \\
SO $_{2}$&$2_{2-0}-2_{1,1}$& 151378.662  &  151379.1(5)   &  18.0(4)   & 1.28(1)   \\
\hline
\multicolumn{6}{c}{NV\,Aur} \\
\hline
SiO &$3-2$                  & 130268.665  & 130268.5(1)     &  17.7(1)    & 13.7(14)  \\
SiS &$8-7$                  & 145227.054  & 145226.9(5)     &  17.1(4)    & 2.98(3)   \\
CS &$3-2$                   & 146969.025  & 146968.5(5)     &  16.3(4)    & 0.77(8) \\
SO & $3_{3}-2_{2}$          & 129138.983  & 129138.7(5)     &  16.4(4)    & 0.67(7)  \\
SO$_{2}$ &$8_{2-6}-8_{1-7}$ & 134004.811  & 134004.9(5)     &  17.0(4)    & 1.4(1)  \\
SO$_{2}$ &$5_{1-5}-4_{0,4}$ & 135696.016  & 135696.1(1)     &  17.1(1)    & 3.57(3)  \\
SO$_{2}$ &$4_{2-2}-4_{1,3}$ & 146605.519  & 146605.6(5)     &  17.0(8)    & 1.14(1)    \\
SO $_{2}$&$2_{2-0}-2_{1,1}$ & 151378.662  & 151379.0(6)     &  17.3(5)   & 0.76(7)    \\
\hline
\multicolumn{6}{c}{V1111\,Oph} \\
\hline
SiO &$3-2$                  & 130268.665  & 130268.6(6) &  15.7(1)   & 21.0(20) \\
SiS &$8-7$                  & 145227.054  & 145227.2(5) &  14.2(7)   & 2.17(2) \\
CS &$3-2$                   & 146969.025  & 146969.4(1) &  14.2(1)   & 1.18(1)\\
SO & $3_{3}-2_{2}$          & 129138.983  & 129138.9(5) &  15.2(3)   & 0.87(8) \\
SO$_{2}$ &$8_{2-6}-8_{1-7}$ & 134004.811  & 134005.0(1) &  14.5(1)   & 1.2(3) \\
SO$_{2}$ &$5_{1-5}-4_{0,4}$ & 135696.016  & 135696.1(5) &  15.9(5)   & 3.89(4)\\
SO$_{2}$ &$4_{2-2}-4_{1,3}$ & 146605.519  & 146605.4(5) &  15.5(6)   & 1.27(1) \\
SO $_{2}$&$2_{2-0}-2_{1,1}$& 151378.662   & 151378.7(5) &  15.3(5)   & 0.77(7) \\
\hline
\multicolumn{6}{c}{RR\,Aql} \\
\hline
SiO &$3-2$                  & 130268.665  & 130268.4(1) & 6.0(10) & 4.86(5) \\
SiS &$8-7$                  & 145227.054  & - & -   & - \\
CS &$3-2$                   & 146969.025  &  - & -  & - \\
SO & $3_{3}-2_{2}$          & 129138.983  & 129138.9(5)  & 8.5(6)    & 0.40(4) \\
SO$_{2}$ &$8_{2-6}-8_{1-7}$ & 134004.811  & 134004.3(1)  & 7.8(1)  & 0.36(3) \\
SO$_{2}$ &$5_{1-5}-4_{0,4}$ & 135696.016  & 135696.2(1)  & 8.0(1) & 1.17(1) \\
SO$_{2}$ &$4_{2-2}-4_{1,3}$ & 146605.519  & 146605.5(1)  & 8.0(1)  & 0.58(1) \\
SO $_{2}$&$2_{2-0}-2_{1,1}$ & 151378.662   & 151379.0(5) & 7.5(4)  & 0.31(3)  \\
\hline
\multicolumn{6}{c}{R\,LMi} \\
\hline
SiO &$3-2$                  & 130268.665 & 130268.6(5)  &  5.8(5)     & 6.23(6)   \\
SiS &$8-7$                  & 145227.054  & - & -     & -    \\
CS &$3-2$                   & 146969.025 & - & -     & - \\
SO & $3_{3}-2_{2}$          & 129138.983  & 129139.1(5)   & 5.5(5)   &  0.18(2)  \\
SO$_{2}$ &$8_{2-6}-8_{1-7}$ & 134004.811  & 134004.9(5) & 4.8(5)     & 0.21(2)  \\
SO$_{2}$ &$5_{1-5}-4_{0,4}$ & 135696.016 & 135694.6(10) & 4.6(10)    & 0.09(2)$^a$   \\
SO$_{2}$ &$4_{2-2}-4_{1,3}$ & 146605.519  & 146606.1(5)  & 5.2(5)  & 0.05(1)$^a$\\
SO $_{2}$&$2_{2-0}-2_{1,1}$ & 151378.662  & - &-  & - \\ 
\hline
\multicolumn{6}{c}{BX\,Cam} \\
\hline
SiO &$3-2$                  & 130268.665   & 130268.6(1)  & 17.3(1)     &  13.2(13)   \\
SiS &$8-7$                  & 145227.054   & 145226.7(5)  & 15.4(6)     &  1.06(3)    \\
CS &$3-2$                   & 146969.025   & 146969.0(5)  & 17.6(5)     &  0.74(7) \\
SO & $3_{3}-2_{2}$          & 129138.983   & 129138.7(5)  & 15.4(5)     &  0.84(17)$^a$  \\
SO$_{2}$ &$8_{2-6}-8_{1-7}$ & 134004.811   & 134004.5(5)  & 14.9(8)      & 0.72(14)$^a$  \\
SO$_{2}$ &$5_{1-5}-4_{0,4}$ & 135696.016   & 135696.3(1)  & 17.0(1)     & 1.47(1)  \\
SO$_{2}$ &$4_{2-2}-4_{1,3}$ & 146605.519   & 146606.0(1)  & 14.9(1)     & 0.49(5)$^a$\\
SO $_{2}$&$2_{2-0}-2_{1,1}$ & 151378.662   & 151378.9(10) & 14.8(10)    & 0.25(5)$^a$ \\ 
\hline
\multicolumn{6}{c}{V1300\,Aql} \\
\hline
SiO &$3-2$                  & 130268.665  & 130268.5(1)  &  14.1(1)   & 7.51(7) \\
SiS &$8-7$                  & 145227.054  & 145227.3(1)  &  14.7(5)  & 10.9(11) \\
CS &$3-2$                   & 146969.025  & 146969.3(5)  &  14.0(5)  & 0.91(18) \\
SO & $3_{3}-2_{2}$          & 129138.983  & 129138.9(4)  &  13.6(4)  & 1.19(1) \\
SO$_{2}$ &$8_{2-6}-8_{1-7}$ & 134004.811  & 134004.8(6)  &  13.0(5)  & 1.16(11) \\
SO$_{2}$ &$5_{1-5}-4_{0,4}$ & 135696.016  & 135696.1(2)  &  13.7(5)  & 4.02(4)  \\
SO$_{2}$ &$4_{2-2}-4_{1,3}$ & 146605.519  & 146605.6(5)  &  13.1(8)  & 1.91(2)\\
SO $_{2}$&$2_{2-0}-2_{1,1}$& 151378.662   & 151378.9(5)  &  13.2(5)  & 1.14(1)  \\
\hline
\multicolumn{6}{c}{R\,Cas} \\
\hline
SiO &$3-2$                  & 130268.665  &  130268.7(5) &  8.3(5)     & 31.9(32)    \\
SiS &$8-7$                  & 145227.054  & 145226.2(10)  & 6.4(10)     & 0.18(4)$^a$  \\
CS &$3-2$                   & 146969.025  & 146969.2(5)  &  6.8(5)     & 0.32(6) \\
SO & $3_{3}-2_{2}$          & 129138.983  & 129138.8(1)  &  7.0(1)     & 0.98(1)   \\
SO$_{2}$ &$8_{2-6}-8_{1-7}$ & 134004.811  & 134004.9(5)  &  6.8(5)     &  0.87(9)  \\
SO$_{2}$ &$5_{1-5}-4_{0,4}$ & 135696.016  & 135696.1(6)  &  8.2(8)     &  0.67(7)   \\
SO$_{2}$ &$4_{2-2}-4_{1,3}$ & 146605.519  & 146606.0(5)  &  7.3(5)     & 0.30(3)  \\
SO $_{2}$&$2_{2-0}-2_{1,1}$ & 151378.662  & 151378.0(10) &  6.7(10)    & 0.18(4)$^a$ \\ 
\hline
\multicolumn{6}{c}{IRC\,-30398} \\
\hline
SiO &$3-2$                  & 130268.665  & 130268.6(1)   &  14.8(5)     & 6.90(7)   \\
SiS &$8-7$                  & 145227.054  & 145227.2(5) &   13.7(5)    & 0.34(7)   \\
CS &$3-2$                   & 146969.025  & 146969.1(5) & 14.0(5) & 0.59(6) \\
SO & $3_{3}-2_{2}$          & 129138.983  & - & -   & - \\
SO$_{2}$ &$8_{2-6}-8_{1-7}$ & 134004.811  & - & -     & -  \\
SO$_{2}$ &$5_{1-5}-4_{0,4}$ & 135696.016  & - &  -  &  -  \\
SO$_{2}$ &$4_{2-2}-4_{1,3}$ & 146605.519  & -  & -  &- \\
SO $_{2}$&$2_{2-0}-2_{1,1}$ & 151378.662  & - &-  & - \\ 
\hline
\multicolumn{6}{c}{TX\,Cam} \\
\hline
SiO &$3-2$                  & 130268.665  & 130268.8(5)  & 17.7(7)  & 36.4(36)\\
SiS &$8-7$                  & 145227.054  & 145226.7(1)  & 17.1(1)   & 7.6(7) \\
CS &$3-2$                   & 146969.025  & 146968.9(1)  & 19.4 (1)  & 8.8(9)\\
SO & $3_{3}-2_{2}$          & 129138.983  & 129139.5(1)  & 16.7(1)   & 1.09(1) \\
SO$_{2}$ &$8_{2-6}-8_{1-7}$ & 134004.811  & 134005.5(10) & 18.6(10)  & 0.60(12) $^a$\\
SO$_{2}$ &$5_{1-5}-4_{0,4}$ & 135696.016  & 135696.5(5)  & 17.9(5)   & 2.29(2) \\
SO$_{2}$ &$4_{2-2}-4_{1,3}$ & 146605.519  & 146606.6(5)  & 19.5(5)  & 0.60(6) \\
SO$_{2}$&$2_{2-0}-2_{1,1}$ & 151378.662   & 151379.1(10) & 19.2(10)  &  0.39(8)$^a$ \\
\hline
\multicolumn{6}{c}{S\,CrB} \\
\hline
SiO &$3-2$                  & 130268.665  & 130268.7(1)  &   4.9(1)    & 2.74(30)   \\
SiS &$8-7$                  & 145227.054  & - &   -    & -    \\
CS &$3-2$                   & 146969.025  & - & - & - \\
SO & $3_{3}-2_{2}$          & 129138.983  & 129139.0(5)  & 4.5(5)   & 0.17(4)$^a$   \\
SO$_{2}$ &$8_{2-6}-8_{1-7}$ & 134004.811  & 134005.6(5) &   4.4(5)  &  0.30(3)$^a$ \\
SO$_{2}$ &$5_{1-5}-4_{0,4}$ & 135696.016  & 135696.4(5) &  5.4(7)   &  0.31(3)$^a$   \\
SO$_{2}$ &$4_{2-2}-4_{1,3}$ & 146605.519  & 146606.7(5)  & 4.7(5)  & 0.18(2)$^a$ \\
SO $_{2}$&$2_{2-0}-2_{1,1}$ & 151378.662  & 151378.5(10) &4.2(10)  & 0.02(1)$^a$  \\ 
\hline
\multicolumn{6}{c}{IRC\,+60169} \\
\hline
SiO &$3-2$                  & 130268.665  & 130268.8(1)  &  13.2(1)   &  3.43(3)  \\
SiS &$8-7$                  & 145227.054  & 145227.8(10) & 15.6(10)  &  0.15(3)$^a$  \\
CS &$3-2$                   & 146969.025 & 146969.5(10)  & 11.3(10)   & 0.11(2)$^a$ \\
SO & $3_{3}-2_{2}$          & 129138.983  & 129138.5(10) & 15.3(10)   & 0.17(3)$^a$  \\
SO$_{2}$ &$8_{2-6}-8_{1-7}$ & 134004.811  & 134004.2(10) & 15.0(10)   & 0.10(2)$^a$ \\
SO$_{2}$ &$5_{1-5}-4_{0,4}$ & 135696.016 & 135696.7(5) &  15.1(1)    &  0.77(7) \\
SO$_{2}$ &$4_{2-2}-4_{1,3}$ & 146605.519  & 146604.5(10)  & 15.0(10) & 0.18(21)$^a$\\
SO $_{2}$&$2_{2-0}-2_{1,1}$ & 151378.662  & 151378.5(10) & 15.3(10) & 0.20(4)$^a$  \\ 
\hline
\multicolumn{6}{c}{R\,Hya} \\
\hline
SiO &$3-2$                  & 130268.665  & 130268.5(5)    & 4.9(5)  & 14.0(14)  \\
SiS &$8-7$                  & 145227.054  &   -   &  -     &  -      \\
CS &$3-2$                   & 146969.025  &   -   &  -     &         \\
SO & $3_{3}-2_{2}$          & 129138.983  &  129140.0(10)  &  3.9(10)  &  0.11(2)$^a$ \\
SO$_{2}$ &$8_{2-6}-8_{1-7}$ & 134004.811  &   -   &  -     &  -        \\
SO$_{2}$ &$5_{1-5}-4_{0,4}$ & 135696.016  &   -   &  -     &  -       \\
SO$_{2}$ &$4_{2-2}-4_{1,3}$ & 146605.519  &   -   &  -     &  -       \\
SO $_{2}$&$2_{2-0}-2_{1,1}$ & 151378.662  &   -   &  -     &  -        \\
\hline
\multicolumn{6}{c}{R\,CrT} \\
\hline
SiO &$3-2$                  & 130268.665  & 130268.5(1) & 10.4(1) & 22.3(20)\\
SiS &$8-7$                  & 145227.054  & - & - & - \\
CS &$3-2$                   & 146969.025  & - & - & - \\ 
SO & $3_{3}-2_{2}$          & 129138.983  & 129139.5(5) & 8.5(4) & 0.82(1) \\
SO$_{2}$ &$8_{2-6}-8_{1-7}$ & 134004.811  & 134004.5(5) & 10.6(6) & 1.3(1)  \\
SO$_{2}$ &$5_{1-5}-4_{0,4}$ & 135696.016  & 135696.4(1) & 9.6(3) & 0.73(7) \\
SO$_{2}$ &$4_{2-2}-4_{1,3}$ & 146605.519  & 146605.2(5) & 10.4(6) & 0.49(9)$^a$\\
SO $_{2}$&$2_{2-0}-2_{1,1}$ & 151378.662  & 151379.0(5)& 10.3(5) & 0.20(4)$^a$ \\
\hline
\multicolumn{6}{c}{O\,Ceti} \\
\hline
SiO &$3-2$                  & 130268.665   & 130268.6(5)  & 3.4(10)  &  1.62(16)   \\
SiS &$8-7$                  & 145227.054   & -            &   -      &  -   \\
CS &$3-2$                   & 146969.025   & 146969.4(5)  &  5.3     &  0.17(3)$^a$ \\
SO & $3_{3}-2_{2}$          & 129138.983   & 129138.7(5)  &  5.1(4)  & 0.22(2)    \\
SO$_{2}$ &$8_{2-6}-8_{1-7}$ & 134004.811   & 134005.3(10) &  4.7(10) & 0.21(4) $^a$ \\
SO$_{2}$ &$5_{1-5}-4_{0,4}$ & 135696.016   & 135696.6(5)  &  2.7(5)  &  0.04(1)$^a$    \\
SO$_{2}$ &$4_{2-2}-4_{1,3}$ & 146605.519   & 146605.6(5)  &  3.7(7)  &  0.05(1)$^a$  \\
SO $_{2}$&$2_{2-0}-2_{1,1}$ & 151378.662  & 151378.7(10)  &  1.5(10) &  0.020(4)$^a$\\ 
\hline
\multicolumn{6}{c}{W\,Hya} \\
\hline
SiO &$3-2$                  & 130268.665  & 130268.6(1)  & 6.3(1) & 26.6(26) \\
SiS &$8-7$                  & 145227.054  & - & - & - \\
CS &$3-2$                   & 146969.025  & 146969.7(10)  & 6.1(10) & 0.31(6)$^a$\\
SO & $3_{3}-2_{2}$          & 129138.983  & 129139.3(5)  & 4.7(1) & 0.33(3)\\
SO$_{2}$ &$8_{2-6}-8_{1-7}$ & 134004.811  & - &- & - \\
SO$_{2}$ &$5_{1-5}-4_{0,4}$ & 135696.016  & - & - &  -\\
SO$_{2}$ &$4_{2-2}-4_{1,3}$ & 146605.519  & - & - & -\\
SO $_{2}$&$2_{2-0}-2_{1,1}$ & 151378.662  & - & - & - \\
\hline
\multicolumn{6}{c}{T\,Cep} \\
\hline
SiO &$3-2$                  & 130268.665  & 130268.7(1)  &  3.8(1)  &  3.20(3)  \\
SiS &$8-7$                  & 145227.054  & - & -     & -    \\
CS &$3-2$                   & 146969.025 & - & -     & - \\
SO & $3_{3}-2_{2}$          & 129138.983  &  - &  -  & -   \\
SO$_{2}$ &$8_{2-6}-8_{1-7}$ & 134004.811  & - & -     & -  \\
SO$_{2}$ &$5_{1-5}-4_{0,4}$ & 135696.016 & - & -     & -   \\
SO$_{2}$ &$4_{2-2}-4_{1,3}$ & 146605.519  &   &   & \\
SO $_{2}$&$2_{2-0}-2_{1,1}$ & 151378.662  & - &-  & - \\ 
\hline
\multicolumn{6}{c}{V1943\,Sgr} \\
\hline
SiO &$3-2$                  & 130268.665  & 130268.6(1) & 4.5(5) & 5.07(5)\\
SiS &$8-7$                  & 145227.054  & - & - & - \\
CS &$3-2$                   & 146969.025  & - & - & - \\ 
SO & $3_{3}-2_{2}$          & 129138.983  & 129138.9(10) & 4.2(10)  & 0.20(4) \\
SO$_{2}$ &$8_{2-6}-8_{1-7}$ & 134004.811  & - & - & - \\
SO$_{2}$ &$5_{1-5}-4_{0,4}$ & 135696.016  & - &- & - \\
SO$_{2}$ &$4_{2-2}-4_{1,3}$ & 146605.519  & - & -  & \\
SO $_{2}$&$2_{2-0}-2_{1,1}$ & 151378.662  & - & - & - \\
\hline
\multicolumn{6}{c}{SW\,Vir} \\
\hline
SiO &$3-2$                  & 130268.665  & 130268.6(1) & 7.5(1)  & 17.4(17) \\
SiS &$8-7$                  & 145227.054  & - & - & -\\
CS &$3-2$                   & 146969.025  & 146969.0(10) & 7.5(8) & 0.33(6)$^a$\\
SO & $3_{3}-2_{2}$          & 129138.983  & 129139.0(5) & 7.1(6) & 0.57(11) \\
SO$_{2}$ &$8_{2-6}-8_{1-7}$ & 134004.811  & - & - & - \\
SO$_{2}$ &$5_{1-5}-4_{0,4}$ & 135696.016  & - & - & -\\
SO$_{2}$ &$4_{2-2}-4_{1,3}$ & 146605.519  & - & - & -\\
SO $_{2}$&$2_{2-0}-2_{1,1}$ & 151378.662  & - & - & - \\
\hline
\multicolumn{6}{c}{AFGL\,292} \\
\hline
SiO &$3-2$                  & 130268.665  & 130268.6(1)     &  6.9(6)     & 3.69(4)  \\
SiS &$8-7$                  & 145227.054  &  -    &   -    &    -    \\
CS &$3-2$                   & 146969.025  & -  &  -   & - \\
SO & $3_{3}-2_{2}$          & 129138.983  & 129138.7(10)    & 5.5(7) & 0.09(2)$^a$    \\
SO$_{2}$ &$8_{2-6}-8_{1-7}$ & 134004.811  &  -    &   -   &   -       \\
SO$_{2}$ &$5_{1-5}-4_{0,4}$ & 135696.016  &  -    &  -     &   -      \\
SO$_{2}$ &$4_{2-2}-4_{1,3}$ & 146605.519  &  -    &   -    &   -      \\
SO $_{2}$&$2_{2-0}-2_{1,1}$ & 151378.662  &   -   &  -     &     -     \\
\hline
\multicolumn{6}{c}{BK\,Vir} \\
\hline
SiO &$3-2$                  & 130268.665  &  130268.6(1)  & 4.2(8)  & 3.3(3) \\
SiS &$8-7$                  & 145227.054  & - &  -  &  -\\
CS &$3-2$                   & 146969.025  &  - & -  & - \\
SO & $3_{3}-2_{2}$          & 129138.983  & -  &  -   &  - \\
SO$_{2}$ &$8_{2-6}-8_{1-7}$ & 134004.811  & -  & -  & -  \\
SO$_{2}$ &$5_{1-5}-4_{0,4}$ & 135696.016  & -  & -  & - \\
SO$_{2}$ &$4_{2-2}-4_{1,3}$ & 146605.519  & -  & -  & - \\
SO $_{2}$&$2_{2-0}-2_{1,1}$ & 151378.662  & - & - & -  \\
\hline
\multicolumn{6}{c}{OH\,26.5+0.6} \\
\hline
SiO &$3-2$                  & 130268.665  & 130268.6(1)  &  14.2(1)     &   6.55(6) \\
SiS &$8-7$                  & 145227.054  & 145227.0(1)  &  12.6(1)     &   1.54(1)  \\
CS &$3-2$                   & 146969.025  & 146968.5(5)  &  15.6(5)      &  0.96(9)     \\
SO & $3_{3}-2_{2}$          & 129138.983  & 129138.6(1)  &  13.1(3)     &   6.29(3) \\
SO$_{2}$ &$8_{2-6}-8_{1-7}$ & 134004.811  & 134005.5(1)  &  14.2(1)     &  4.15(4) \\
SO$_{2}$ &$5_{1-5}-4_{0,4}$ & 135696.016  & 135696.1(1)  &  14.3(4)     &  8.53(8)  \\
SO$_{2}$ &$4_{2-2}-4_{1,3}$ & 146605.519  & 146605.4(4)  &  14.4(7)     &  3.79(4)    \\
SO $_{2}$&$2_{2-0}-2_{1,1}$ & 151378.662  & 151378.6(5)  &  14.1(4)     &  2.71(3)\\ 
\hline
\multicolumn{6}{c}{Ep\,Aqr} \\
\hline
SiO &$3-2$                  & 130268.665  & 130268.0(5)    &  8.1(10)   &  22.5(22)      \\
SiS &$8-7$                  & 145227.054  &   -   &  -     &   -     \\
CS &$3-2$                   & 146969.025  & 146969.0(10)   &   0.9(5)    &    0.030(6)$^a$ \\
SO & $3_{3}-2_{2}$          & 129138.983  & 129138.8(5)    &  3.0(10)     &  1.16(23)   \\
SO$_{2}$ &$8_{2-6}-8_{1-7}$ & 134004.811  & 134004.8(5)    &  2.2(10)   &   0.38(8)  \\
SO$_{2}$ &$5_{1-5}-4_{0,4}$ & 135696.016  & 135696.0(5)    &  2.3(10)     &  0.13(1)  \\
SO$_{2}$ &$4_{2-2}-4_{1,3}$ & 146605.519  & 146605.4(5)   &  3.2(10)     &  0.15(2) \\
SO $_{2}$&$2_{2-0}-2_{1,1}$ & 151378.662  & 151378.6(5)   &  3.7(10)     &  0.08(2)  \\    
\hline
\multicolumn{6}{c}{X\,Her} \\
\hline
SiO &$3-2$                  & 130268.665  & 130268.4(5)  & 6.5(5)   & 10.2(10) \\
SiS &$8-7$                  & 145227.054  & -  &  -  & - \\
CS &$3-2$                   & 146969.025  & -  &  -  &  - \\
SO & $3_{3}-2_{2}$          & 129138.983  & 129138.8(5)   & 2.8(5)   & 0.52(5) \\
SO$_{2}$ &$8_{2-6}-8_{1-7}$ & 134004.811  & -   & -   & - \\
SO$_{2}$ &$5_{1-5}-4_{0,4}$ & 135696.016  & -  & -  & - \\
SO$_{2}$ &$4_{2-2}-4_{1,3}$ & 146605.519  & -  &  - & - \\
SO $_{2}$&$2_{2-0}-2_{1,1}$ & 151378.662  & -  &  -  & -  \\
\hline
\end{longtable}
\tablefoot{
Numbers in parentheses are 1$\sigma$ uncertainties in units of the last digits. \\
$^a$ Marginal detection.}}
\end{document}